\documentclass[final,onecolumn]{IEEEtran}

\usepackage{psfrag,epsfig,graphics}
\usepackage{amsmath,amssymb,multirow,etoolbox}
\usepackage{subfigure}
\usepackage[below]{placeins}
\usepackage{cite}
\usepackage[ruled,vlined]{algorithm2e}

\usepackage{relsize}

\usepackage{color}  

\newcommand{\cb}[1]{{\boldsymbol{#1}}}
\newcommand{\cp}[1]{\ifmmode {\mathcal{#1}}\else ${\mathcal{#1}}$\fi}

\newcommand{\beps}{\boldsymbol{\epsilon}}
\newcommand{\bpsi}{\boldsymbol{\psi}}
\newcommand{\bphi}{\boldsymbol{\phi}}

\newcommand{\f}{\boldsymbol{f}}
\newcommand{\ba}{\boldsymbol{a}}
\newcommand{\bu}{\boldsymbol{u}}
\newcommand{\bp}{\boldsymbol{p}}
\newcommand{\bh}{\boldsymbol{h}}
\newcommand{\bg}{\boldsymbol{g}}
\newcommand{\br}{\boldsymbol{r}}

\newcommand{\bw}{\boldsymbol{w}}
\newcommand{\bx}{\boldsymbol{x}}

\newcommand{\bv}{\boldsymbol{v}}

\newcommand{\bB}{\boldsymbol{B}}
\newcommand{\bC}{\boldsymbol{C}}
\newcommand{\bK}{\boldsymbol{K}}

\newcommand{\bG}{\boldsymbol{G}}

\newcommand{\bH}{\boldsymbol{H}}

\newcommand{\bA}{\boldsymbol{A}}
\newcommand{\cAT}[1]{\boldsymbol{\cal{A}}_{#1}^{\top}}
\newcommand{\cA}{\boldsymbol{\cal{A}}}

\newcommand{\cC}{\boldsymbol{\cal{C}}}
\newcommand{\cCT}{\boldsymbol{\cal{C}}^\top}

\newcommand{\bR}{\boldsymbol{R}}

\newcommand{\bT}{\boldsymbol{T}}
\newcommand{\bU}{\boldsymbol{U}}

\newcommand{\bSig}{\boldsymbol{\Sigma}}
\newcommand{\bsig}{\boldsymbol{\sigma}}

\newcommand{\bGam}{\boldsymbol{\Gamma}}
\newcommand{\bI}{\boldsymbol{I}}

\newcommand{\N}[1]{\cp{N}_{#1}}

\newcommand{\tr}{\text{trace}}
\newcommand{\vc}{\text{vec}}
\newcommand{\col}{\text{col}}

\newtheorem{theorem}{Theorem}

\newtheorem{assumption}{Assumption}
\newtheorem{lemma}{Lemma}
\newtheorem{corollary}{Corollary}

\begin{document}

\title{Diffusion LMS over Multitask Networks}
\author{Jie Chen$^\dag$, \IEEEmembership{Member, IEEE}, C{\'e}dric Richard$^\dag$, \IEEEmembership{Senior Member, IEEE} \\
Ali H. Sayed$^\ddag$, \IEEEmembership{Fellow Member, IEEE}
\thanks{The work of C. Richard was partly supported by the Agence Nationale pour la Recherche, France, (ODISSEE project, ANR-13-ASTR-0030). The work of A. H. Sayed was supported in part by NSF grants CCF-1011918 and ECCS-1407712. A short and preliminary version of this work appears in the conference publication~\cite{chen2013performance}.}
 \\ \vspace{0.5cm}
\small{\linespread{0.2} $^\dag$ Universit{\'e} de Nice Sophia-Antipolis, UMR CNRS 7293, Observatoire de la C{\^{o}}te d'Azur \\
Laboratoire Lagrange, Parc Valrose, 06102 Nice - France \\
phone: (33) 492 076 394 \hspace{0.5cm} \hspace{0.5cm} fax:
(33) 492 076 321 \\ dr.jie.chen@ieee.org \hspace{0.5cm}
cedric.richard@unice.fr}
\vspace{0.3cm}\\
\small{\linespread{0.2} $^\ddag$ Electrical Engineering Department \\
University of California, Los Angeles, USA \\
phone: (310) 267 2142 \hspace{0.5cm} \hspace{0.5cm} fax:
(310) 206 8495 \\
sayed@ee.ucla.edu}
}

\maketitle

\vspace{-1.5cm}
\begin{abstract}

The diffusion LMS algorithm has been extensively studied in recent years. This efficient strategy allows to address distributed optimization problems over networks in the case where nodes have to collaboratively estimate a single parameter vector. Nevertheless, there are several problems in practice that are multitask-oriented in the sense that the optimum parameter vector may not be the same for every node. This brings up the issue of studying the performance of the diffusion LMS algorithm when it is run, either intentionally or unintentionally, in a multitask environment. In this paper, we conduct a theoretical analysis on the stochastic behavior of diffusion LMS in the case where the single-task hypothesis is violated. {We analyze the competing factors that influence the performance of diffusion LMS in the multitask environment, and which allow the algorithm to continue to deliver performance superior to non-cooperative strategies in some useful circumstances. We also propose} an unsupervised clustering strategy that allows each node to select, via adaptive adjustments of combination weights, the neighboring nodes with which it can collaborate to estimate a common parameter vector. Simulations are presented to illustrate the theoretical results, and to demonstrate the efficiency of the proposed clustering strategy.

\end{abstract}


\section{Introduction}

Distributed adaptive estimation is an attractive and challenging problem that allows a collection of interconnected nodes to perform preassigned tasks from streaming measurements, such as parameter estimation.  Although centralized strategies may  benefit from information collected throughout a network, in most cases, distributed strategies are more robust to solve inference problems in a collaborative and autonomous manner~\cite{Sayed2014Proc}.

Most recent efforts in the study of distributed estimation problems have focused on scenarios where the  network is employed to collectively estimate a single parameter vector.  Several strategies have been proposed for this purpose for sequential data processing over networks, including consensus {strategies~\cite{Tsitsiklis1984,Xiao2004,Braca2008TSP,Nedic2009,Kar2009,Braca2010,Dimakis2010,Srivastava2011}}, incremental strategies~\cite{Bertsekas1997,Nedic2001,Rabbat2005,Blatt2007,Lopes2007incr}, and diffusion strategies~\cite{Lopes2008diff,Cattivelli2010diff}.
Diffusion strategies are particularly attractive due to their enhanced adaptation performance and wider stability ranges when constant step-sizes are used to enable continuous learning~\cite{Tu2012}. For this reason, we focus on this class of strategies in the remainder of the article. These strategies estimate a common parameter vector by minimizing, in a distributed manner, a global criterion that aggregates neighborhood cost functions. Nodes exchange information locally and cooperate only with their neighbors, without the need for sharing and requiring any global information. The resulting networks benefit from the temporal and spatial diversity of the data and end up being endowed with powerful learning and tracking abilities~\cite{Tu2012,ChenUCLA2012}. 
{The performance} of the corresponding adaptive networks have been extensively studied in the literature, under favorable and unfavorable conditions such as model non-stationarities and imperfect communication~\cite{Khalili2012,Zhao2012impe}. This framework has also been extended by considering more general cost functions and data models~\cite{ChenUCLA2012,ChenUCLA2013,Gharehshiran2013,Chouvardas2011set,sayed2014adaptation}, by incorporating additional regularizers~\cite{Liu2012,Chouvardas2012,Lorenzo2013spar}, or by expanding its use to other scenarios~\cite{chainais2013distributed,chainais2013learning,Predd2006WSN,honeine2008regression}.

The working hypothesis for these earlier studies on diffusion LMS strategies is that the nodes cooperate with each other to estimate a single parameter vector. We shall refer to problems of this type as \emph{single-task} problems. However, many problems of interest happen to be \emph{multitask}-oriented in the sense that there are multiple optimum parameter vectors to be inferred simultaneously and in a collaborative manner. The multitask learning problem is relevant in several machine learning formulations~\cite{chen2009mtl,Chapelle2011mtl,zhou2011mtl}. {In the distributed estimation context, which is the focus of this work, there exist many  applications where either agents are subject to data measurements arising from different models, or they are sensing data that varies over the spatial domain. Only a handful of works have considered problem formulations that deal multitask scenarios. A brief summary follows.}

{For instance, if different groups of agents within a network happen to be tracking different moving targets, then all agents within the same cluster would be interested in estimating the same parameter vector (say, the vector that describes the location of their target). If the targets are moving in formation, then their location vectors would be related to each other and, therefore, cooperation among clusters would be beneficial. {In a second example~\cite{chen2013multitask}, we consider agents engaged in cooperative spectrum sensing over cognitive radio networks. These networks involve two types of users: primary users and secondary users. Secondary users are allowed to detect and occupy temporarily unused spectral bands provided that they do not cause interference to primary users. Therefore, secondary users need to estimate the aggregated spectrum transmitted by all primary users, as well as local interference profiles. This multitask estimation problem requires cooperation between nodes because noncooperative strategies
would lead to local spectral profiles that are  subject to hidden node effects~\cite{chen2013multitask}. In another example,} a network may be deployed to estimate the spatially-varying temperature profile over a certain region, where the parameters that determine how the temperature varies across the agents may be space-dependent~\cite{Abdolee2014}. In {another} example, the works~\cite{Bertrand2010P1,Bertrand2011}}
{describe a multitask estimation algorithm over a fully connected broadcasting network. {These works} assume that the node-specific parameter vectors to estimate lie in a common latent signal subspace {and exploit this property} to compress information and reduce communication costs. {Another scenario is described {in~\cite{Bogdanovic2013,Bogdanovic2014}, where incremental and diffusion strategies are} used to solve a distributed estimation problem} with nodes that simultaneously estimate local and global parameters. {In~\cite{Chen2014Subspace}, the parameter space is decomposed into two orthogonal subspaces with one of them being common to all nodes.}}

{In all these previous examples, it is assumed beforehand that the nodes have some prior knowledge about clustering or about the parameter space, such as which agents belong to a particular cluster or how the parameter space is subdivided. The aim of the current work is  different and does not assume prior information about the tasks or clusters. It then becomes critical to assess the performance limits of diffusion strategies when used, knowingly or unknowingly, in an multitask environment.} Due to inaccurate modeling, or minor differences between tasks neglected intentionally, there may be situations in which the diffusion LMS algorithm is applied {to multitask scenarios}. When these situations occur, the distributed implementation will lead to biased results that may be acceptable depending on the application at hand. This biased solution may still be beneficial compared to purely non-cooperative strategies provided that the local optimums are sufficiently close to each other. 

These observations motivate us to examine  the performance of the diffusion LMS strategy when it is run, either intentionally or unintentionally, in a multitask environment.  In this respect, we shall analyze the performance of the diffusion LMS in terms of its mean weight deviation and mean-square error in the case when the single-task hypothesis is violated. {We shall also identify and analyze the competing factors that influence the performance of diffusion LMS in the multitask environment, and which allow this algorithm to continue to deliver performance superior to non-cooperative strategies in some useful circumstances.}
We shall also propose an unsupervised clustering strategy that allows each node to select, via adaptive adjustments of combination weights, the neighboring nodes with which it should collaborate to improve its estimation accuracy. In the related work~\cite{chen2013multitask}, we formulated the multitask problem directly over networks with connected clusters of nodes. In that work, the clusters are assumed to be known beforehand and no clustering is proposed. We then derived extended diffusion strategies that enable adaptation and learning under these conditions. In the current work, on the other hand, the clusters are not assumed to be known. It then becomes necessary to examine how this lack of information influences performance. It also becomes necessary to endow the nodes with the ability to identify and form appropriate clusters to enhance performance.  {One clustering strategy was proposed in the earlier work~\cite{Zhao2012}; its performance is dependent on the initial conditions used by the nodes to launch their adaptation rules. In this work, we  propose a more robust clustering strategy.}

 {\bf Notation.} Boldface small letters $\bx$ denote vectors. All vectors are column vectors. Boldface capital letters $\cb{X}$ denote matrices. The superscript $(\cdot)^\top$ represents the transpose of a matrix or a vector. Matrix trace is denoted by trace$\{\cdot\}$, Kronecker product is denoted by $\otimes$, and expectation is denoted by $\mathbb{E}\{\cdot\}$. Identity matrix of size $N\times N$ is denoted by $\bI_N$, and {the all-one vector of length $N$ is denoted by $\cb{1}_N$.} We denote by $\N{k}$ the set of node indices in the neighborhood of node $k$, including $k$ itself, and $|\N{k}|$ its cardinality. The operator $\col\{\cdot\}$ stacks its vector arguments on the top of each other to generate a connected vector. The other symbols will be defined in the context where they are used.

\section{Multitask problems and diffusion LMS}

\subsection{Modeling assumptions and Pareto solution}

We consider a connected network composed of $N$ nodes. The problem is to estimate  $L\times 1$ unknown vectors $\bw_k^\star$ at each node~$k$ from collected  measurements.  Node $k$  has access to temporal wide-sense stationary measurement sequences $\{d_k(n), \bx_k(n)\}$, with~$d_k(n)$ denoting  a scalar zero-mean reference signal, and $\bx_k(n)$ an $L\times 1$ regression vector with a positive definite covariance matrix $\bR_{x,k}=\mathbb{E}\{\bx_k(n)\bx_k^\top(n)\} > 0$. The data at node $k$ are assumed to be related via the linear regression model:
\begin{equation}
           \label{eq:datamodel}
            d_k(n)=\bx_k^\top(n)\,\bw_k^\star + z_k(n)
\end{equation}
where $z_k(n)$ is a zero-mean i.i.d. additive noise at node $k$ and time $n$. Noise $z_k(n)$ is assumed to be independent of any other signals and has variance $\sigma_{z,k}^2$.  Let $J_k(\bw)$ denote the mean-square-error cost at node $k$, namely,
\begin{equation}
           J_k(\bw) = \mathbb{E} \left\{ |d_k(n) - \bx^\top_k(n)\,\bw|^2\right\}.
\end{equation}
It is clear from~\eqref{eq:datamodel} that each $J_k(\bw)$ is minimized at $\bw_k^\star$. Depending on whether the minima of all the $J_k(\bw)$ are achieved at the same location or not, referred to as tasks, the distributed learning problem can be single-task or multitask oriented~\cite{chen2013multitask}.

In a single-task network, all nodes have to estimate the same parameter vector $\bw^\star$. That is, in this case we have that
\begin{equation}
	\bw_k^\star = \bw^\star,\quad \forall k \in \{1, ..., N\}.
\end{equation}
Diffusion LMS strategies for the distributed estimation of $\bw^\star$ under this scenario were derived in~\cite{Lopes2008diff,Cattivelli2010diff,Sayed2013intr,Sayed2014Proc} by seeking the minimizer of the following aggregate cost function:
\begin{equation}
	\label{eq:Jglob}
	J^{\text{glob}}(\bw) = \sum_{k=1}^N J_k(\bw)
\end{equation}
in a cooperative manner in order to improve  estimation accuracy. {In a multitask network, on the other hand, each node needs to determine its own parameter vector $\bw^\star_k$. In~\cite{chen2013multitask}, we assume that {the} parameter vectors at two connected nodes $k$ and $\ell$ {may satisfy certain similarity properties, such as being close to each other in some Euclidean norm sense}. Nodes can also be interested in simultaneously estimating some parameters of local interest as well as parameters of global interest~\cite{Bogdanovic2014,Chen2014Subspace}. Cooperation between these nodes can therefore be beneficial to infer $\bw^{\star}_k$ and $\bw^{\star}_\ell$. A possible way to exploit and model relationships among tasks is to formulate optimization problems with appropriate co-regularizers~\cite{chen2013multitask}. An alternative is to build on the principle that the node hypothesis spaces partially {overlap~\cite{Bogdanovic2014,Chen2014Subspace}.  These formulations, however, require some prior knowledge about how tasks are related to each other. In this work, we do not assume the availability of any prior information; in particular, nodes do not know which other nodes share similar objectives.}}
{Now since} each cost function $J_k(\bw)$ may not be minimized at the same location, the minimizer of the aggregate cost~\eqref{eq:Jglob} can be shown to correspond to a Pareto optimum solution  for the multi-objective optimization problem~\cite{ChenUCLA2013,boyd2004convex,sayed2014adaptation}. 
{Diffusion} LMS thus leads to a compromise for the entire network, {and we would like to examine how its performance is affected when used in a multitask scenario.} 

\subsection{Diffusion LMS}

The diffusion LMS algorithm was originally designed for minimizing the cost function~\eqref{eq:Jglob} in an adaptive and distributed manner~\cite{Lopes2008diff,Cattivelli2010diff,Sayed2013diff,Sayed2013intr}. Let $\bw_k(n)$ denote the estimate of the minimizer of~\eqref{eq:Jglob} at node $k$ and time instant $n$. The general structure of the algorithm consists of the following~steps:
\begin{align}
	\bphi_k(n)  \!&=\! \sum_{\ell\in\N{k}} a_{1,\ell k}\, \bw_\ell (n)   \label{eq:diff1}\\
\hspace{-2mm}	\bpsi_k\!(n\!+\!1) \!&=\! \bphi_k\!(n) \!+\! \mu_k\!\!\! \sum_{\ell\in\N{k}} \!\!c_{\ell k}   \, \bx_\ell(n)\! \big[d_\ell(n) \!\!-\!\! \bx_\ell^\top(n)\bphi_k(n)\big]   \label{eq:diff2} \\
\hspace{-2mm}	\bw_k\!(n\!+\!1) \!&=\! \sum_{\ell\in\N{k}} a_{2,\ell k} \, \bpsi_{\ell} (n+1).   \label{eq:diff3}
\end{align}
The non-negative coefficients $a_{1,\ell k}$, $a_{2,\ell k}$ and $c_{\ell k}$ are the ($\ell, k$)-th entries of two left-stochastic matrices, $\bA_1$ and $\bA_2$, and a right-stochastic matrix $\bC$, that is,
\begin{equation}
	\label{eq:cstAC}
	\bA_1^\top\cb{1}_N = \cb{1}_N, \;  \bA_2^\top\cb{1}_N = \cb{1}_N, \; \bC\cb{1}_N=\cb{1}_N
\end{equation}
{and satisfy}
\begin{equation}
		a_{1, \ell k} = 0, \; a_{2, \ell k} = 0, \;  c_{\ell k} =0 \quad \text{if}  \quad \ell \notin \N{k}.
\end{equation}
Several adaptive strategies can be obtained as special cases of~\eqref{eq:diff1}--\eqref{eq:diff3} through appropriate selections of $\bA_1$, $\bA_2$ and $\bC$. For instance, setting $\bA_1 = \bI_N$ yields the so-called adapt-then-combine (ATC) diffusion LMS. Setting $\bA_2 = \bI_N$ leads to the combine-then-adapt (CTA) diffusion LMS. By setting $\bA_1 = \bA_2  = \bC = \bI_N$, the algorithm degenerates to non-cooperative LMS that will be considered in the sequel for comparison purposes.

When applying ATC diffusion LMS without {gradient information} exchange, that is, with $\bC=\bI_N$, the agents converge toward the Pareto optimum with a bias of the order $\cp{O}(\mu_{\max})$, where $\mu_{\max}$ denotes the largest step-size parameter across all nodes~\cite{ChenUCLA2013}. In this paper, rather than {focusing} on this convergence point that can be perceived as a compromise, we shall study analytically how diffusion LMS~\eqref{eq:diff1}--\eqref{eq:diff3} behaves in a multitask environment in relation to the optimum vectors
$\bw_k^\star$. Moreover, in order to generalize the analysis, we shall consider drifting optimums around a fixed value $\bw_k^\star$, namely,
\begin{equation}
          \label{eq:optimum.tv}
          \bw_k^\star(n) = \bw_k^\star + \beps_k(n)
          \vspace{-2mm}
\end{equation}
where $\beps_k(n)$ is a zero-mean random perturbation independent of any other signal, with zero-mean and covariance matrix~$\sigma_{\epsilon,k}^2 \, \bI_L$. 
{Under~\eqref{eq:optimum.tv}, model}~\eqref{eq:datamodel} is replaced by
\vspace{-2mm}
 \begin{equation}
           \label{eq:datamodelvrt}
            d_k(n)=\bx_k^\top(n)\,\bw_k^\star(n) + z_k(n).
\end{equation}


 \section{Performance analysis of diffusion LMS for multitask networks}

We collect the information from across the network into block vectors and matrices. In particular, we denote by $\bw(n)$, $\bw^\star$ and $\bw^\star(n)$ the block weight estimate vector {\eqref{eq:defbw}}, the block optimum mean weight vector  {\eqref{eq:defbws}}, and the instantaneous block optimum weight vector {\eqref{eq:defbwn}}, all of size $LN\times 1$, that is,
\begin{align}
             \bw(n) &= \col\{\bw_1(n), \dots, \bw_N(n)\} \label{eq:defbw}  \\
             \bw^\star &= \col\{\bw_1^\star, \dots, \bw_N^\star\} \label{eq:defbws}  \\
             \bw^\star(n) &= \col\{\bw_1^\star(n), \dots, \bw_N^\star(n) \}  \label{eq:defbwn}.
\end{align}
The weight error vector $ \bv_k(n)$ for each node $k$ at iteration $n$ is defined by
\begin{equation}
           \bv_k(n)=\bw_k(n)-\bw_k^\star(n).
\end{equation}
Let
\begin{equation}
          \bv^\star_k(n)=\bw_k(n)-\bw_k^\star
\end{equation}
be the weight error vector between $\bw_k(n)$ and the fixed weight vector $\bw_k^\star$. The following relation holds
\begin{equation}
	\label{eq:v.tv}
	\begin{split}
	\bv_k(n) 
                       = \bv^\star_k(n) - \beps_k(n).
       \end{split}
\end{equation}
This relation allows us to derive recursions with respect to $\bv_k^\star(n)$, and then get back to $\bv_k(n)$. The weight error vectors $\bv_k(n)$ and  $\bv^\star_k(n)$ are also stacked on top of each other to get the block weight error vectors:
\begin{align}
        \bv(n) &= \col\{\bv_1(n), \dots, \bv_N(n)\}  \\
        \bv^\star(n) &= \col\{\bv_1^\star(n), \dots, \bv_N^\star(n)\}.
\end{align}
To perform the theoretical analysis,  we introduce the following independence assumption.

\assumption (Independent regressors) The regression vectors $\bx_k(n)$ arise from a zero-mean random process that is temporally {(over $n$)} stationary, white, and independent over space {(over $k$)} with $\bR_{x,k} = \mathbb{E}\{\bx_k(n)\,\bx^\top_k(n)\}>0$. \hfill$\blacksquare$

 A direct consequence is that $\bx_k(n)$ is independent of $\bv_\ell(m)$ for all $\ell$ and $m\leq n$. Although not true in general, this assumption is commonly used for analyzing adaptive constructions because it allows to simplify the derivation without constraining  the conclusions. {There are several results in the adaptation literature that show that performance results that are obtained under the above independence assumptions match well the actual performance of the algorithms when the step-sizes are sufficiently small. (see, e.g.,~\cite[App. 24.A]{sayed2008adaptive} and \cite[Chs. 10-11]{sayed2014adaptation} and the many references therein).}

\subsection{Mean weight behavior analysis}

Subtracting optimum vectors $\bw_k^\star$ from both sides of the first step of diffusion LMS, namely equation~\eqref{eq:diff1}, gives
\begin{equation}
	\label{eq:phi.err.k}
	\bphi_k(n) -\bw_k^\star= \sum_{\ell\in\N{k}} a_{1,\ell k}\, \bw_\ell (n) -\bw_k^\star.
\end{equation}
Defining  $\cA_1= \bA_1\otimes \bI_L$ and using $\bw_k^\star=\bw_k(n)-\bv_k^\star(n)$, expression~\eqref{eq:phi.err.k} can be expressed in {block-based form:}
\begin{equation}
	\label{eq:phi.err}
	\bphi(n) -\bw^\star= \cAT1\, \bv^\star(n) + (\cAT1-\bI_{NL}) \bw^\star.
\end{equation}
Note that the term $(\cAT1-\bI_{NL})\,\bw^\star$, which does not appear for single-task networks, is inherited from the {multitask context}\footnote{{In the single-task context, and since all nodes share the same optimum, say $\bw_1^\star$, we can write $\bw^\star = \col\{\bw_1^\star, \dots, \bw_1^\star\}$. Consequently, $\cAT1\bw^\star = (\bA_1\otimes \bI_L) \, (\cb{1}_N \otimes \bw_1^\star) = (\bA_1\cb{1}_N) \otimes \bw_1^\star = \bw^\star$, where the last step is due to the fact that $\bA_1$ is left-stochastic. This result leads to $(\cAT1-\bI_{NL})\,\bw^\star=\cb{0}_{NL}$.}}.
The estimation error in the second step~\eqref{eq:diff2} of diffusion LMS can be rewritten as
\begin{equation}
	\label{eq:err.v}
	\begin{split}
         	  {	d_\ell(n)-\bx_\ell^\top(n)\bphi_k(n)  }
                                                                              = z_\ell(n) - \bx_\ell^\top(n)\,(\bphi_k(n) - \bw^\star_\ell ) +\bx_\ell^\top(n)\,\beps_\ell(n).
	\end{split}
\end{equation}
For single-task networks, $(\bphi_k(n) - \bw^\star_\ell)$ in the above expression reduces to $(\bphi_k(n) - \bw^\star_k)$ since $\bw^\star_k = \bw^\star_\ell$ for all $k$, $\ell$. In the multitask context, we can establish the following relationship:
\begin{equation}
	\begin{split}
		\bphi_k(n) - \bw^\star_\ell & = (\bphi_k(n)-\bw_k^\star) + (\bw_k^\star-\bw_\ell^\star)  \\
			& = (\bphi_k(n)-\bw_k^\star) + \bu^\star_{k\ell}
	\end{split}
\end{equation}
where $\bu^\star_{k\ell}$  is the difference between the fixed weight vectors $\bw_k^\star$ and $\bw_\ell^\star$.  Incorporating this expression into \eqref{eq:err.v} yields:
\begin{equation}
         \begin{split}
	  &d_\ell(n)-\bx_\ell^\top(n)\bphi_k(n) =z_\ell(n) \!-\! \bx_\ell^\top(n) \,  (\bphi_k(n)\!-\!\bw_k^\star) \!-\! \bx_\ell^\top(n)\, \bu^\star_{k\ell} \!+\! \bx_\ell^\top(n)\, \beps_\ell(n).
	\end{split}
\end{equation}
Subtracting $\bw^\star_k$ from both sides of equation~\eqref{eq:diff2} and using the above relation, we have
\begin{equation}
	\label{eq:psi.errk}
	\begin{split}
		&\bpsi_k(n+1)- \bw_k^\star\\ 
		&= (\bphi_k(n)-\bw_k^\star) - \mu_k \sum_{\ell\in\N{k}} c_{\ell k}\,
		\bx_\ell(n)\bx_\ell^\top(n)\, (\bphi_k(n)-\bw_k^\star) \\
		\quad & + \mu_k \sum_{\ell\in\N{k}} c_{\ell k} \, \bx_\ell(n)\,z_\ell(n) -\mu_k \sum_{\ell\in\N{k}}c_{\ell k}\, \bx_\ell(n)\bx_\ell^\top(n)\, \bu^\star_{k\ell}  + \mu_k \sum_{\ell\in\N{k}}
		c_{\ell k}\,\bx_\ell(n)\bx_\ell^\top(n)\, \beps_\ell(n).
	\end{split}
\end{equation}
Let us  introduce the following $N\times N$ block diagonal matrices with blocks of size $L\times L$:
\begin{align}
	\bU &= \text{diag} \{\mu_1\bI_L, \dots, \mu_N\bI_L\}   \\	
	\bH(n) & = \text{diag}\Big\{  \sum_{\ell\in{\N{k}}} c_{\ell {k} }\,\bx_\ell(n)\bx_\ell^\top(n) \Big\}_{k=1}^N,
\end{align}
and the following vectors of length $NL$:
\begin{align} 
	\bh_u(n) & = \col\Big\{  \sum_{\ell\in\N{k}} c_{\ell k }\,\bx_\ell(n)\bx_\ell^\top(n)\, \bu^\star_{{k\ell}}\Big\}_{k=1}^N  \\
	\bh_\epsilon(n) & = \col \Big\{  \sum_{\ell\in\N{k}} c_{\ell k }\,\bx_\ell(n)\bx_\ell^\top(n)\,\beps_\ell(n){\Big\}_{k=1}^N} \\
	\bp_{zx}(n) &= \col\Big\{\sum_{\ell\in\N{k}}  c_{\ell k}\, \bx_\ell(n)\,z_\ell(n)\Big\}_{k=1}^N.
\end{align}
Using the above notation and equations~\eqref{eq:phi.err} and~\eqref{eq:psi.errk}, the block weight error vector $\bpsi(n+1)$ can be expressed as
\begin{equation}
	\label{eq:psi.v}
	\begin{split}
	\bpsi(n+1)- \bw^\star
	 &= ( \bI_{NL} \!-\! \bU\bH(n))\big( \cAT1\, \bv^\star(n) \!+\! (\cAT1\!-\!\bI_{NL}) \bw^\star\big) + \bU\bp_{zx}(n) - \bU(\bh_u(n) - \bh_\epsilon(n)).
	\end{split}
\end{equation}
Let $\cA_2 = \bA_2 \otimes \bI_L$. The combination step~\eqref{eq:diff3} of diffusion LMS leads to
\begin{equation}
         \label{ea:wApsi}
	\bw(n+1) ={\cAT{2}}\;\bpsi(n+1).
\end{equation}
Subtracting $\bw^\star$ from both {sides of~\eqref{ea:wApsi}}, we get
\begin{equation}
          \label{eq:w.psi}
          \bv^\star(n+1) = \cAT2 (\bpsi(n+1)-\bw^\star) + (\cAT2 - \bI_{NL})\,\bw^\star.
\end{equation}
Again, note that the term $(\cAT2-\bI_{NL})\,\bw^\star$ does not appear for single-task networks and is inherited from the multitask context. {Combining}~\eqref{eq:psi.v} and~\eqref{eq:w.psi}, we obtain the update relation for $\bv^\star(n)$ in a single expression as follows:
\begin{equation}
	\label{eq:v}
	\begin{split}
		\bv^\star(n+1) 
			&= {\cAT2} \left(\bI_{NL} - \bU\bH(n)\right)\cAT1\bv^\star(n) 	+ {\cAT2}\bU\bp_{zx}(n)-\cAT2\,\bU(\bh_u(n) - \bh_\epsilon(n)) \\
					  &+ \! \left({\cAT2}(\bI_{NL}(n) \!-\! \bU\bH(n))(\cAT1\!-\!\bI_{NL})+(\cAT2 \!-\! \bI_{NL})\right)\bw^\star.
	\end{split}
\end{equation}
In order to make the presentation clearer, we use the following notation for terms in expression~\eqref{eq:v}:
\begin{align}
	\bB(n)  &=  {\cAT2}\left( \bI_{NL} - \bU\bH(n) \right)\cAT1     \\
	\bg(n)  &= {\cAT2}\bU\bp_{zx}(n)  \label{eq:bgn} \\
	\br(n)  
	&= \underbrace{\cAT2\bU\bh_u(n)}_{\br_u(n)} - \underbrace{\cAT2\bU\bh_\epsilon(n)}_{\br_\epsilon(n)} 	-\!\underbrace{\left({\cAT2}(\bI_{NL} \!\!-\!\! \bU\bH(n))(\cAT1\!\!-\!\!\bI_{NL})  \!+\! (\cAT2 \!\!-\!\! \bI_{NL})\right)\!\bw^\star}_{\br_w(n)}.  \label{eq:brn}
\end{align}
{Then}, recursion~\eqref{eq:v} can be rewritten as
\begin{equation}
	\label{eq:rec.bv}
	\bv^\star(n+1) = \bB(n)\,\bv^\star(n) + \bg(n) - \br(n).
\end{equation}
{The non-zero driving term $\br(n)$, arising from the multitask scenario and the random perturbations $\beps_k(n)$, introduces a further level of complexity in the algorithm analysis, especially in the mean-square error behavior one. This analysis reduces to the traditional analysis of diffusion LMS by setting $\br(n)=0$.} Let $\bH$ be the expected value of $\bH(n)$ given by
\begin{equation}
        \label{eq:H}
	\bH= \text{diag}\left\{ \bR_1, \, \dots, \,  \bR_N \right\}
\end{equation}
in terms of the neighborhood covariance matrices:
\begin{equation}
	\bR_k = \sum_{\ell\in\N{k}} c_{\ell k}\, \bR_{x,\ell}.
\end{equation}
Let $\bh_u$ be the expected value $\mathbb{E}\{\bh_u(n)\}$, that is,
\begin{equation}
	\bh_u =\col\Big\{ \sum_{\ell\in\N{1}} c_{\ell 1}\bR_{x,\ell}\,\bu^\star_{1\ell}, \, \dots, \,
	\sum_{\ell\in\N{N}} c_{\ell N} \bR_{x,\ell}\,\bu^\star_{N\ell} \Big\}.
\end{equation}
The independence assumption (Assumption 1), and the statistical properties of noise $z_k(n)$ and perturbations $\beps_k(n)$, lead us to the following expected values for $\bB(n)$, $\bg(n)$ and $\br(n)$:
\begin{align}
	\bB & =  \cAT2\left(\bI_{NL} - \bU\bH\right)\cAT1 \\
	\bg & = 0 \\
	\br  & =  \underbrace{\cAT2\,\bU\,\bh_u}_{\br_u}-\!\underbrace{\left(\cAT2(\bI_{NL} \!-\! \bU\bH)(\cAT1\!-\!\bI_{NL})\!+\!(\cAT2\!-\!\bI_{NL})\right)\bw^\star}_{\br_w}  \label{eq:Er}
\end{align}
where $\br_u$, $\br_\epsilon$ and $\br_w$ denote the expected values of $\br_u(n)$, $\br_\epsilon(n)$ and $\br_w(n)$, respectively. Note that the expected value $\br$ is expressed as $\br=\br_u-\br_w$ because $\br_\epsilon=0$. Taking the expectation of both sides of \eqref{eq:v}, {and observing that $\bB(n)$ and $\bv^\star(n)$ are independent under Assumption 1}, we get
\begin{equation}
	\label{eq:Ev}
	\mathbb{E}\{\bv^\star(n+1)\} = \bB \mathbb{E}\{\bv^\star(n)\} - \br_u + \br_w.
\end{equation}
Moreover, equation~\eqref{eq:v.tv} tells us that
\begin{equation}
               \mathbb{E}\{\bv(n+1)\} = \mathbb{E}\{\bv^\star(n+1)\}.
\end{equation}

\theorem \textbf{(Stability in the mean)} Assume data model~\eqref{eq:datamodel} and Assumption 1 hold. Then, for any initial condition, the diffusion LMS strategy \eqref{eq:diff1}--\eqref{eq:diff3} applied to multitask networks asymptotically converges in the mean if the step-sizes are chosen to satisfy
\begin{equation}
	\label{eq:stepsize1}
	0<  \mu_k < \frac{2}{\lambda_{\max}\{\bR_k\}}, \qquad k=1,\dots, N
\end{equation}
where $\lambda_\text{max}\{\cdot\}$ denotes the maximum eigenvalue of its matrix argument. In that case, it follows from~\eqref{eq:Ev} that the asymptotic mean bias is given by
\begin{equation}
	\label{eq:bias}
	\mathbb{E}\{\bv(\infty)\}=\left(\bI_{NL} - \bB\right)^{-1}(-\br_u + \br_w).
\end{equation}
\begin{proof}
Since the last two  terms on the RHS of~\eqref{eq:Ev} are constant, the convergence of this recursion requires that the matrix $\bB$ be stable. As shown in~\cite{Sayed2013intr}, {because $\cA_1$ and $\cA_2$ are left-stochastic, the spectral norm of $\cAT2(\bI_{NL} - \bU\bH)\cAT1$ is upper bounded by the spectral norm of $\bI_{NL} - \bU\bH$. The former is thus stable if the latter is stable}. This yields condition~\eqref{eq:stepsize1} {considering that $\bH$ is a block diagonal matrix of the form~\eqref{eq:H}.}
\end{proof}

 Inspecting expression \eqref{eq:bias}, we observe that the bias of diffusion LMS originates from the multiple local optimums $\bw^\star_k(n)$ and information {exchange among neighbors. This means that, even though the algorithm converges toward the Pareto optimum over multitask networks~\cite{ChenUCLA2013}, the bias~\eqref{eq:bias} can be large if the distance between the $\bw^\star_k(n)$ is large, and if nodes cooperate to estimate them.}

\subsection{Mean-square error behavior analysis}

By Assumption~1, equation~\eqref{eq:rec.bv}, {and definition~\eqref{eq:bgn} of $\bg(n)$ where $z_k(n)$ is a zero-mean noise independent of any other signal}, the mean-square of the weight error vector $\bv^\star(n+1)$ weighted by any positive semi-definite matrix $\bSig$ satisfies the following relation:
\begin{equation}
	\label{eq:EvSig}
	\begin{split}
	\mathbb{E}\{\|\bv^\star(n+1)\|^2_{\bSig}\} 
	&= \mathbb{E}\{\|\bv^\star(n)\|^2_{\bSig'}\} + \tr \left\{ \bSig\,\mathbb{E}\{\bg(n)\bg^\top(n)\} \right\} +  \mathbb{E}\{\|\br(n)\|^2_{\bSig}\} -2 \mathbb{E}\{ \br^\top(n)\,\bSig\,  \bB(n)\,\bv^\star(n)\}
	\end{split}
\end{equation}
with {$\|\bx\|^2_{\bSig} = \bx^\top \bSig\,\bx$, and} $\bSig' = \mathbb{E}\{ \bB^\top(n)\bSig\,\bB(n)\}$. The freedom in selecting $\bSig$ will allow us to derive several performance metrics. Let
\begin{equation}
	\begin{split}
	\bG 	& = \mathbb{E}\{\bg(n)\bg^\top(n)\}  \\
		& = \cAT2\bU\cCT\text{diag}\{\sigma_{z,1}^2 \bR_{x,1},\dots,\sigma_{z,N}^2 \bR_{x,N}\}\cC\bU\cA_2
	\end{split}
\end{equation}
where $\cC = \bC\otimes \bI_L$. For the sake of clarity, let us introduce
\begin{equation}
            \f(\br(n), \bSig, \bv^\star(n)) =  \|\br(n)\|^2_{\bSig}  - 2 \br^\top(n)\bSig \bB(n) \bv^\star(n).
\end{equation}
Relation~\eqref{eq:EvSig} can then be written as
 \begin{equation}
          \label{eq:EvSig2}
          \begin{split}
             \mathbb{E}\{\|\bv^\star(n+1)\|^2_{\bSig}\} &= \mathbb{E}\{\|\bv^\star(n)\|^2_{\bSig'}\} +  \tr\left\{ \bSig\bG\right\}  +\mathbb{E}\{\f(\br(n), \bSig, \bv^\star(n))\}.
           \end{split}  
\end{equation}
Let $\vc\{\cdot\}$ denote the operator that stacks the columns of a matrix on top of each other. Vectorizing both matrices $\bSig$ and $\bSig'$ by $\bsig=\vc\{\bSig\}$ and $\bsig' = \vc\{\bSig'\}$, it can be checked that
\vspace{-2mm}
\begin{equation}
	\bsig' = \bK\,\bsig
	\vspace{-2mm}
\end{equation}
where $\bK$ is the $(NL)^2 \times (NL)^2$ matrix given by
\begin{equation}
	\label{eq:K}
	\bK = {\mathbb{E}\{\bB^\top(n) \otimes \bB^\top(n)\}}. 
\end{equation}
{We can rewrite $\bK$ as {$\bB^\top\!\otimes \bB^\top+\cp{O}(\mu_{\max}^2)$, with an error term that depends on the square of the (maximum) step-size entry (see~\cite[Section 6.5]{Sayed2013intr} and~\cite[Ch. 10]{sayed2014adaptation})}. It is sufficient for the exposition {in this work} to focus on the case of sufficiently small step-sizes where terms involving higher powers of {the} step-sizes can be ignored. Therefore, we continue our discussion by letting
\begin{equation}
             \label{eq:Kapp}
             \bK = \bB^\top \otimes \bB^\top
\end{equation}
Let us now examine the term $\mathbb{E}\{\f(\br(n), \bSig, \bv^\star(n)) \}$. Consider first the weighted norm $\mathbb{E}\{\|\br(n)\|^2_{\bSig}\}$:
\begin{equation}
	\label{eq:rSigapp}
	\begin{split}
		\mathbb{E}\{\|\br(n)\|^2_{\bSig}\}
		 = \mathbb{E}\{ \br_u^\top(n) \, \bSig \, \br_u(n) \}
			+  \mathbb{E}\{ \br_\epsilon^\top(n) \, \bSig \, \br_\epsilon(n) \}   
		+  \mathbb{E}\{ \br_w^\top(n) \, \bSig \, \br_w(n) \} - 2\, \mathbb{E}\{  \br_u^\top(n) \, \bSig \, \br_w(n) \}.
	\end{split}
\end{equation}
We note that the stochastic components in $\br_u(n)$, $\br_\epsilon(n)$ and $\br_w(n)$ depend on the square of the step-sizes. We can write
\begin{equation}
	\mathbb{E}\{\|\br(n)\|^2_{\bSig}\} = \|\br\|^2_{\bSig} + {\cp{O}(\mu_{\max}^2)}.
\end{equation}
Likewise, we can write}
\begin{equation}
	\label{eq:rSigvapp}
	\begin{split}
		\mathbb{E}\{\br^\top(n)\,\bSig\,\bB(n)\,\bv^\star(n)\} = \br^\top\,\bSig\,\bB\,\mathbb{E}\{\bv^\star(n)\}+ {\cp{O}(\mu_{\max}^2)}.
	\end{split}
\end{equation}
By ignoring the higher-order terms for small step-sizes, we can continue the presentation by considering:
\begin{equation}
	\label{eq:f}
	\mathbb{E}\{\f(\br(n), \bSig, \bv^\star(n)) \} = \f(\br, \bSig, \mathbb{E}\{\bv^\star(n)\}).
\end{equation}
In this way, relation~\eqref{eq:EvSig2} can be approximated as follows:
\begin{equation}
	\label{eq:EvSig3}
	\begin{split}
		\mathbb{E}\{\|\bv^\star(n+1)\|^2_{\bsig}\} = \mathbb{E}\{\|\bv^\star(n)\|^2_{\bK\bsig}\} +  \vc(\bG^\top)^\top\bsig + \f(\br, \bsig, \mathbb{E}\{\bv^\star(n)\}).
	\end{split}
\end{equation}
where we are using the notation $\|\cdot\|_{\bSig}^2$ and $\|\cdot\|_{\bsig}^2$ interchangeably to refer to the same square weighted norm using $\bSig$ or its vector representation.

\theorem\textbf{(Mean-square stability)} Assume model~\eqref{eq:datamodel} and Assumption 1 hold. Assume that the step-sizes $\{\mu_k\}$ are sufficiently small such that condition~\eqref{eq:stepsize1} is satisfied and approximations~\eqref{eq:Kapp} and~\eqref{eq:f} are justified by ignoring higher-order powers of $\{\mu_k\}$. Then, the diffusion LMS strategy \eqref{eq:diff1}--\eqref{eq:diff3} applied over multitask networks is mean-square stable if the matrix $\bK$ is stable. Under approximation~\eqref{eq:Kapp}, the stability of $\bK$ is guaranteed for sufficiently small step-sizes that also satisfy~\eqref{eq:stepsize1}.
\begin{proof}
Iterating~\eqref{eq:EvSig3} starting from $n=0$, we find that
\vspace{-1mm}
\begin{equation}
                 \mathbb{E}\{\|\bv^\star(n\!+\!1)\|^2_{\bsig}\} = \|\bv^\star(0)\|^2_{\bK^{n+1}\bsig} +
                 \vc(\bG^\top)^\top \sum_{i=0}^n \bK^i \bsig   + \sum_{i=0}^n \f(\br, \bK^i\bsig, \mathbb{E}\{\bv^\star(n-i)\})           \label{eq:EVn0}   
\end{equation}
with the initial condition $\bv^\star(0)=\bw(0)-\bw^\star$. Provided that matrix $\bK$ is stable, {the terms on the RHS of~\eqref{eq:EVn0}  converge either to zero, or to bounded values. The algorithm is then mean-square stable for sufficiently small step-sizes. }
\end{proof}

\corollary \textbf{(Transient MSD)} Consider sufficiently small step-sizes $\mu_k$ that ensure mean and mean-square stability, and let $\bSig = \frac{1}{N}\bI_{NL}$. Then, the {mean-square deviation (MSD)} learning curve of the diffusion LMS algorithm  in a multitask environment, defined by $\zeta(n)={\mathbb{E}\{\|\bv(n)\|^2\}}/{N}$, evolves according to the following recursion for~$n\geq 0$
\begin{equation}
               \label{eq:zeta}
                  \zeta(n) = \zeta^\star(n) + \frac{L}{N}\sum_{k=1}^N\sigma_{\epsilon,k}^2
 \end{equation}
where $\zeta^\star(n)$ is evaluated as follows
%
%
\begin{equation}
	\begin{split}\label{eq:TransMSD1}
		\zeta^\star(n+1)  &= \zeta^\star(n) 
			{+ \Big(}  \big(\vc\{\bG^\top\}\big)^\top \bK^{n} \bsig_I +  \|\br\|^2_{\bK^n\bsig_I}   -\|\bv^\star(0)\|^2_{(\bI_{(NL)^2}-\bK)\bK^n\bsig_I} \\
		&\hspace{0.3cm} -2 \,\big(\bGam(n) + (\bB\,\mathbb{E}\{\bv^\star(n)\})^\top \otimes \br^\top \big)\, \bsig_I\Big)   
	\end{split}
\end{equation}
\begin{equation}
	\label{eq:TransMSD2}
	\begin{split}
             \hspace{-3cm}   \bGam(n+1) = \bGam(n) \bK + (\bB\,\mathbb{E}\{\bv^\star(n)\})^\top \otimes \br^\top) (\bK-\bI_{(NL)^2})
	\end{split}
\end{equation}
with $\bsig_I=\vc\{\frac{1}{N}\bI_{NL}\}$, $\zeta^\star(0) = \frac{1}{N}\|\bv^\star(0)\|^2$, $\bGam(0)  = \cb{0}_{1\times (NL)^2}$.

{
\begin{proof}
Comparing~\eqref{eq:EVn0} at instants $n+1$ and $n$, we can relate $\mathbb{E}\{\|\bv(n+1)\|^2_{\bsig}\}$ to $\mathbb{E}\{\|\bv(n)\|^2_{\bsig}\}$:
\begin{equation}
	\label{eq:EVrec_sig}
	\begin{split}
		\mathbb{E}\{\|\bv^\star(n+1)\|^2_{\bsig}\} 
			&= \mathbb{E}\{\|\bv^\star(n)\|^2_{\bsig}\} - \mathbb{E}\{\|\bv^\star(0)\|^2_{(\bI_{(NL)^2}-\bK)\bK^n\bsig}\} \\
                 	&+\mu^2\,\, \vc(\bG^\top)^\top\, \bK^{n} \bsig + \mu^2\eta^2 \|\br\|^2_{\bK^n\bsig} -2\mu\,\eta\left( (\bB\,\mathbb{E}\{\bv^\star(n)\}\otimes \br)^\top + \bGam(n)\right)\bsig
	\end{split}
\end{equation}
where
\begin{equation}
	\label{eq:Gam}
	\begin{split}
        &\bGam(n)
        = \sum_{i=1}^n  \left(\bB \mathbb{E}\{\bv^\star(n-i)\}\otimes \br\right)^\top\bK^i + \sum_{i=0}^{n-1}  \left(\bB \mathbb{E}\{\bv^\star(n-i-1)\}\otimes \br\right)^\top \bK^i.
	\end{split}
\end{equation}
We can then rewrite \eqref{eq:EVrec_sig}--\eqref{eq:Gam} as \eqref{eq:TransMSD1}--\eqref{eq:TransMSD2}.
\end{proof}}

\corollary\textbf{(Steady-state MSD)}  If the step-sizes are sufficiently small to ensure mean and mean-square-error convergences, then the steady-state MSD for diffusion LMS in a multitask environment is given by
\begin{align}
        \text{MSD}^\text{network} 	&=\frac{1}{N}\, \big(\vc\{\bG^\top\}\big)^\top\,   (\bI_{(NL)^2} - \bK)^{-1} \vc\{\bI_{NL}\} \nonumber \\
        						&+\!\f \Big(\br, \!\frac{1}{N}(\bI_{(NL)^2} \!-\! \bK)^{-1} \vc\{\bI_{NL}\}, \mathbb{E}\{\bv^\star(\infty)\}\Big)  +\! \frac{L}{N}\sum_{k=1}^N\sigma_{\epsilon,k}^2  \label{eq:MSD}
\end{align}
with $\mathbb{E}\{\bv(\infty)\}$ determined by~\eqref{eq:bias}.
\begin{proof} The steady-state MSD is given by the limit
\begin{equation}
      \label{eq:MSD.def}
     \begin{split}
      \text{MSD}^\text{network} &= \lim_{n\rightarrow\infty}\frac{1}{N}\mathbb{E}\{\|\bv(n)\|^2\} \\
                          &= \lim_{n\rightarrow\infty}\frac{1}{N}\mathbb{E}\{\|\bv^\star(n)\|^2\} +  \frac{L}{N}\sum_{k=1}^N\sigma_{\epsilon,k}^2.
      \end{split}
\end{equation}
Recursion~\eqref{eq:EvSig3} with $n\rightarrow\infty$ yields
\begin{equation}
        \label{eq:stablepoint}
        \begin{split}
        &\lim_{n\rightarrow\infty}\mathbb{E}\{\|\bv^\star(n)\|^2_{(\bI_{(NL)^2}-\bK)\bsig}\} =   \big(\vc\{\bG^\top\}\big)^\top\,\bsig + \f(\br,\bsig, \mathbb{E}\{\bv^\star(\infty)\}).
        \end{split}
\end{equation}
In order to use~\eqref{eq:stablepoint} in~\eqref{eq:MSD.def}, we select $\bsig$ to satisfy:
\begin{equation}
          \label{eq:sig_MSD}
          (\bI_{(NL)^2} - \bK) \,\bsig = \frac{1}{N}\vc\{\bI_{NL}\}.
\end{equation}
This leads to expression~\eqref{eq:MSD}.
\end{proof}

{The transient and steady-state MSD for any single node $k$ can be obtained by setting $\bSig = \text{diag}\{\cb{O}_N, \dots, \cb{I}_L, \dots, \cb{O}_N\}$ in Corollaries 1 and 2, with the identity matrix $\cb{I}_L$ at the $k$-th diagonal block and the all-zero matrix $\cb{O}_N$ at the others. }

The steady-state MSD can be expressed in an alternative form, which will facilitate the performance analysis. Since $\bK$ is stable when the network is mean-square stable, we can write
\begin{equation}
	\label{eq:expansion}
                    (\bI_{(NL)^2} - \bK)^{-1} =  \sum_{i=0}^\infty \bK^i  =   \sum_{i=0}^\infty (\bB^\top\otimes \bB^\top)^i.
\end{equation}
Consider now the following formula involving the trace of a product of matrices and the Kronecker product~\cite{Abadir2005matrix}
\begin{equation}
          \label{eq:relationKron}
           \tr\{\cb{X}_1^\top\, \cb{X}_2\, \cb{X}_3\,\cb{X}_4^\top\} = \vc\{\cb{X}_1\}^\top (\cb{X}_4\otimes \cb{X}_2)\, \vc\{\cb{X}_3\}
\end{equation}
where $\cb{X}_1$ to $\cb{X}_4$ denote matrices with compatible sizes. Using expansion \eqref{eq:expansion} with \eqref{eq:relationKron}, the first term on the RHS of~\eqref{eq:MSD} can be expressed as follows
\begin{equation}
        \begin{split}
        &\frac{1}{N}\, \big(\vc\{\bG^\top\}\big)^\top\,   (\bI_{(NL)^2} - \bK)^{-1} \vc\{\bI_{NL}\} = \frac{1}{N}\, \sum_{j=0}^{\infty}\tr\{\bB^j \, \bG \, \bB^{j \top}\}.
        \end{split}
\end{equation}
Similarly, the second term on the RHS of~\eqref{eq:MSD} can be written~as
\begin{align}
	&\f \Big(\br, \frac{1}{N}(\bI_{(NL)^2} - \bK)^{-1} \vc\{\bI_{NL}\}, \mathbb{E}\{\bv^\star(\infty)\}\Big) \nonumber\\
	&=\vc\{\br\br^\top\}^\top \frac{1}{N}(\bI_{(NL)^2} \!-\! \bK)^{-1} \vc\{\bI_{NL}\}  - 2 \vc\{  (\bB \mathbb{E}\{\bv^\star(\infty)\}\,\br^\top)^\top\}  \frac{1}{N}(\bI_{(NL)^2} - \bK)^{-1} \vc\{\bI_{NL}\} \nonumber \\
	&=\frac{1}{N}\, \sum_{j=0}^\infty \tr\{\bB^j\, [\br\br^\top - 2 \, \bB \mathbb{E}\{\bv^\star(\infty)\}\, \br^\top]\,  \bB^{j\top} \} \nonumber \\
	&{\mathop{=}^{(48)}  \frac{1}{N} \sum_{j=0}^\infty \tr\{\bB^j [\bI_{NL} \!+\! 2  \bB (\bI-\bB)^{-1}] \br\br^\top \bB^{j\top} \}. }
\end{align}
Finally, we can  express the steady-state MSD~\eqref{eq:MSD} as
\begin{align}
        &\text{MSD}^\text{network}  =  \frac{L}{N}\sum_{k=1}^N\sigma_{\epsilon,k}^2 +  \frac{1}{N}\!\sum_{j=0}^\infty  \text{trace} \left\{ \bB^j  (\bG \!+\! (\bI_{NL} \!+\! 2\bB(\bI_{NL}\!-\!\bB)^{-1})\br\br^\top){\bB^{j}}^\top\!\right\}\!.  \label{eq:MSDexp}
\end{align}
In the sequel, this formulation will allow us to compare the performance of different algorithms.

\subsection{Performance comparison with non-cooperative LMS}

We shall now compare the performance of the ATC and CTA diffusion LMS algorithms with the non-cooperative LMS strategy when applied to a multitask network. We consider the case of uniform step-sizes, $\mu_k=\mu$, for a meaningful comparison. Diffusion LMS degenerates to non-cooperative LMS by setting
\begin{equation}
	\bC = \bI_N, \quad \bA_1 = \bI_N,  \quad  \bA_2 = \bI_N
\end{equation}
from which the performance of the latter can be easily derived. In this case, matrices $\bB$ and $\bG$ reduce to
\begin{align}
	\bB_{\text{lms}} &= \bI_{NL} - \mu\bH   \\
	\bG_{\text{lms}} &=  \mu^2 \,  \text{diag}\{  \sigma_{z,1}^2 \bR_{x,1}, \dots,   \sigma_{z,N}^2 \bR_{x,N} \}
\end{align}
where we use the subscript LMS for clarity. Note that in this case we have $\br_w(n) = 0$. In addition, since $\cp{N}_k=\{k\}$ and $\bu^\star_{kk}=0$, we have $\br_u(n)=0$. This implies that $\br=0$. The steady-state MSD for non-cooperative LMS is then given by:
\begin{equation}
	\text{MSD}^\text{network}_\text{lms} = \frac{1}{N} \sum_{j=0}^\infty  \text{trace} \left(\bB_{\text{lms}}^j\,\bG_{\text{lms}}\,
	{\bB_{\text{lms}}^j}^{\hspace{-3mm}\top}\,\right)+\frac{L}{N}\sum_{k=1}^N \sigma_{\epsilon,k}^2.
\end{equation}
It is useful to note that the matrices $\bB$ and $\bG$ for diffusion LMS can be expressed in terms of $\bB_{\text{lms}}$ and $\bG_{\text{lms}}$:
\begin{equation}
	\begin{split}
	\bB 		&= {\cAT2} \left( \bI_{NL} - \mu\bH \right) \cAT1 
			= {\cAT2}\, \bB_{\text{lms}}\,\cAT1 \\ \vspace{-5mm}
	\bG		&= \mu^2\,\cAT2\,\cCT \, \text{diag}\{\sigma_{z,1}^2 \bR_{x,1},\dots,\sigma_{z,N}^2 \bR_{x,N} \} \,\cC \cA_2 \\
			&= \cAT2 \, \cCT \, \bG_{\text{lms}} \,\cC \, \cA_2
	\end{split}
\end{equation}
with $\cb{A}_1=\bI_{N}$ for the ATC diffusion strategy, and $\cb{A}_2=\bI_{N}$ for the CTA diffusion strategy. Using the series expansions for $\text{MSD}^\text{network}$ and $\text{MSD}^\text{network}_\text{lms}$, the difference between the MSDs for non-cooperative LMS and diffusion LMS is given by
\begin{equation}
	\label{eq:MSDcmp}
	\begin{split}
		&\text{MSD}^\text{network}_\text{lms} - \text{MSD}^\text{network} \\
		 = &\frac{1}{N} \sum_{j=0}^\infty  \text{trace} \left\{ \bB_{\text{lms}}^j\,\bG_{\text{lms}}\,
		{\bB_{\text{lms}}^j}^{\hspace{-3mm}\top}\, - \bB^j\,\bG\, {\bB^{j}}^\top \right\}   - \!\frac{1}{N} \sum_{j=0}^\infty  \text{trace} \left\{\bB^j (\bI_{NL} \!+\! 2\bB(\bI_{NL}\!-\!\bB)^{-1})\,
		\br\br^\top { {\bB^{j}}^\top} \right\}\!.
	\end{split}
\end{equation}
Note that the first term is the difference in performance between the non-cooperative LMS strategy and the cooperative diffusion strategy. It was first analyzed in~\cite{Sayed2013intr} and, because it is not specific to the multitask context, it is denoted by $\Delta\text{MSD}^\text{network}$. Only the second term, which depends on $\br$, is specific to the multitask scenario. Thus, it is denoted by $\Delta\text{MSD}^\text{network}_\text{multi}(\br)$. Therefore,
\begin{equation}
	\label{eq:MSDcmp2}
	\text{MSD}^\text{network}_\text{lms} - \text{MSD}^\text{network} = \Delta\text{MSD}^\text{network} - 	\Delta\text{MSD}^\text{network}_\text{multi}(\br).
\end{equation}

In order to obtain analytical results that allow some understanding of the algorithm behavior, we further assume that the matrices $\bC$, $\bA_1$ and $\bA_2$ in the diffusion implementation are doubly stochastic, and the regression covariance matrices are uniform across the agents, that is, $\bR_{x,k} = \bR_x$. With these assumptions, it was shown in~\cite[{Sec. 7}]{Sayed2013intr} that the first term $\Delta\text{MSD}^\text{network}$ on the RHS of~\eqref{eq:MSDcmp} is always nonnegative, namely,
\begin{equation}
	\text{trace} \left\{ \bB_{\text{lms}}^j\,\bG_{\text{lms}}\,{\bB_{\text{lms}}^j}^{\hspace{-3mm}\top}\, - \bB^j\,\bG\, {\bB^{j}}^\top \right\} \geq 0.
\end{equation}
We need to check under which conditions the second term $\Delta\text{MSD}^\text{network}_\text{multi}(\br)$ on the RHS of equation~\eqref{eq:MSDcmp} is nonnegative so that it can be viewed as a degradation factor caused by the multitask scenario. Introduce the symmetric matrix $\cb{Z} = \bI_{NL} \!+\! 2\bB(\bI_{NL}\!-\!\bB)^{-1}=2\,(\bI_{NL}\!-\!\bB)^{-1}\!-\!\bI_{NL}$. We find that
\begin{equation}
	\label{eq:trace1}
	\begin{split}
		\Delta\text{MSD}^\text{network}_\text{multi}(\br)
		&=
		{\frac{1}{N}\,\br^\top(\bI_{NL} - \bB^\top\bB)^{-1}\,\cb{Z}\,\br}.
	\end{split}
\end{equation}
We conclude that expression \eqref{eq:trace1} is non-negative for all $\br$ if, and only if, $(\bI_{NL} - \bB^\top\bB)^{-1}\,\cb{Z}$ is a symmetric positive semidefinite matrix. Now, we show that this condition is met for a large class of information exchange protocols. Assume, for instance, that either $\bA_1$ or $\bA_2$ is symmetric, depending on whether the focus is on the CTA or ATC strategy. Recalling conditions $\bR_{x,k} = \bR_x$ and $\mu_k=\mu$ for uniform data profile, it then holds that $\bB$ is a symmetric matrix. {It can be further verified that $(\bI_{NL} - \bB^2)^{-1}$ and $\cb{Z}$ are positive definite when $\bB$ is stable.} Now, we verify that the product $(\bI_{NL} - \bB^2)^{-1}\cb{Z}$ is a symmetric positive semidefinite {matrix.}

\lemma \textbf{(Positivity of a matrix product)} \cite[{Fact 8.10.11}]{Bernstein2009}  Given two symmetric positive semidefinite matrices $\cb{X}_1$ and $\cb{X}_2$ with compatible sizes. Then, $\cb{X}_1\cb{X}_2$ is symmetric positive semidefinite if, and only if, $\cb{X}_1\cb{X}_2$ is normal, that is, if it satisfies: $(\cb{X}_1\cb{X}_2)(\cb{X}_1\cb{X}_2)^\top=(\cb{X}_1\cb{X}_2)^\top(\cb{X}_1\cb{X}_2)$. \hfill$\blacksquare$

By setting $\cb{X}_1=(\bI_{NL} - \bB^2)^{-1}$ and $\cb{X}_2=\cb{Z}$, observe that $\cb{X}_1\cb{X}_2$ is {symmetric. It} then holds that $(\bI_{NL} - \bB^2)^{-1}\,\cb{Z}$ is normal. By Lemma~1, $(\bI_{NL} - \bB^2)^{-1}\,\cb{Z}$ is a symmetric positive semidefinite matrix, which means that $\Delta\text{MSD}^\text{network}_\text{multi}(\br)$ is nonnegative under the conditions specified above. It follows that this term can be viewed as a degradation factor caused by the cooperation of nodes performing different estimation tasks, which can be expressed as $\frac{1}{N}\|\br\|^2_{(\bI_{NL} - \bB^2)^{-1}\cb{Z}}$. We summarize the results in the following statement.

\theorem \textbf{(Non-cooperative vs. cooperative strategies)}  
Consider the same setting of Theorems 1 and 2, with the additional requirement that the conditions for a uniform data profile hold. {The adaptive ATC or CTA diffusion strategies outperform the non-cooperative strategy if, and only if,
\begin{equation}
	\label{eq:condition.vs.noncoop}
	\Delta\text{MSD}^\text{network} - \Delta\text{MSD}^\text{network}_\text{multi}(\br) \geq 0.
\end{equation}
Given doubly stochastic $\cb{A}$ and $\cb{C}$, the gain $\Delta\text{MSD}^\text{network}$ in performance between the cooperative diffusion strategy and the non-cooperative LMS strategy, which is independent of $\br$, is nonnegative.
{Furthermore, by} assuming that $\bA$ is symmetric, then the degradation in performance $\Delta\text{MSD}^\text{network}_\text{multi}(\br)$ caused by the multitask environment is positive. It is given by}
\begin{equation}
	\Delta\text{MSD}^\text{network}_\text{multi}(\br) = \frac{1}{N}\|\br\|^2_{(\bI_{NL} - \cb{B}^2)^{-1}\cb{Z}}
\end{equation}
where $\cb{Z} = 2\,(\bI_{NL}-\cb{B})^{-1}-\bI_{NL}$.  \hfill$\blacksquare$

Although condition~\eqref{eq:condition.vs.noncoop} allows to determine whether using the diffusion LMS is beneficial for multitask learning compared to the non-cooperative LMS strategy, it cannot be easily exploited to estimate appropriate combination coefficients because of its complexity and the need to handle dynamic problems. The aim of the next section is to derive an efficient strategy to estimate these coefficients.

\section{Node clustering via combination matrix selection}

We  now derive a clustering strategy where each node $k$ can adjust the combination weights $a_{\ell k}$ in an online manner, for $\ell \in \N{k}$, in order to adapt to multitask environments. It is sufficient to focus on the adapt-then-combine diffusion LMS defined by steps~\eqref{eq:diff2} and~\eqref{eq:diff3}. For ease of presentation, the corresponding algorithm is summarized below:
\begin{equation}
	\label{eq:ATC}
	\left\{
	\begin{split}
	& \bpsi_k(n\!+\!1)	\!=\! \bw_k(n) \!+\! \mu_k \!\!\sum_{\ell\in\N{k}} \!\!c_{\ell k}  \bx_\ell(n) \big[d_\ell(n) \!-\! \bx_\ell^\top(n)\bw_k(n)\big] \\
	& \bw_k(n\!+\!1) \!=\! \sum_{\ell\in\N{k}}  a_{\ell k} \, \bpsi_\ell(n+1)
	\end{split} \right.
\end{equation}
where $a_{\ell k}$ is used instead of $a_{2,\ell k}$.  As shown in the previous section, running~\eqref{eq:ATC} in a multitask environment leads to biased results.  We now discuss how to cluster nodes in order to reduce this effect.

\subsection{Clustering via matrix $\bA$ adjustments}

{Following}~\cite{Zhao2012}, we suggest to adjust matrix $\bA$ in an online manner via MSD optimization. At each instant $n$, the {instantaneous} MSD at node $k$ is given by
\begin{equation}
	\label{eq:MSDinst}
	\mathbb{E}\{\|\bv^\star_k(n+1)\|^2\} = E\Big\{\| \bw^\star_k - \sum_{\ell\in\N{k}}a_{\ell k}\,\bpsi_\ell(n+1)\|^2\Big\}.
\end{equation}
{Computation of this quantity requires the knowledge of $\bw^\star_k$}. Because the matrix $\bA$ is assumed left-stochastic, this expression can be rewritten~as
\begin{equation}
	\label{eq:MSDinst2}
       \begin{split}
        &\mathbb{E}\{\|\bv^\star_k(n+1)\|^2)\} = \!\!\sum_{\ell\in\N{k}} \!\sum_{p\in\N{k}} \!\! a_{\ell k}\, a_{p k}\,
        E\!\left\{[\bw^\star_k \!-\! \bpsi_\ell(n+1)]^\top[\bw^\star_k \!-\! \bpsi_p(n+1)] \right\}.
        \end{split}
\end{equation}
Let $\cb{\Psi}_k$ be the matrix at each node $k$ with $(\ell,p)$-th entry defined as
\begin{equation}
	[\cb{\Psi}_k]_{\ell p} = \left\{
	\begin{array}{l}
		\mathbb{E}\left\{ [\bw^\star_k - \bpsi_\ell(n+1)]^\top[\bw^\star_k - \bpsi_p(n+1)] \right\}, \\	
		                     \hspace{4.3cm}  \ell, p  \in \N{k} \\
		0, 	\hspace{4cm} \text{otherwise}.
	\end{array}
	\right.
\end{equation}
Let $\ba_{k} = [a_{1k}, \dots, a_{Nk}]^\top$. {Minimizing~\eqref{eq:MSDinst2} for node $k$ at time $n$, subject to left-stochasticity of $\bA$ and $a_{\ell k}=0$ for $\ell\notin\N{k}$, can be formulated as follows:}
\begin{equation}
	\label{opt.A2.full}
	\begin{split}
		&\min_{\ba_{k}} \quad \ba_{k}^\top \, \cb{\Psi}_k \, \ba_{k} \\
		& \text{subject to} 			\quad \cb{1}_N^\top\, \ba_{k} = 1,\quad  a_{\ell k}\geq 0,  \\
		&\phantom{\text{subject to}} 	\quad a_{\ell k}=0 \, \text{  if  }\,  \ell\notin\N{k}.
	\end{split}
\end{equation}
{Generally, {it is not possible to} solve this problem at each node $k$ since $\bw^\star_k$ and $\cb{\Psi}_k$ are unknown. We suggest to use an approximation for $\bw_k^\star$, to approximate matrix $\cb{\Psi}_k$ by an instantaneous value, and to drop its off-diagonal entries in order to make the problem tractable and have a closed-form solution (see \eqref{eq:a2sol}). The resulting problem is as follows:}
\begin{equation}
	\label{opt.A2.apprx}
	\begin{split}
		&\min_{\ba_{k}} \quad \sum_{\ell=1}^N a_{\ell k}^2 \,  \|\widehat{\bw}^\star_k - \bpsi_\ell(n+1))\|^2 \\
		& \text{subject to} 			\quad \cb{1}_N^\top\, \ba_{k} = 1,\quad  a_{\ell k}\geq 0,  \\
		&\phantom{\text{subject to}} 	\quad a_{\ell k}=0 \, \text{ if }\,  \ell\notin\N{k}
	\end{split}
\end{equation}
with $\widehat{\bw}_k^{\star}$ some approximation for $\bw_k^{\star}$. {The objective function shown above has the natural interpretation of penalizing the combination weight $a_{\ell k}$ assigned by node $\ell$ to node $k$ if the local estimate at node $\ell$ is far from the objective at node $k$.} The solution to this problem is given by{\footnote{{To achieve this result, discard the non-negativity constraint first, and write the {Lagrangian} function with respect to the equality constraint only. The solution to this simplified problem is given by~\eqref{eq:a2sol}. Observe that it satisfies the non-negativity constraint $a_{\ell k}\geq 0$. Consequently, \eqref{eq:a2sol} is also the solution to problem~\eqref{opt.A2.apprx}.}}}
\begin{equation}
	\label{eq:a2sol}
	a_{\ell k}(n+1) = \frac{\|\widehat{\bw}^\star_k - \bpsi_\ell(n+1))\|^{-2}}{\sum_{j\in\N{k}}\|\widehat{\bw}^\star_k
	- \bpsi_j(n+1))\|^{-2}},  \text{ for } \ell\in\N{k}.
\end{equation}
Let us now {construct an} approximation for $\bw^\star_k$ to be used in~\eqref{eq:a2sol}. In order to reduce the MSD bias that results from the cooperation of nodes performing distinct estimation tasks, one strategy is to use the local one-step approximation:
\begin{equation}
	\label{eq:update0}
	\widehat{\bw}^\star_k(n+1) = \bpsi_k(n+1) - \mu_k\, \nabla J_k(\bw)\big|_{\bw=\bpsi_k(n+1)}.
\end{equation}
Since the true gradient of $J_k(\bw)$ at $\bpsi_k(n+1)$ is not available in an adaptive implementation, we can approximate it by using the instantaneous value $\cb{q}_k(n) \triangleq e_k(n)\bx_k(n)$ with $e_k(n)=[d_k(n) - \bx_k^\top(n)\bpsi_k(n+1)]$.
This yields the following approximation:
\begin{equation}
	\label{eq:update}
	\widehat{\bw}^\star_k(n+1) = \bpsi_k(n+1) + \mu_k\,\cb{q}_k(n).
\end{equation}
Substituting this expression into \eqref{eq:a2sol},  we get the combination rule
\begin{equation}
	\label{eq:a2alg}
	\begin{split}
	a_{\ell k}(n\!+\!1) &\!=\! \frac{\|\bpsi_k(n+1) + \mu_k\, \cb{q}_k(n)- \bpsi_\ell(n+1)\|^{-2}}{\sum_{j\in\N{k}}\|\bpsi_k(n+1)
	+ \mu_k \,\cb{q}_k(n) - \bpsi_j(n+1)\|^{-2}}, \\
	  &  \text{ for } \ell\in\N{k}.
	\end{split}
\end{equation}
This rule  admits a useful interpretation. On the one hand, as mentioned above, it relies on the local estimate \eqref{eq:update0} in order to reduce the MSD bias effect caused by the cooperation of neighboring nodes estimating distinct parameter vectors. On the other hand, consider the inverse of the numerator of rule~\eqref{eq:a2alg}:
\begin{equation}
	\label{eq:num.combrule}
        \begin{split}
          &\|\bpsi_k(n+1) + \mu_k\, \cb{q}_k(n)- \bpsi_\ell(n+1)\|^2   \\
          &=\|\bpsi_\ell(n+1)-\bpsi_k(n+1)\| ^2+ 2[\bpsi_\ell(n+1)-\bpsi_k(n+1)]^\top[-\mu_k\,\cb{q}_k(n)] + \mu_k^2\,\|\cb{q}_k(n)\|^2.
         \end{split}
\end{equation}
The first term $\|\bpsi_\ell(n+1)-\bpsi_k(n+1)\| ^2$ on the RHS accounts for the distance of the current estimates between nodes $k$ and~$\ell$; this term tends to decrease the combination weight $a_{\ell k}(n+1)$ if this distance is large, and to limit information exchange. Now, consider the first-order Taylor series expansion of $J_k(\bw)$ at $\bpsi_k(n+1)$:
\begin{equation}
	J_k(\bpsi) \approx J_k(\bpsi_k(n+1)) - [\bpsi-\bpsi_k(n+1)]^\top\cb{q}_k(n).
\end{equation}
The second term $[\bpsi_\ell(n+1)-\bpsi_k(n+1)]^\top[-\mu_k\,\cb{q}_k(n)]$ on the RHS of~\eqref{eq:num.combrule} is proportional to $J_k(\bpsi_\ell(n+1))-J_k(\bpsi_k(n+1))$. This term also tends to decrease the combination weight $a_{\ell k}(n+1)$ if $J_k(\bpsi_\ell(n+1))>J_k(\bpsi_k(n+1))$. Indeed, in this case, it is not recommended to promote the combination of models $\bpsi_k(n+1)$ and $\bpsi_\ell(n+1)$ because the latter induces an increase of the cost function. Finally, $\mu_k^2\,\|\cb{q}_k(n)\|^2$ is the same for all $\ell\in\N{k}$. To summarize this discussion, the combination rule \eqref{eq:a2alg}  considers the closeness of the local estimate to the neighboring estimates, and the local slope of the cost function, to adjust the combination weights. This tends to promote information exchange between nodes that estimate the same optimum parameter vector, and thus to reduce the MSD bias and improve the estimation accuracy. 

\subsection{Algorithm}
The flexibility of multitask networks may be exploited by considering distinct cost functions for each node. This raises the issue of sharing information via the exchange matrix $\bC$, which can be simply set to the identity.
However, the time-variant combination matrix $\bA(n)$ determined by~\eqref{eq:a2alg} describes how each agent combines the parameter vectors transmitted by its neighbors as a function of the estimated contrast between tasks. An additional way to exploit this information is that each agent uses the reciprocity principle defined by
\begin{equation}
        \label{eq:CA}
	\bC(n+1) = \bA^\top(n+1).
\end{equation}
{The rationale underlying this principle is that the magnitude of $a_{\ell k}$ reflects the similarity of the estimation tasks performed by nodes $k$ and $\ell$, as it is perceived by node $k$. It is reasonable that node $\ell$ should use this information, and {scale} the local cost function accordingly. The smaller $a_{\ell k}$ is, the smaller $c_{k \ell}$ should be because nodes $k$ and $\ell$ do not address the same estimation problem. Other strategies, in the spirit of~\eqref{eq:a2alg}, may be considered to estimate the coefficients $c_{k \ell}$. 
{Moreover}, we found that using the normalized gradient $\cb{q}_k(n)/(\|\cb{q}_k(n)\|+\xi)$, with $\xi$ a small positive number to avoid division by zero, prevents premature convergence due to over-corrections.
%
%
The ATC diffusion algorithm with adaptive clustering defined by time-variant combination matrices $\bA(n)$ and $\bC(n)$ is summarized in Algorithm~\ref{algo:diffLMS.mult}.} {Considering that no prior information on clusters is available, we suggest to initialize the combination matrices $\bA(0)$ and $\bC(0)$ with $\bI_N$. During simulations, we did not experience convergence issues with other initial settings, provided that $\bA(0)$ and $\bC(0)$ are left-stochastic and right-stochastic, respectively. Further analysis can help guide more informed choices for the combination policies.}

\begin{algorithm}
\textbf{Initialization:} {Set $\bC(0) = \bI_N$ and $\bA(0) = \bI_N$. \\ Set $\bw_k(0) = 0$ for all $k= 1, ..., N$.}
\newline\textbf{Algorithm:}\hspace{2ex} At each time instant $n \geq 1$, and for each node $k$, update $\bpsi_k(n+1)$:
\begin{equation}
	\begin{split}
        &\bpsi_k(n+1) \\
        &\,\,=\ \bw_k(n) + \mu_k \sum_{\ell\in\N{k}} \!\! c_{\ell k}(n)\!\big[d_\ell(n)-\bx_\ell^\top\!(n)\bw_k(n)\big]\bx_\ell(n)
        \end{split}
\end{equation}
Update the combination coefficients:
\begin{equation}
             \begin{split}
                   & \cb{q}_k(n) = [d_k(n) - \bx_k^\top(n)\bpsi_k(n+1)]\,\bx_k(n)  \\
                   & \text{Optional: normalize } \cb{q}_k(n), \text{ i.e., use } \cb{q}_k(n)/(\|\cb{q}_k(n)\|+\xi) \\ 
                   &  a_{\ell k}(n\!+\!1)\! =\! \frac{\|\bpsi_k(n\!+\!1) \!+\! \mu_k \cb{q}_k(n)
                   	\!-\! \bpsi_\ell(n\!+\!1))\|^{-2}}{\sum_{j\in\N{k}}\|\bpsi_k(n\!+\!1) \!+\! \mu_k\! \cb{q}_k(n) \!-\! \bpsi_j(n\!+\!1))\|^{-2}}
	   \end{split}
\end{equation}
Optional: \quad $c_{k \ell}(n+1) = a_{\ell k}(n+1)$ \\
Combine weights:
\begin{equation}
              \bw_k(n+1) = \sum_{\ell\in\N{k}}  a_{\ell k}(n+1) \, \bpsi_k(n+1)
\end{equation}
\caption{ATC Diffusion LMS with adaptive clustering for multitask problems} \label{algo:diffLMS.mult}
\end{algorithm}

\section{Simulations}

In this section, we report simulation results that validate the algorithm and the theoretical results. The ATC diffusion LMS algorithm is considered. All nodes were initialized with zero parameter vectors $\bw_k(0)$. All simulated curves were obtained by averaging over 100 runs, {since this gave sufficiently smooth curves to check consistency with theoretical results\footnote{Matlab source code is available at {http://www.jie-chen.com.}}.

\subsection{Model validation}

\label{sec:Simu_I}
{For the validation}, we consider a network consisting of 8 nodes with interconnections shown in Fig.~\ref{fig:simu1}(a). The parameter vectors to be estimated are of length $L = 2$. The optimum mean vectors are uniformly distributed on a circle of radius $r$ centered at $\bw_o$, that is,
\begin{equation}
	\begin{split}
		& \bw^\star_k = \bw_o + r \left(\begin{array}{c} \cos \theta_k \\ \sin\theta_k\end{array}\right) \\
		& \theta_k = 2\pi(k-1)/N+{\pi}/{8}.
	\end{split}
\end{equation}
The regression inputs $\bx_{k}(n)$ were zero-mean $2\times 1$ random vectors governed by a Gaussian distribution with covariance matrices {$\bR_{x,k} = \sigma_{x,k}^2\,\bI_L$.} The background noises $z_k(n)$ were i.i.d. zero-mean Gaussian random variables, and independent of any other signal. {The variances $\sigma_{x,k}^2$ and $\sigma_{z,k}^2$ are depicted in Fig.~\ref{fig:simu1_var}.} We considered the ATC diffusion LMS with measurement diffusion governed by a uniform matrix $\bC$ such {that $c_{\ell k} = |\N{\ell}|^{-1}$ for all $k \in \N{\ell}$.} The combination matrix $\bA$ simply averaged the estimates from the neighbors, namely, $a_{\ell k} = |\N{k}|^{-1}$ for $\ell \in \N{k}$. For all nodes, the step-sizes were set to $\mu_k = 0.01$.

\begin{figure}[!t]
  \subfigure[Network topology. ]{
 \label{fig:simu1_topo}
   	\begin{minipage}[c]{.48\linewidth}
   		\centering
      		\includegraphics[trim = 0.5mm -15mm 0mm 0mm, clip, scale=0.4]{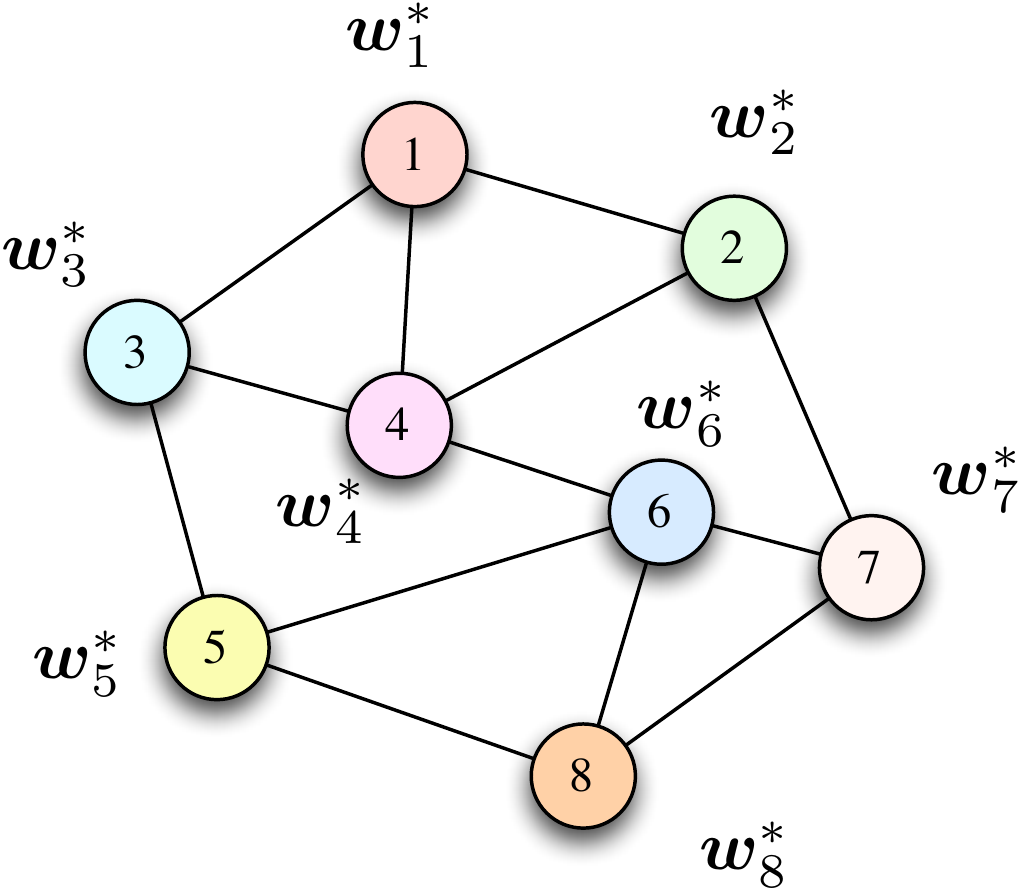}
   	\end{minipage}} 
 \subfigure[\,Input variances (top) and noise variances (bottom).]{
 \label{fig:simu1_var}
   		\begin{minipage}[c]{.48\linewidth}
   		\hspace{-5mm}
      		\includegraphics[trim =30mm 13mm 15mm 0mm, clip, scale=0.30]{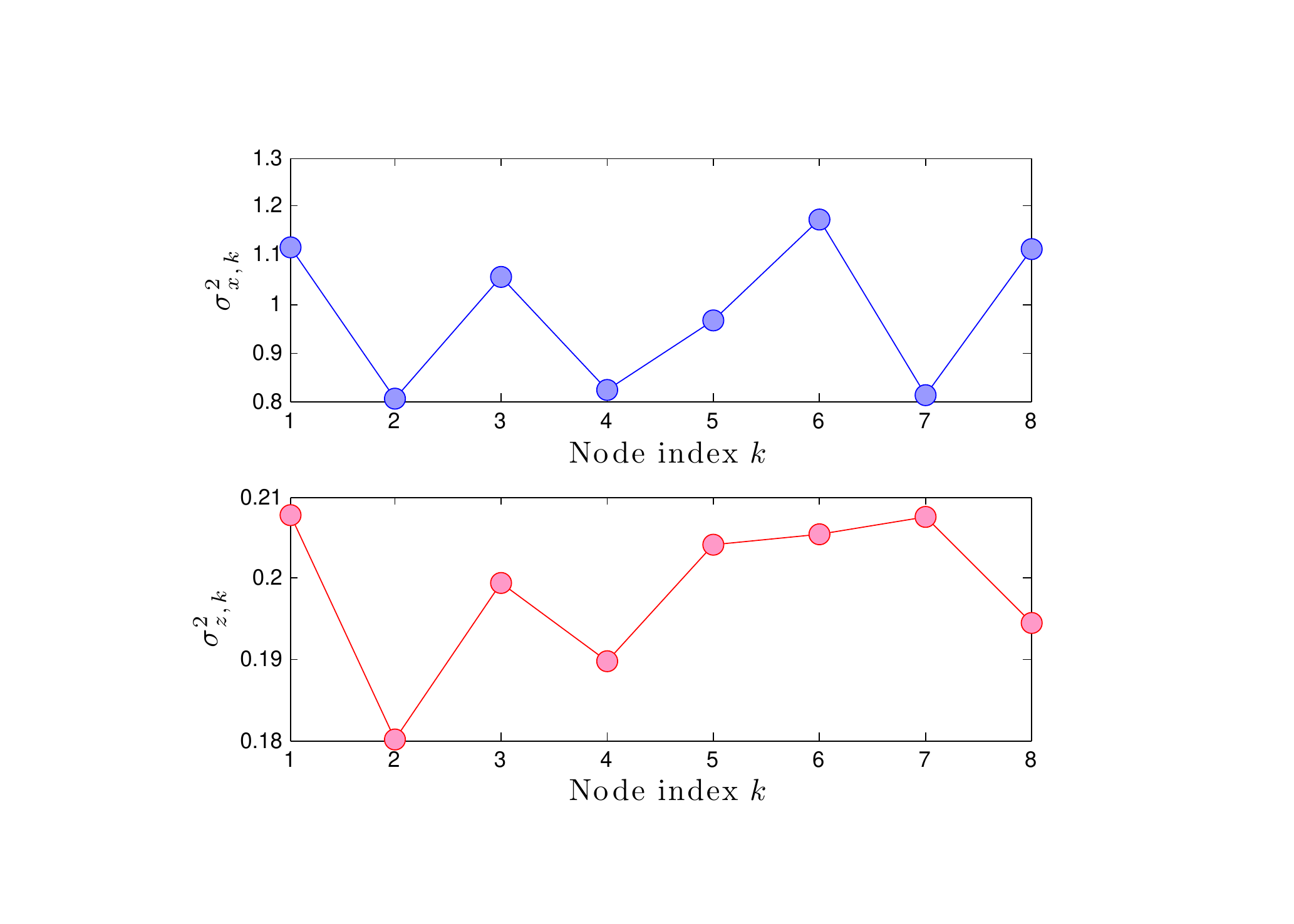}
   	\end{minipage}}
	\vspace{-4mm}
	\caption{(a) Network studied in Section~\ref{sec:Simu_I}, with $8$ nodes. (b) Input signal and noise
	variances for each sensor node.}
	\label{fig:simu1}
	\vspace{-5mm}
\end{figure}

 \begin{figure*}[!t]
      		\includegraphics[trim = 20mm 10mm 20mm 10mm, clip, scale=0.38]{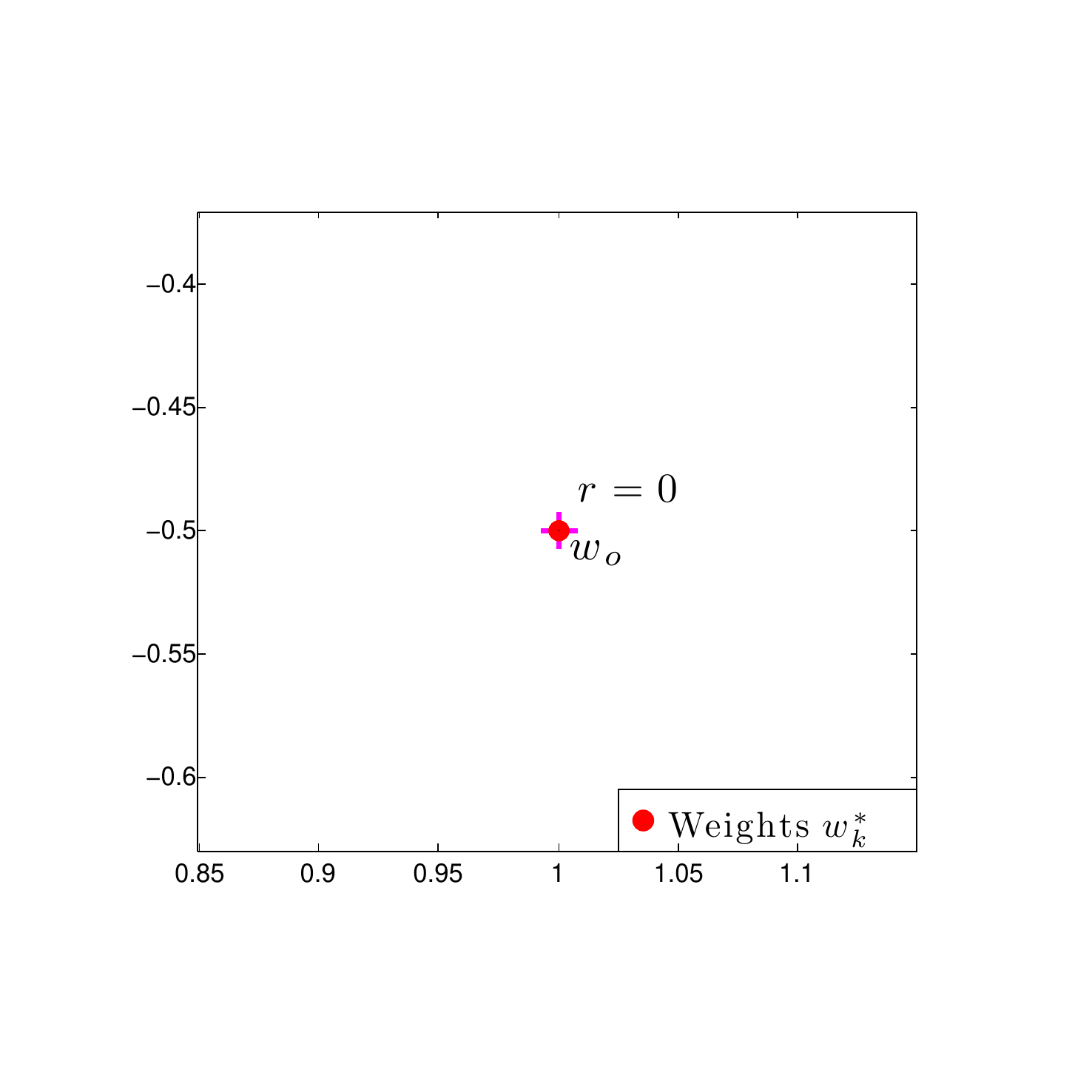} \;
		\includegraphics[trim = 20mm 10mm 20mm 10mm, clip, scale=0.38]{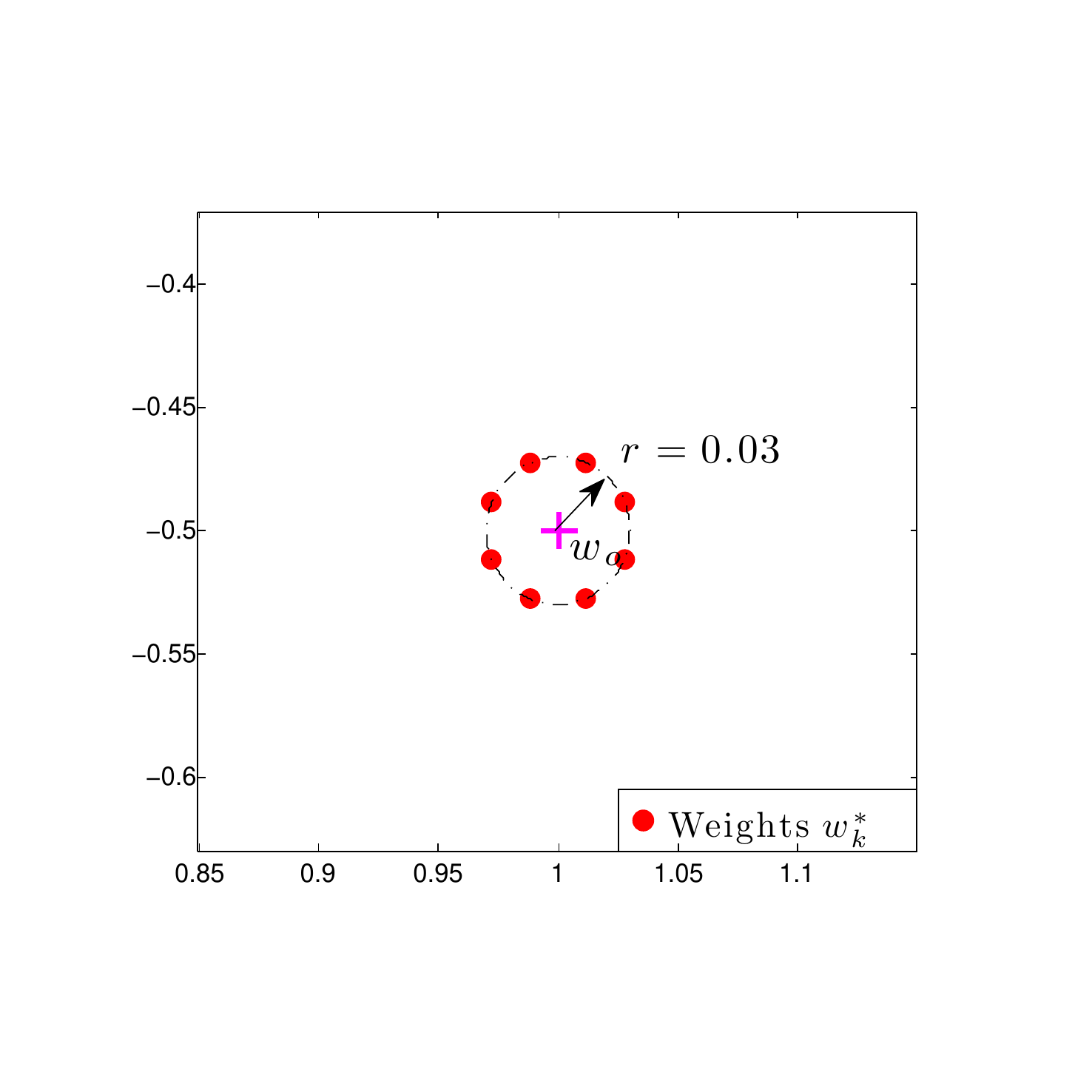}  \;
		\includegraphics[trim = 20mm 10mm 20mm 10mm, clip, scale=0.38]{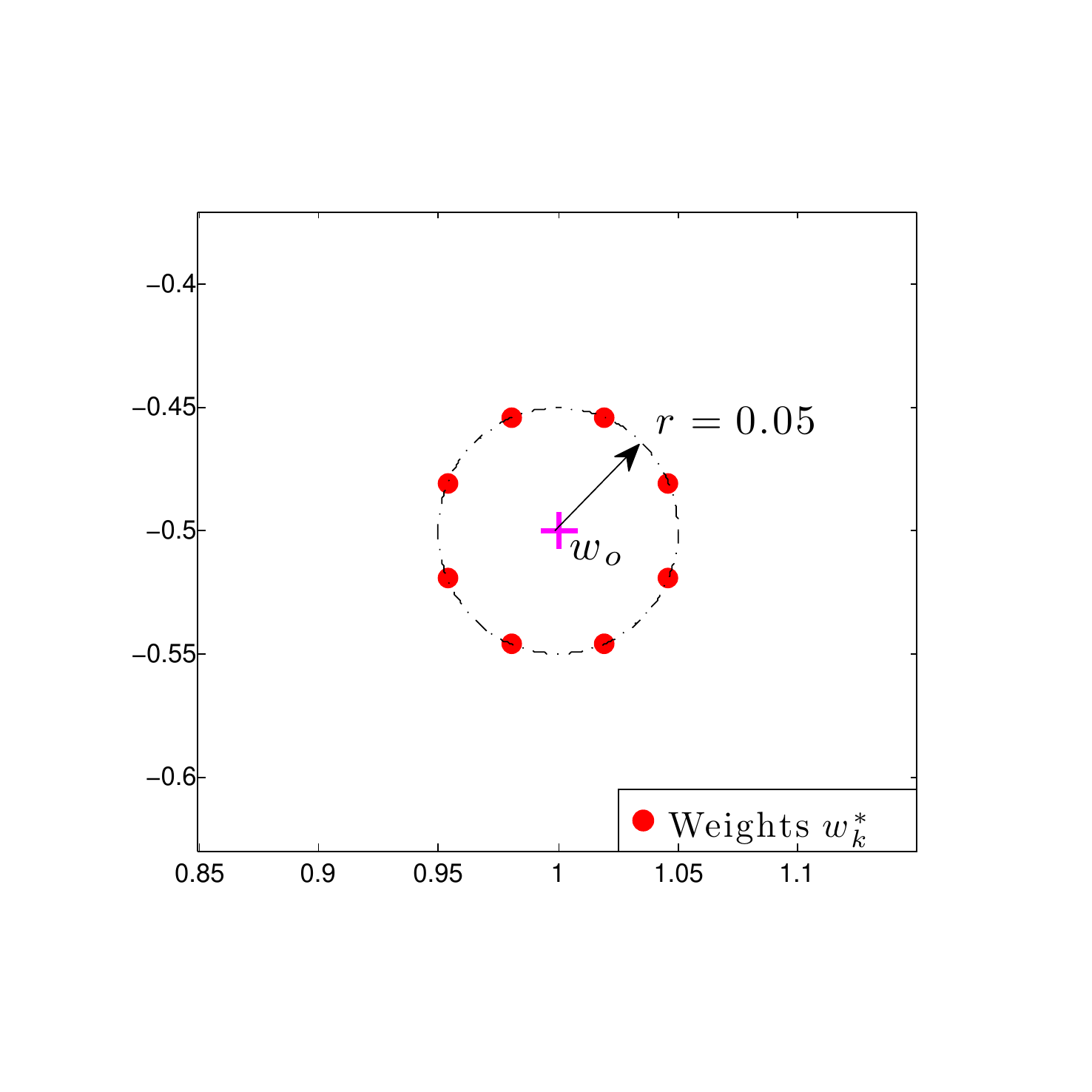}  \;
     		\includegraphics[trim = 20mm 10mm 20mm 10mm, clip, scale=0.38]{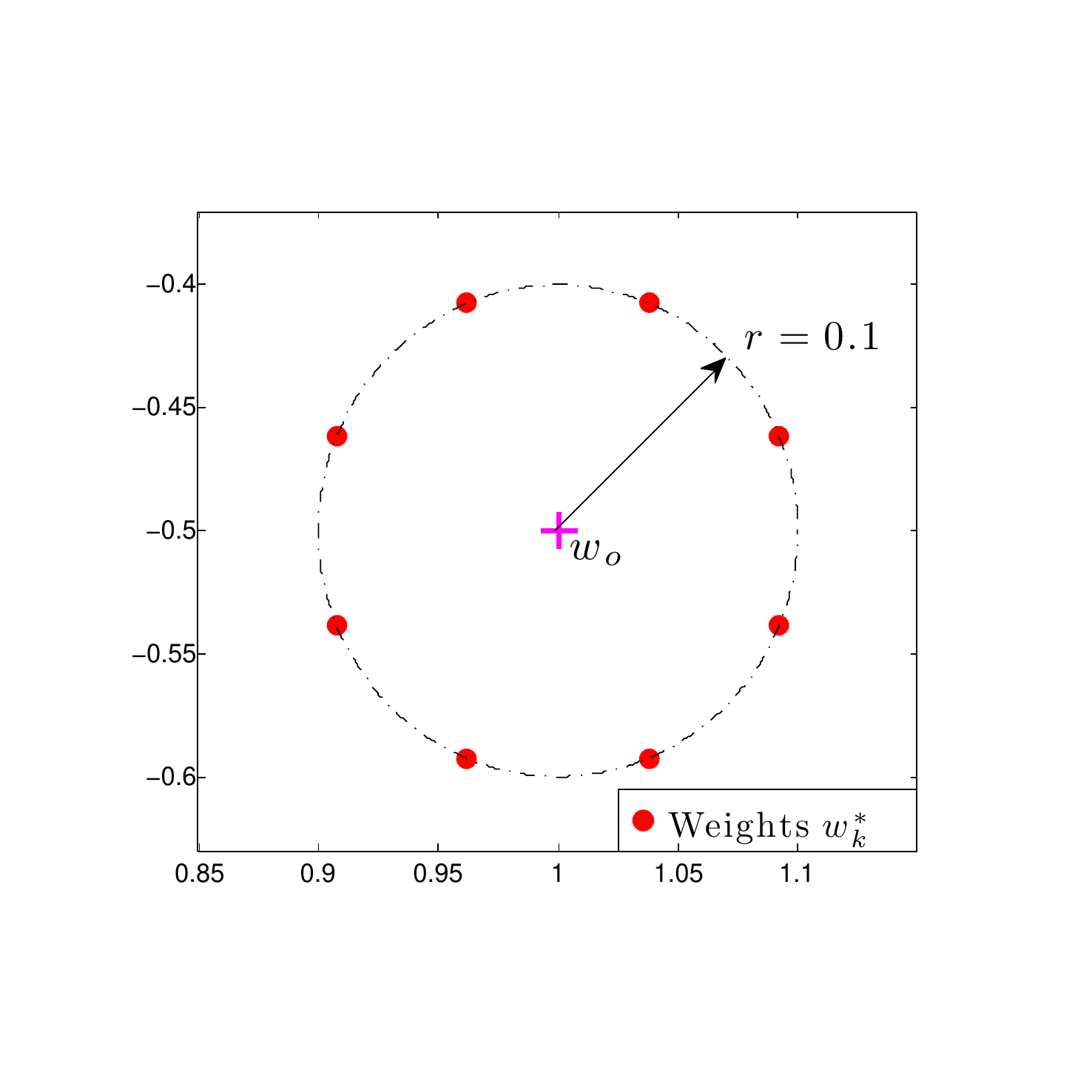}    \;
         \vspace{-8mm}
	\caption{Node coefficients centered at $\bw_o=[1, -0.5]^\top$ with $r=0$,  $r=0.03$, $r=0.05$ and $r=0.1$ (from left to right).}
	\label{fig:coeff1a}
	\vspace{-5mm}
\end{figure*}

\subsubsection{Stationary optimums}
We first check the convergence  analysis with stationary parameter vectors, that is, $\sigma_{\epsilon,k}^2$ = 0 for all nodes. Four groups of  coefficient vectors, centered at $\bw_o=[1, -0.5]^\top$ with $r = 0$, $r=0.03$, $r=0.05$ and $r=0.1$ were considered, as illustrated in Fig.~\ref{fig:coeff1a}. 
Note that the case $r=0$ corresponds to the single-task network where $\bw_k^\star = \bw_o$ for each node. Running ATC diffusion LMS with these four settings, we obtained the MSD curves shown in Fig.~\ref{fig:simu1a_MSD}. {Observe that the theoretical and simulated transient MSD curves are accurately superimposed.} The non-cooperative LMS algorithm was also considered. Since the average steady-state MSD of the non-cooperative LMS algorithm {over all nodes} is approximately given by~\cite{sayed2008adaptive,Sayed2013diff}:
\begin{equation}
	\text{MSD}^\text{network}_\text{lms} = \frac{1}{N}\sum_{k=1}^N\frac{\mu_k\,\sigma_{z,k}^2\;L}{2},
\end{equation}
then the MSD behavior with the different settings is almost the same, provided the other parameters remain unchanged. Consequently, the theoretical MSD curve for the non-cooperative LMS algorithm is only provided for $r=0.05$. It can be observed that diffusion LMS can still be advantageous over non-cooperative LMS if the differences between local optimum weight vectors are sufficiently small, $r=0$ and $r=0.03$ in this simulation. However, when the contrast between the tasks increases, diffusion LMS provides lower performance than non-cooperative LMS due to the bias introduced by the algorithm, $r=0.05$ and $r=0.1$ {in this simulation}. 


\begin{figure}[!t]
         \centering
	\includegraphics[trim = 15mm 5mm 0mm 5mm, clip, scale=0.55]{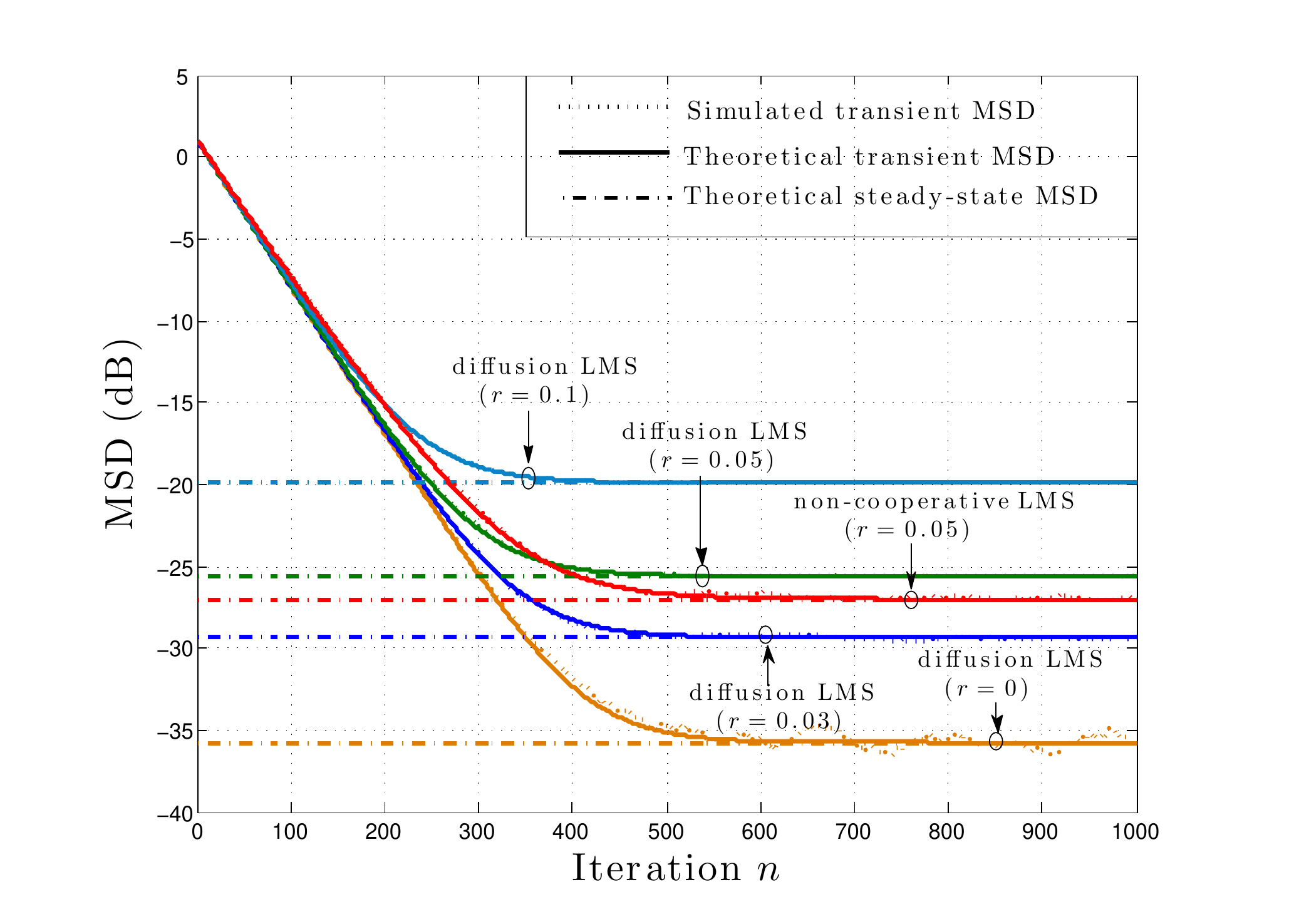}
         \vspace{-7mm}
	\caption{Network MSD behavior for the deterministic case. Theoretical MSD curves were obtained by {Corollary~1}
	and steady-state MSD values were obtained by {Corollary~2}. Simulated and theoretical transient MSD curves are perfectly superimposed.}
	\label{fig:simu1a_MSD}
	\vspace{-5mm}
\end{figure}


\subsubsection{Randomly perturbed optimums}
We now consider the network described previously with $r=0$ so that the differences between the optimum weight vectors $\bw_k^\star(n)$  arise from the random perturbations $\beps_k(n)$. Keeping all the other parameters unchanged, the variance of these perturbations was successively set to
$\sigma_{\epsilon}^2 =0$, $0.01$, $0.05$ and $0.1$ for all the agents. MSD curves for diffusion LMS and non-cooperative LMS are provided in Fig.~\ref{fig:simu1b_MSD}. It can be observed that diffusion LMS always outperformed its non-cooperative counterpart. This experiment shows the advantage provided by cooperation. The relative performance gain becomes smaller as $\sigma_\epsilon^2$ increases because weight lags caused by random perturbations dominate the estimation error.

\begin{figure}[!t]
          \centering
	\includegraphics[trim = 15mm 5mm 0mm 5mm, clip, scale=0.55]{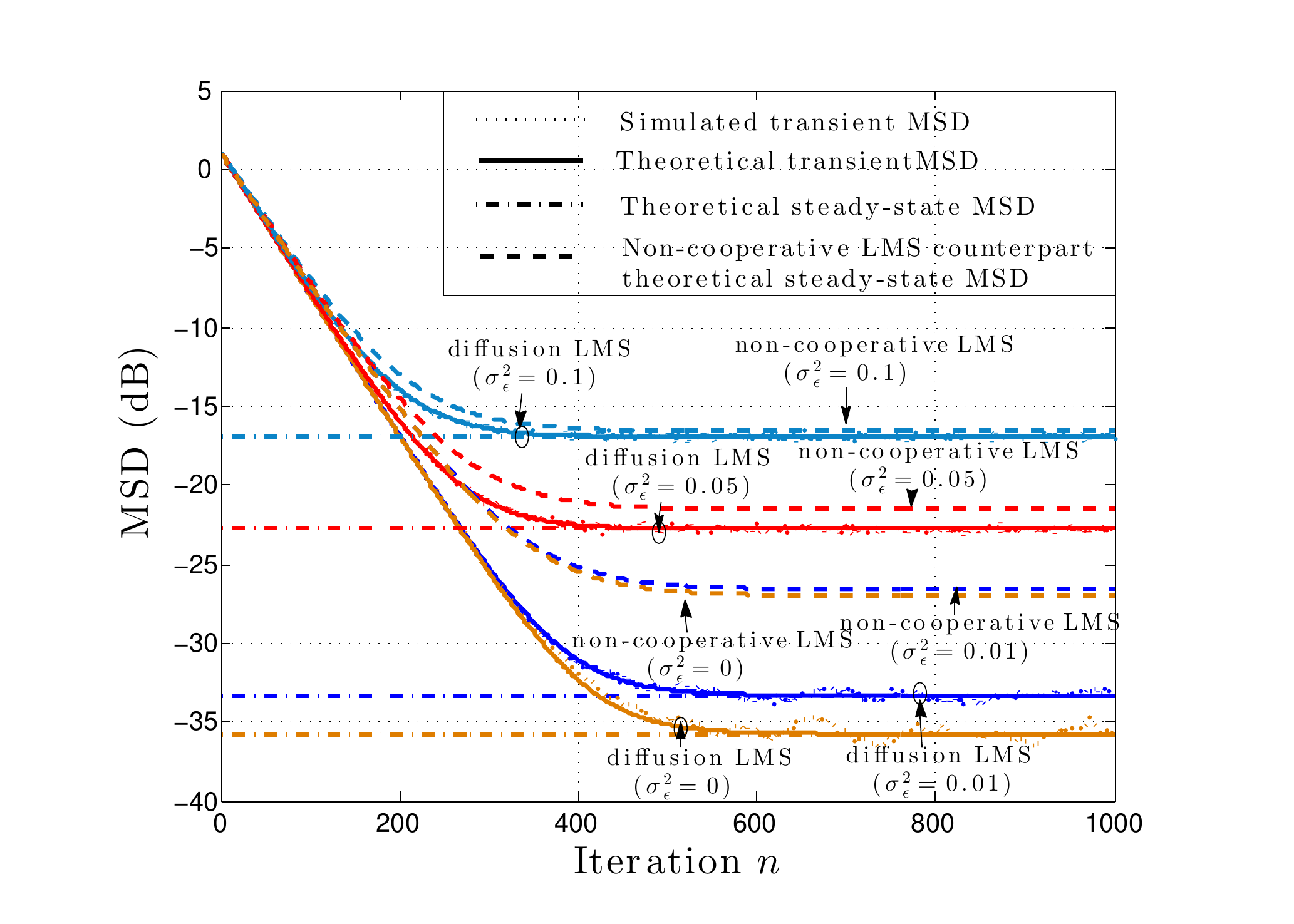}
         \vspace{-5mm}
	\caption{Network MSD behavior for the perturbation-only case. Theoretical MSD curves were obtained by {Corollary~1} and steady-state
	MSD values were obtained by {Corollary~2}. Note that simulated and theoretical transient MSD curves are  superimposed.}
	\label{fig:simu1b_MSD}
	\vspace{-5mm}
\end{figure}

\subsubsection{Correlated in time inputs}
This simulation example illustrates the accuracy of models \eqref{eq:zeta}--\eqref{eq:MSD} for inputs correlated in time. We considered regression vectors
\begin{equation}
          \bx_k(n) = [x_k(n) \; x_k(n-1)]^\top
\end{equation} 
with a first-order AR model given by
\begin{equation}
          x_k(n) =  0.5\,x_k(n-1)+\sqrt{(1-0.5^2)\sigma_{x,k}^2}\,w_k(n).
\end{equation}
The parameters $\sigma^2_{x,k}$ were set as in Fig.~\ref{fig:simu1_var}. The noise $w_k(n)$ was i.i.d. and drawn from a zero-mean Gaussian distribution with variance $\sigma_w^2=1$, so that
\begin{equation*}
	\bR_{x,k} = \sigma_{x,k}^2\left(\begin{array}{cc}1& 0.5\\0.5& 1\end{array}\right)
\end{equation*}
The diffusion LMS algorithm was tested in the following experimental settings:
\begin{equation}
	\begin{split}
		{\cp{S}}_1 &: \{r=0.01, \sigma_\epsilon=0.010\} \\
		{\cp{S}}_2 &: \{r=0.05, \sigma_\epsilon=0.015\} \\
		{\cp{S}}_3 &: \{r=0.10, \sigma_\epsilon=0.015\}.
	\end{split}
\end{equation}
Although Assumption 1 is not valid, observe in Fig.~\ref{fig:simu1c_MSD} that the theoretical and simulated transient MSD curves are superimposed. This illustrates the accuracy of the {analysis when the step-sizes are sufficiently small}.
\begin{figure}[!h]
          \centering
	\includegraphics[trim = 5mm 0mm 0mm 10mm, clip, scale=0.55]{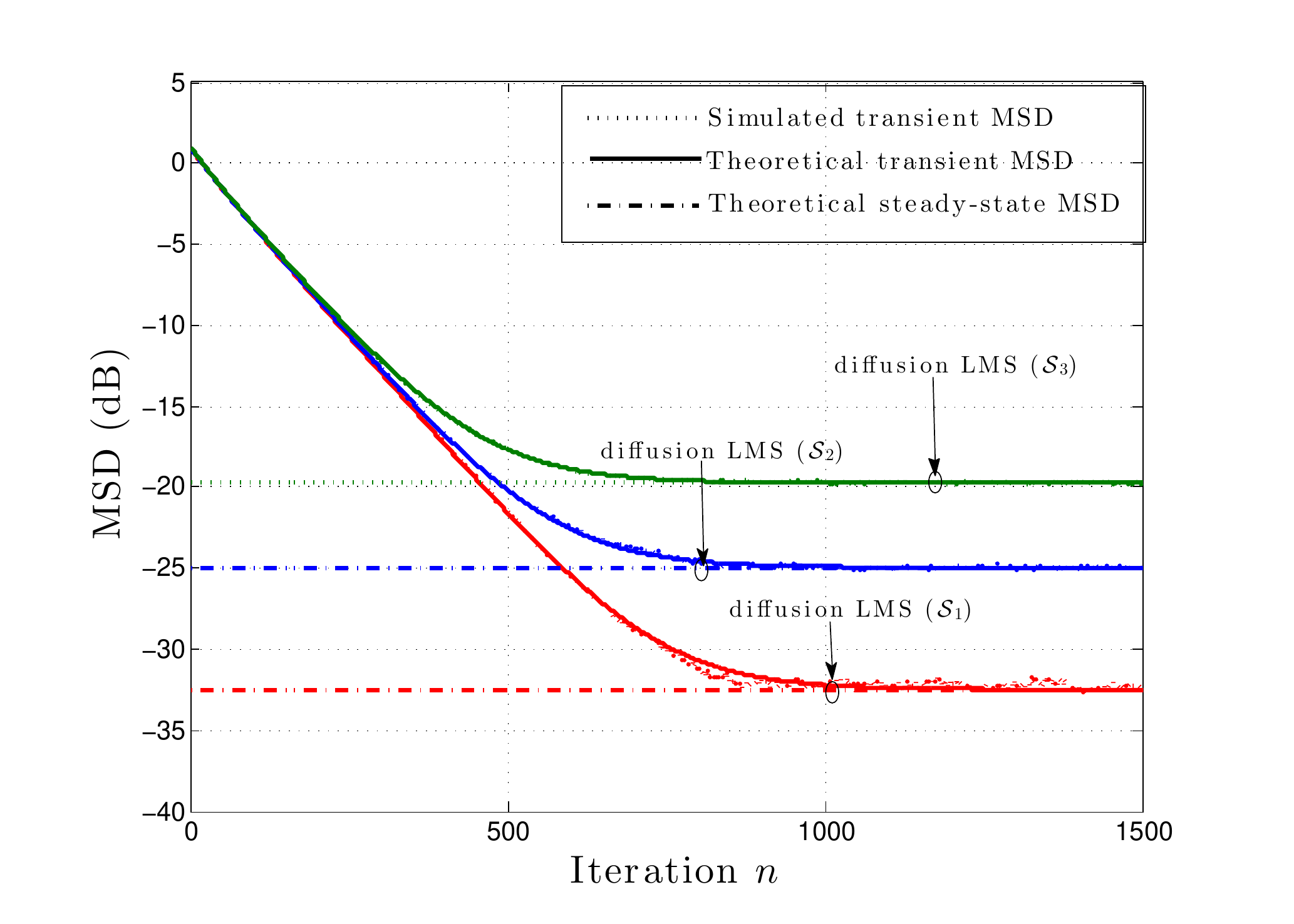}
	\vspace{-5mm}
	\caption{Network MSD behavior for correlated inputs.
	Theoretical MSD curves were obtained by Corollary~1 and steady-state MSD values were obtained by Corollary~2.
	Simulated curves and theoretical curves are accurately superimposed.}
	\label{fig:simu1c_MSD}
\end{figure}

\subsection{Adaptive clustering in multitask networks}

We shall now illustrate the performance of diffusion LMS with adaptive clustering in a multitask environment. Our approach is compared with the strategy introduced in~\cite{Zhao2012}. For the latter, as suggested in~\cite{Zhao2012}, the so-called smoothing factor $\gamma$ was set to $0.1$. A stationary problem is first considered. Next, a dynamic problem with time-varying clusters is introduced in order to confirm the reliability of our approach.

\subsubsection{Stationary environment}
\label{sec:Simu_cluster_st}

Consider the network of $16$ agents depicted in Fig.~\ref{fig: multi_topo_real}.  The regression inputs $\bx_{k}(n)$ were zero-mean $2\times 1$ random vectors governed by a Gaussian distribution with covariance matrices {$\bR_{x,k} = \sigma_{x,k}^2\,\bI_L$}.  The background noises $z_k(n)$ were i.i.d. zero-mean Gaussian random variables, independent of any other signals. {The  variances $\sigma_{x,k}^2$ and $\sigma_{z,k}^2$ are depicted in Fig.~\ref{fig:varzx2}.} The scenario under study is a multitask problem with a cluster {structure. 
Nodes} $1$ to $4$ belong to the first cluster. Nodes $5$ to $9$ are in the second cluster. Nodes $10$ to $14$ compose the third cluster, and nodes $15$ and $16$ are in the fourth cluster. The parameter vectors to be estimated are as follows:
\begin{equation}
	\bw_k^\star = \left\{
	\begin{array}{lllr}
		&[0.5 \,-0.4]^\top     &  k = 1,\dots, 4  		& \quad\text{Cluster 1}  \\
		&[0.6 \,-0.2]^\top     &  k = 5,\dots, 9   	& \quad\text{Cluster 2}  \\
		&[0.3 \,-0.3]^\top     &  k = 10,\dots, 14   	& \quad\text{Cluster 3}  \\
		&[-0.8 \;\;0.5]^\top   &  k = 15, 16           	& \quad\text{Cluster 4}
	\end{array}
	\right.
\end{equation}
{Note that the distances between {the} optimum parameter vectors for clusters 1, 2 and 3 are much smaller than those with respect to cluster 4, which acts as an outlier.} 
The following algorithms were considered for estimating the four optimum parameter vectors: 1) diffusion LMS with a uniform combination matrix~$\bA$, 2) non-cooperative LMS, 3) diffusion LMS with the clustering strategy introduced in~\cite{Zhao2012}, 4) diffusion LMS with our clustering strategy, with $\bC = \bI$ and $\bC(n)=\bA^\top(n)$. The step-size was set to $\mu=0.01$ for all  nodes.
\begin{figure}[!t]
	\centering
 	\subfigure[Network topology]{
		\includegraphics[trim = 5mm 5mm 5mm 0mm, clip, scale=0.4]{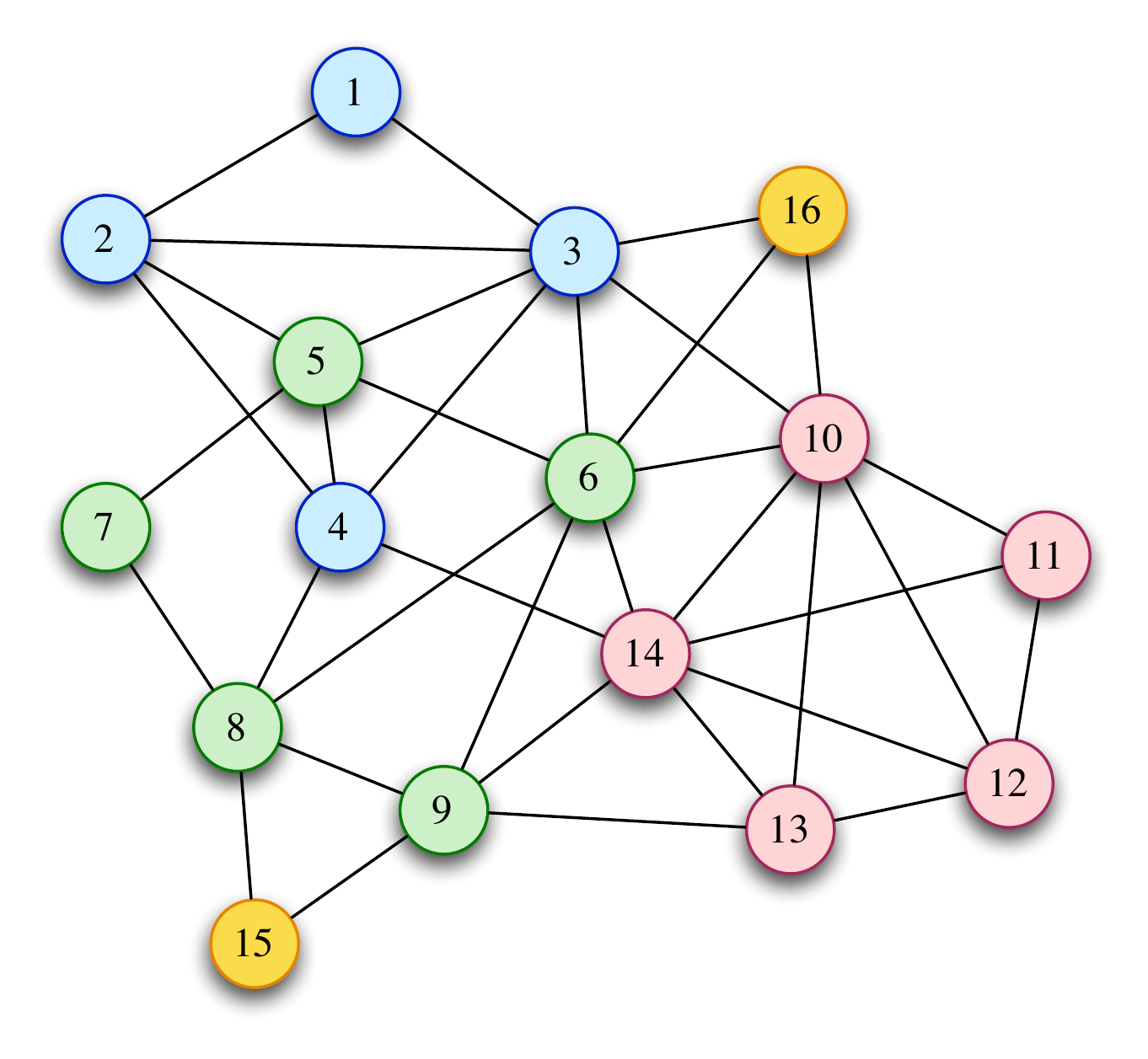}
		\label{fig: multi_topo_real}
		\vspace{-5mm}} \qquad
	\subfigure[Input variances (top) and noise variances (bottom)]{
		\includegraphics[trim =35mm 15mm 42mm 0mm, clip, scale=0.42]{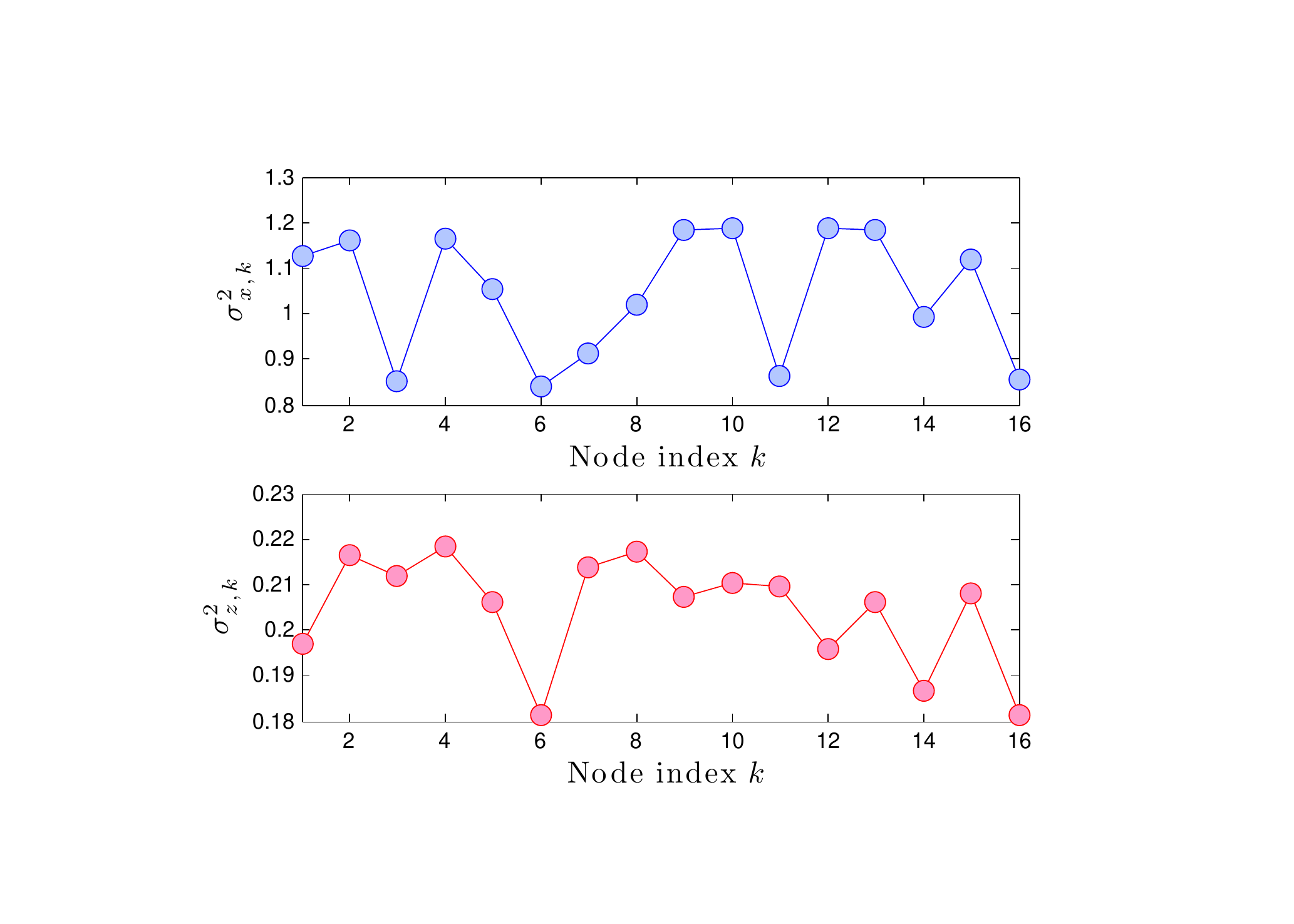}
		\label{fig:varzx2}
		\vspace{-7mm}}
	\caption{Network topology in Section~\ref{sec:Simu_cluster_st} and associated input variances and noise variances.}
	\label{fig: multi_topo}
        \vspace{-4mm}
\end{figure}

{Figure~\ref{fig:St_MSD} illustrates} the MSD convergence behavior for these algorithms. Due to large bias of the estimated weights, diffusion LMS with a uniform combination matrix had large MSD. Non-cooperative LMS performed much better as it provides unbiased estimates. The proposed algorithm with $\bC=\bI$ achieved better performance, and $\bC(n)=\bA^\top(n)$ led to additional performance gain due to information exchange. Finally, in order to provide a straightforward but visually-meaningful clustering result, we averaged the combination matrix $\bA$ over the last 100 iterations of a single realization, and we considered that $a_{\ell k} > 0.05$ represents a one-way connection from $\ell$ to $k$. The estimated relationships between nodes provided in Fig.~\ref{fig:multi_topo_est} perfectly match the ground truth configuration.


\begin{figure}[!t]
	\centering
	\subfigure[MSD behavior.]{
	\includegraphics[trim = 5mm 0mm 0mm 0mm, clip, scale=0.4]{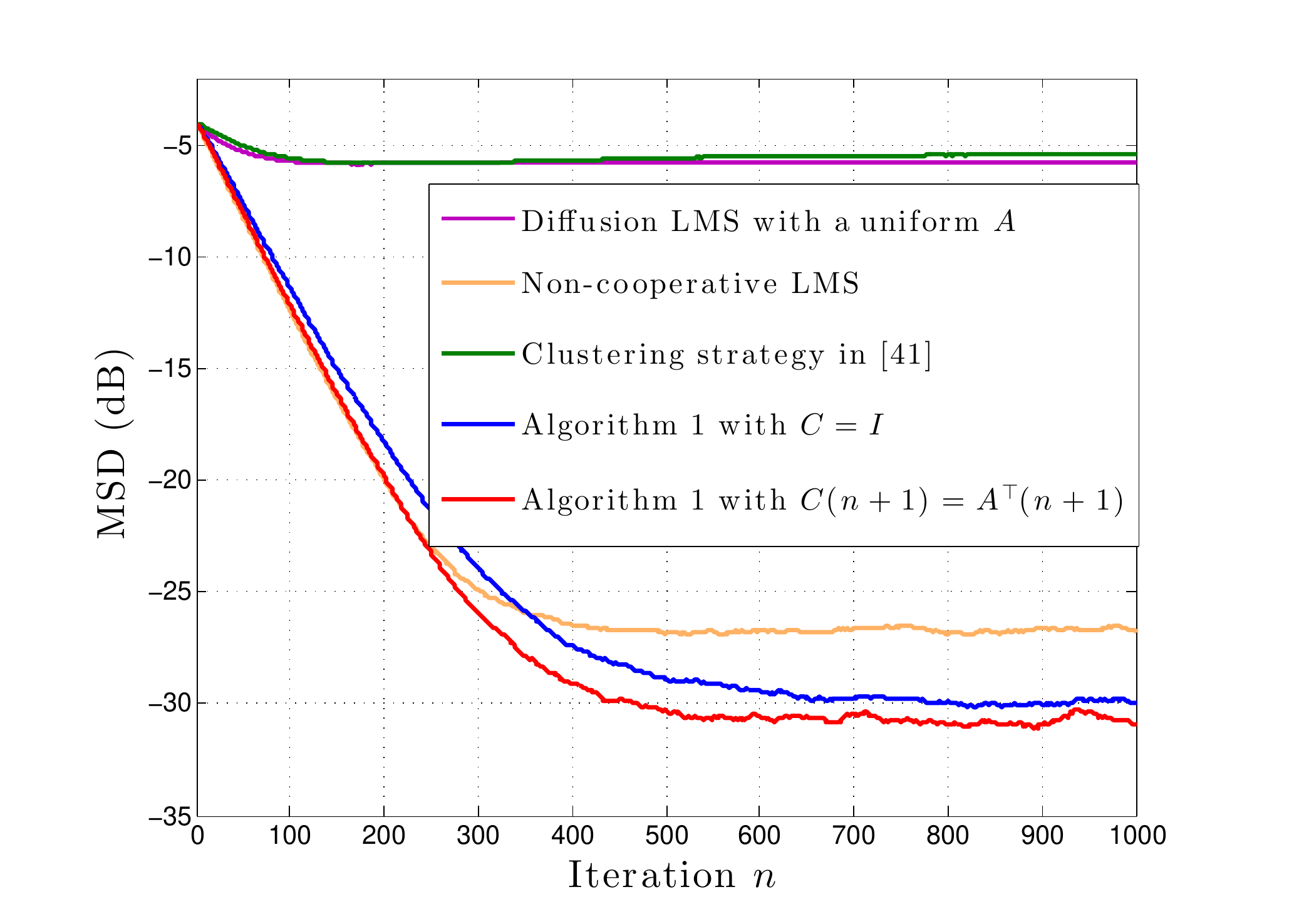}    	
	\label{fig:St_MSD} \vspace{-5mm}}
	\subfigure[Estimated cluster structure.]{
	\includegraphics[trim = 0mm 5mm 0mm 0mm, clip, scale=0.5]{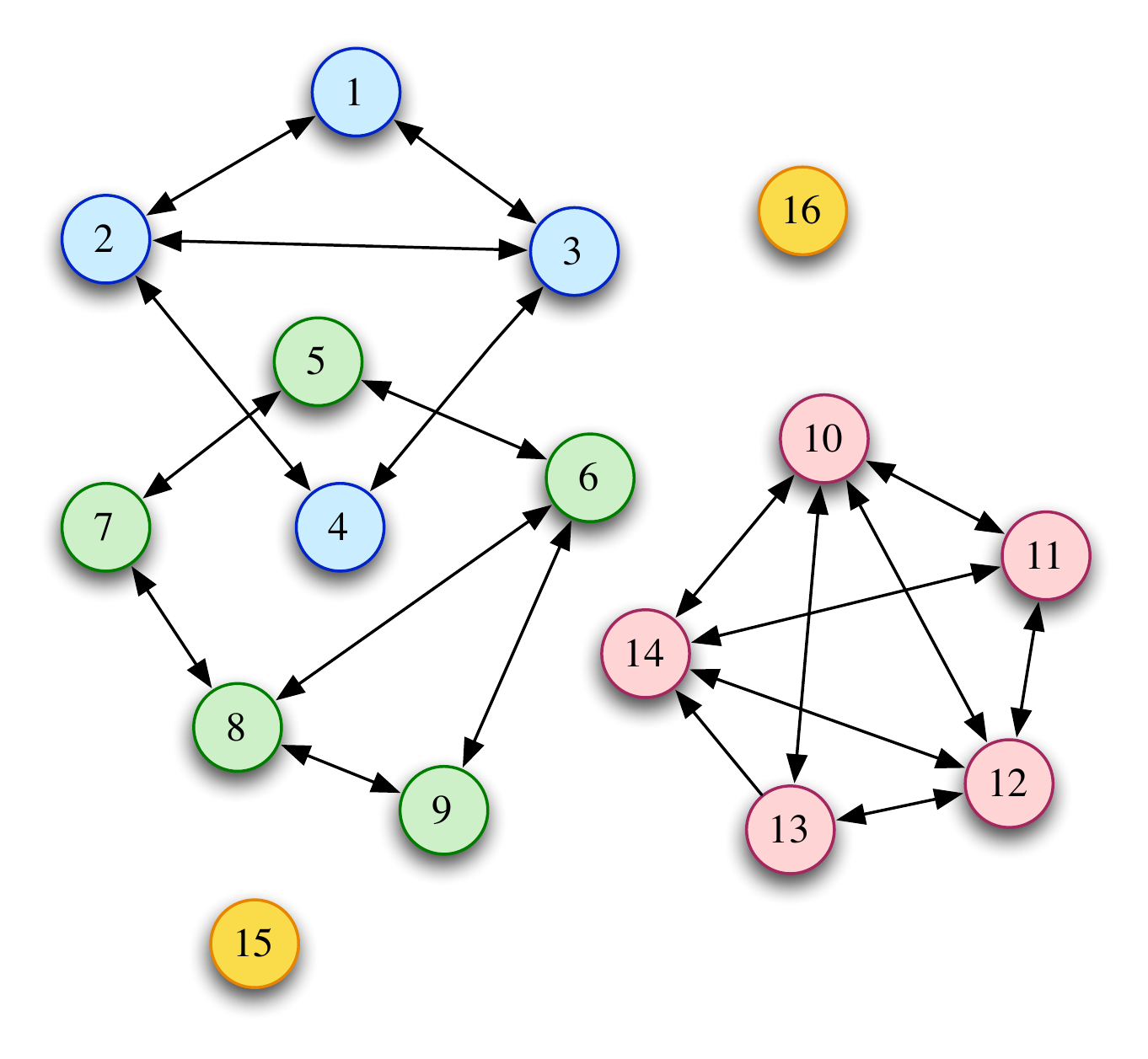}
	\label{fig:multi_topo_est}   \vspace{-8mm}}				
	\caption{Network MSD comparison in a stationary multitask environment, and estimated cluster structure by the proposed algorithm (averaged over the last 100 instants in one realization). {The weight iterates for~\cite{Zhao2012}  were initialized at the same value. If random well-separated initial conditions are used across the nodes, then the performance of~\cite{Zhao2012}  becomes similar to that of the non-cooperative solution in the above plot.}}
	 \vspace{-5mm}
\end{figure}

\subsubsection{Non-stationary environment}

Consider now a more complex environment where clusters vary over time. Four stationary stages and three transient episodes were modeled in this experiment. Properties of input signals and noise were the same as those in the stationary case considered above. From instant $n=1$ to $1000$, the network consisted of one cluster with a unique optimum parameter vector to estimate. From $n=1501$ to $2500$, nodes were split into two clusters with two different optimums. From $n = 3001$ to $4000$, nodes were split again to give four clusters. Finally, from instant $n=4501$, nodes were aggregated into one cluster with {another} unique parameter vector to {estimate}. 
Transient episodes were designed with linear interpolation between each steady-state stage over a period of $500$ time samples. Taking, for example, the first component of the weight vector of node $1$ over the time interval $1$ to $2500$, the time-variant optimum $w_{1,1}^\star(n)$  is expressed by
\begin{equation}
             w_{1,1}^\star(n) \!=\!\left\{
             \begin{array}{ll}
             0.3, &  n = 1, \dots, 1000\\
             0.3 + \frac{0.5-0.3}{500} (n-1000), &  n = 1001,\dots, 1500\\
             0.5, & n = 1501, \dots, 2500.
             \end{array} \right.
\end{equation}
Cluster structures and optimum parameter vectors are illustrated in Fig.~\ref{fig:Nst_topo} and~\ref{fig:Nst_weights}, respectively.

The same four algorithms as before were considered for comparison. {Figure~\ref{fig:Nst_weights_est} shows their mean weight behavior. Conventional diffusion LMS with a uniform combination matrix made all nodes converge to the same Pareto optimum during all phases. The non-cooperative LMS estimated the optimum weight vectors without bias. The algorithm presented in~\cite{Zhao2012}, which generally performs well for well-separated tasks and well-separated initial {random} weights, did not perform well with this setting. Our algorithm showed the same convergence behavior for $\bC = \bI$ and $\bC(n)=\bA^\top(n)$. Only the case $\bC(n)=\bA^\top(n)$ is presented here due to space limitation. It can be observed that, for each node, the parameter vector converged properly in accordance to the original cluster structures represented in Fig.~\ref{fig:Nst_topo}.} MSD learning curves are shown in Fig.~\ref{fig:Nst_MSD}. Transient stages can be clearly observed on both weight behavior and MSD behavior curves. Diffusion LMS enforced the weight vectors estimated by each agent to converge to the same solution at each stage. As a consequence, the MSD learning curve shows poor performance due to large bias. Non-cooperative LMS converged without bias towards the optimum parameter vectors. The algorithm introduced by~\cite{Zhao2012} showed some ability to conduct clustering but did not provide satisfactory results during transient episodes. During stages 1 and 4, it worked as well as diffusion LMS. However, during stages 2 and 3, it only performed slightly better than diffusion LMS. The proposed algorithm was able to track the system dynamic with correct clustering and appropriate convergence in the mean-square sense.

\begin{figure*}[!t]
	\centering
	\includegraphics[trim = 5mm 0mm 5mm 0mm, clip, scale=0.3]{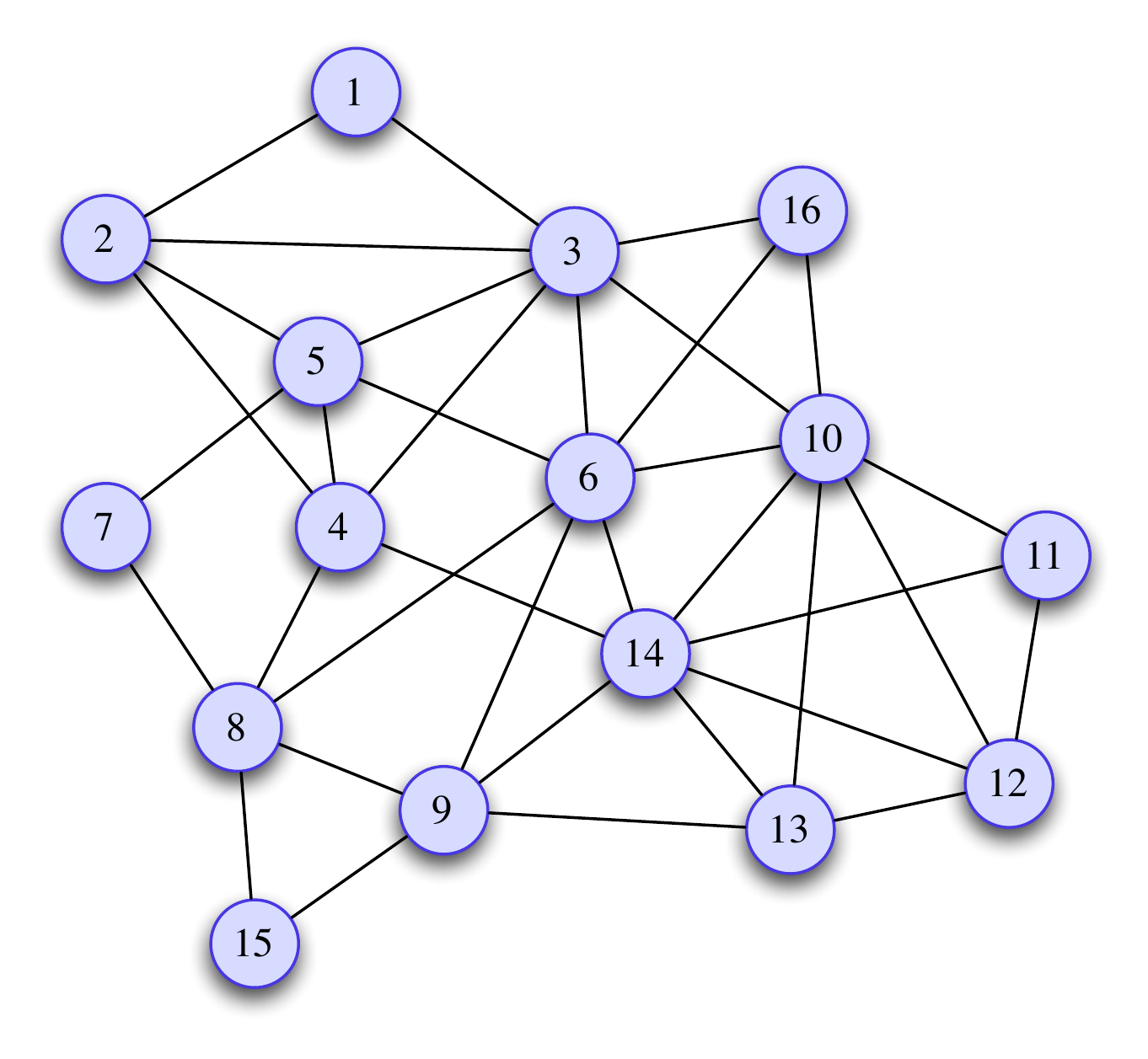}
	\includegraphics[trim = 8mm -50mm 8mm 0mm, clip, scale=0.3]{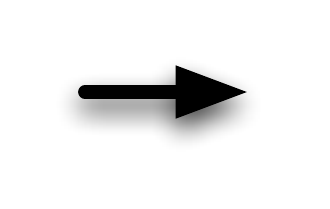}
	\includegraphics[trim = 5mm 0mm 5mm 0mm, clip, scale=0.3]{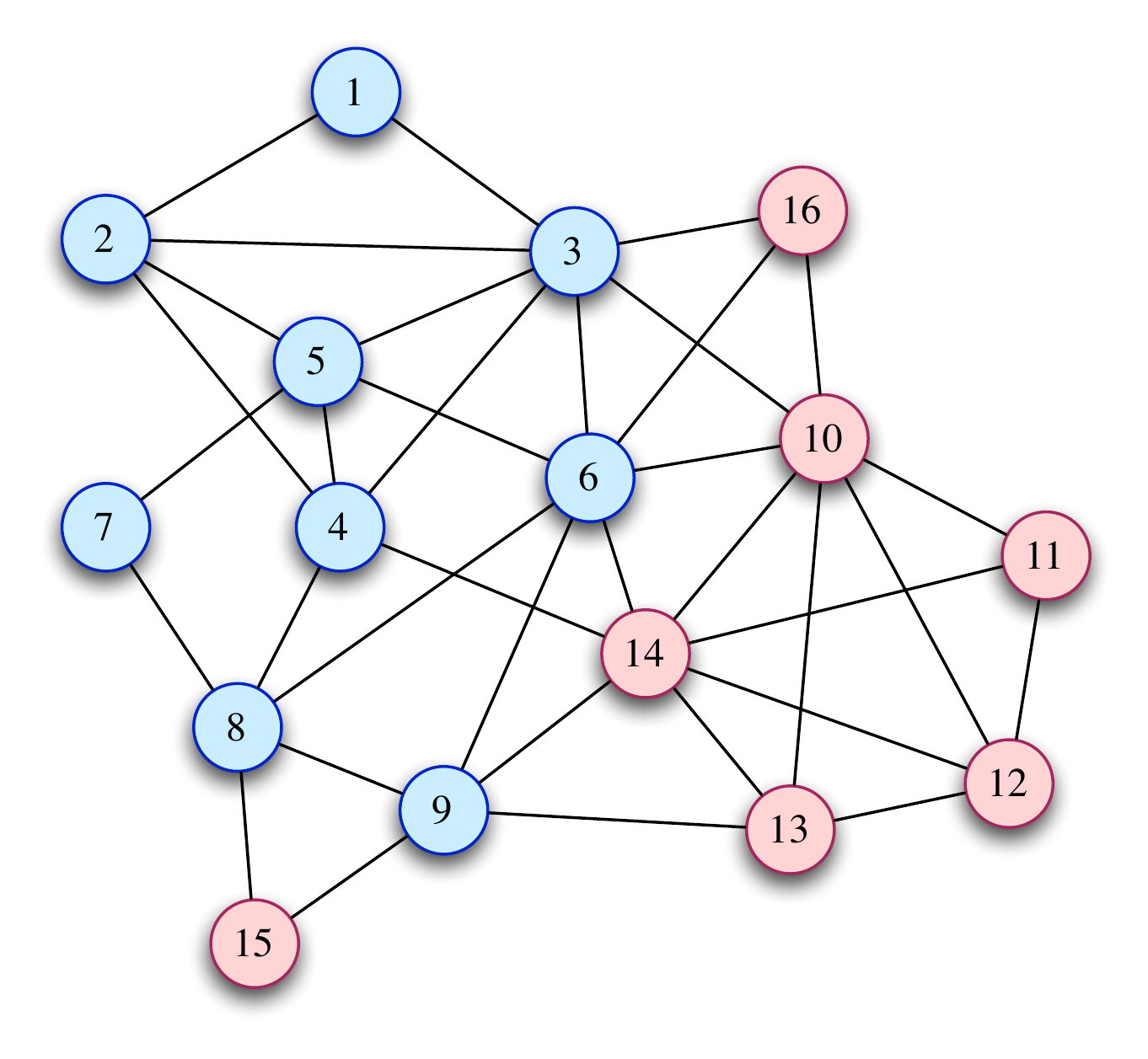}
	\includegraphics[trim = 8mm -50mm 8mm 0mm, clip, scale=0.3]{arrow.pdf}
	\includegraphics[trim = 5mm 0mm 5mm 0mm, clip, scale=0.3]{Multitask_4tks.pdf}
	\includegraphics[trim = 8mm -50mm 8mm 0mm, clip, scale=0.3]{arrow.pdf}
	\includegraphics[trim = 5mm 0mm 5mm 0mm, clip, scale=0.3]{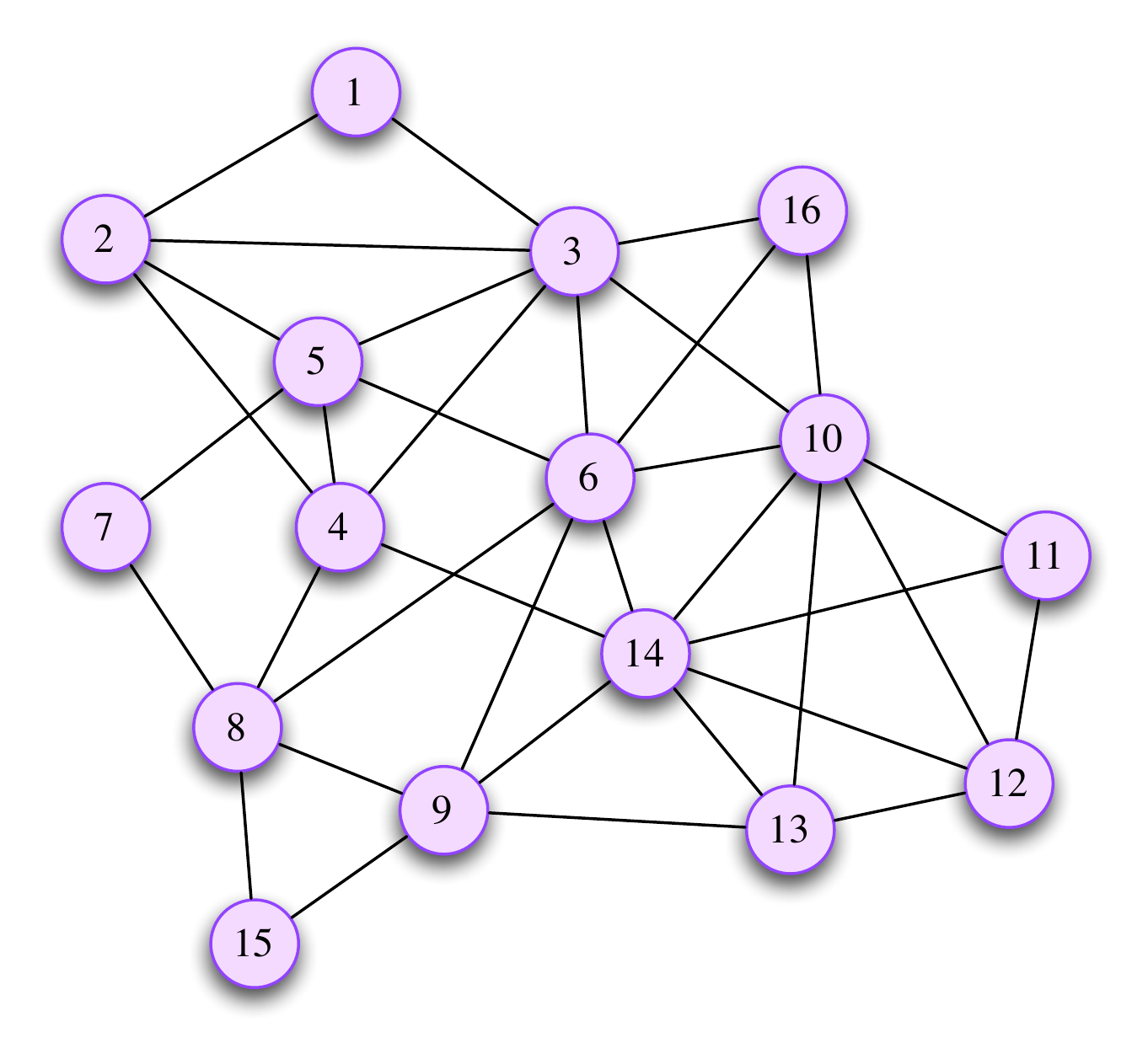}
	\vspace{-5mm}
	\caption{Evolution of cluster structures of the network (1 cluster $\rightarrow$ 2 clusters $\rightarrow$ 4 clusters $\rightarrow$ 1 cluster).}
	\label{fig:Nst_topo}
\end{figure*}

\begin{figure}[!t]
	\centering
	\includegraphics[trim = 10mm 3mm 10mm 10mm, clip, scale=0.5]{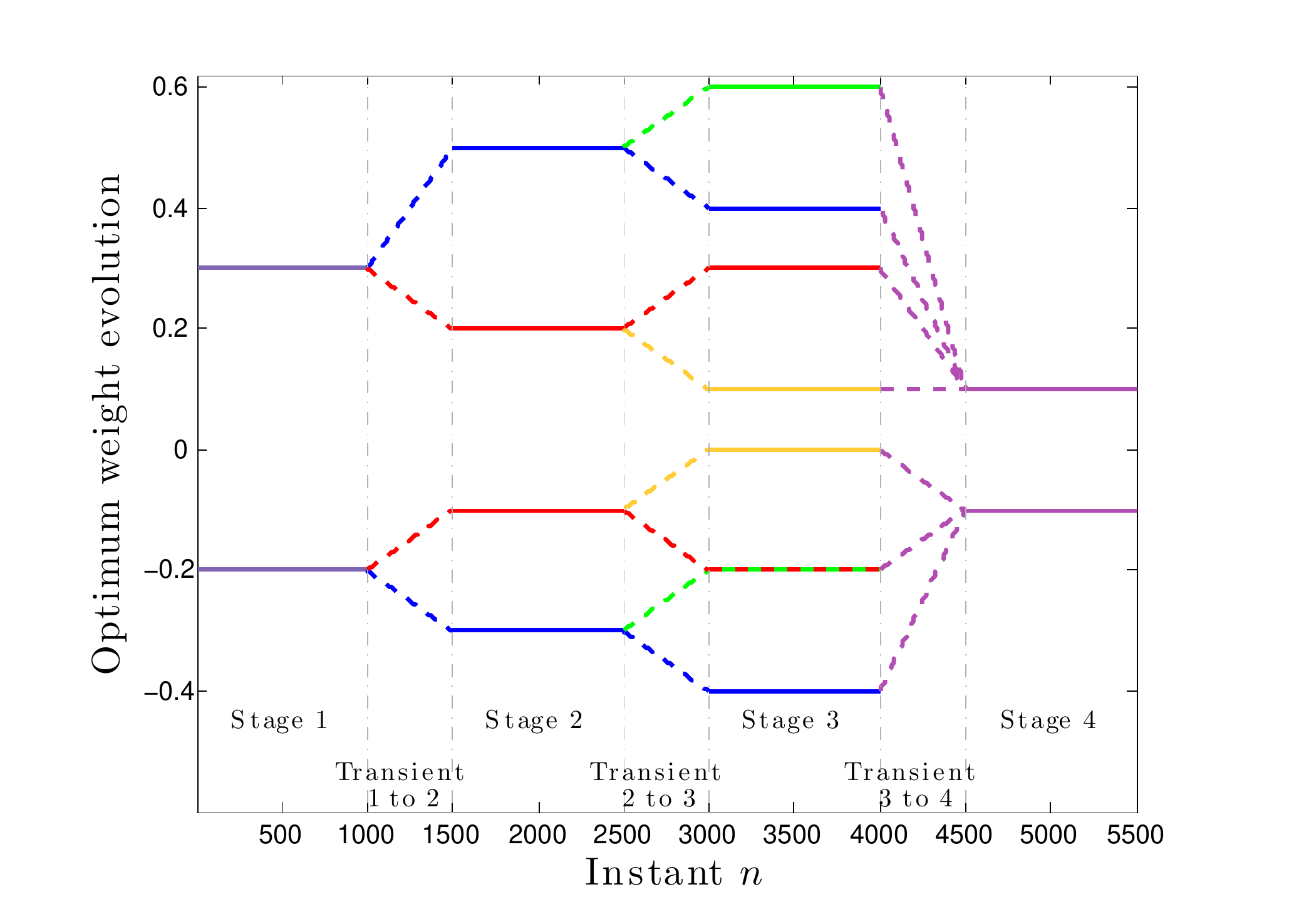}
	\vspace{-4mm}
	\caption{Evolution of clusters over time. Colors are consistent with those of clusters in Fig.~\ref{fig:Nst_topo}.
	Dashed lines represent optimums during transient episodes.}
	\label{fig:Nst_weights}
	\vspace{-2mm}
\end{figure}

\begin{figure*}[!t]  \addtolength{\subfigcapskip}{-0.2in}
	\centering
          \subfigure[Diffusion LMS with uniform $\bA$, $\bC$.]{\hspace{-1mm}
	\includegraphics[trim = 15mm -10mm 22mm 10mm, clip, scale=0.25]{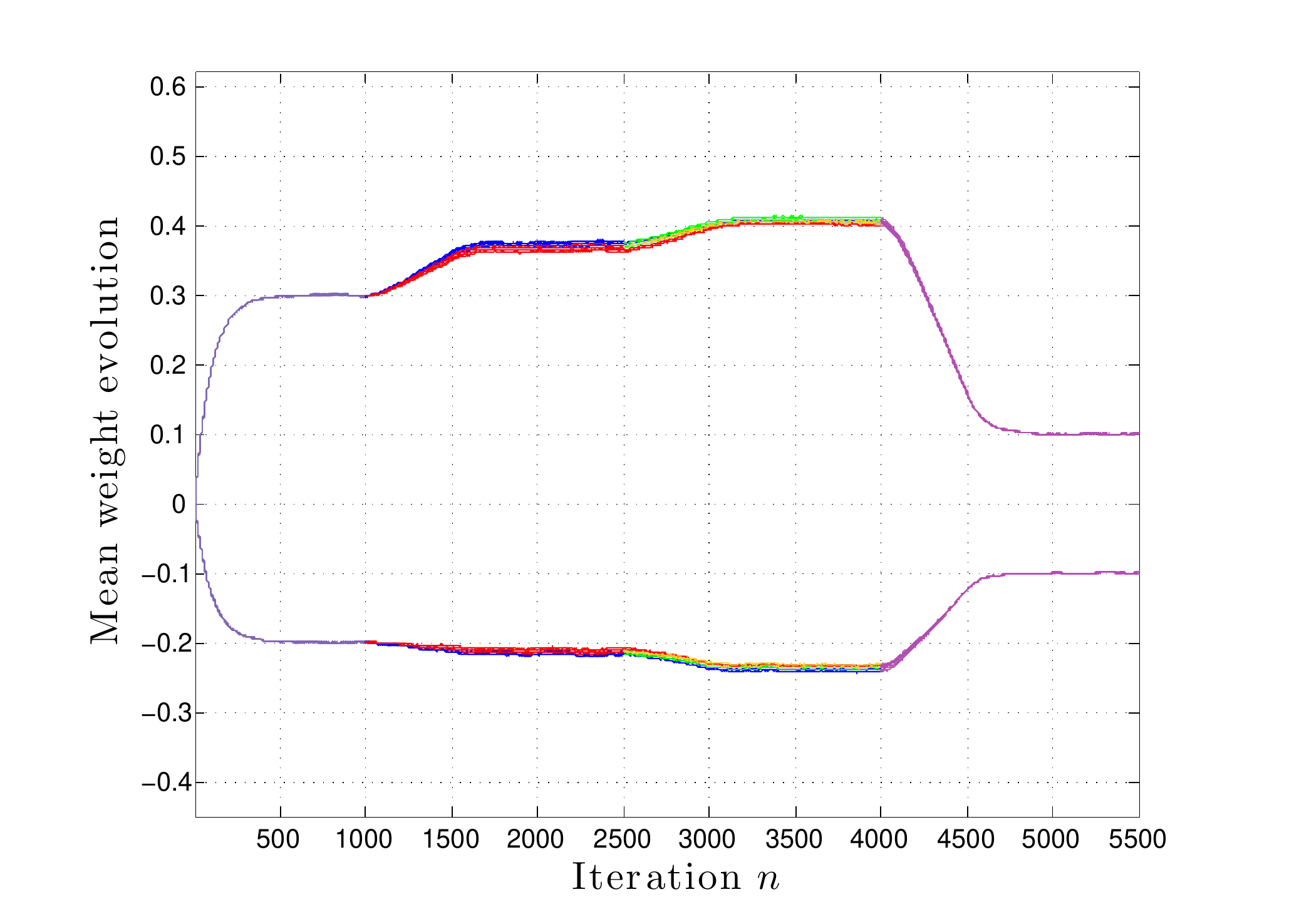} }
         \subfigure[Non-cooperative LMS.]{\hspace{-1mm}
	\includegraphics[trim = 15mm -10mm 22mm 10mm, clip, scale=0.25]{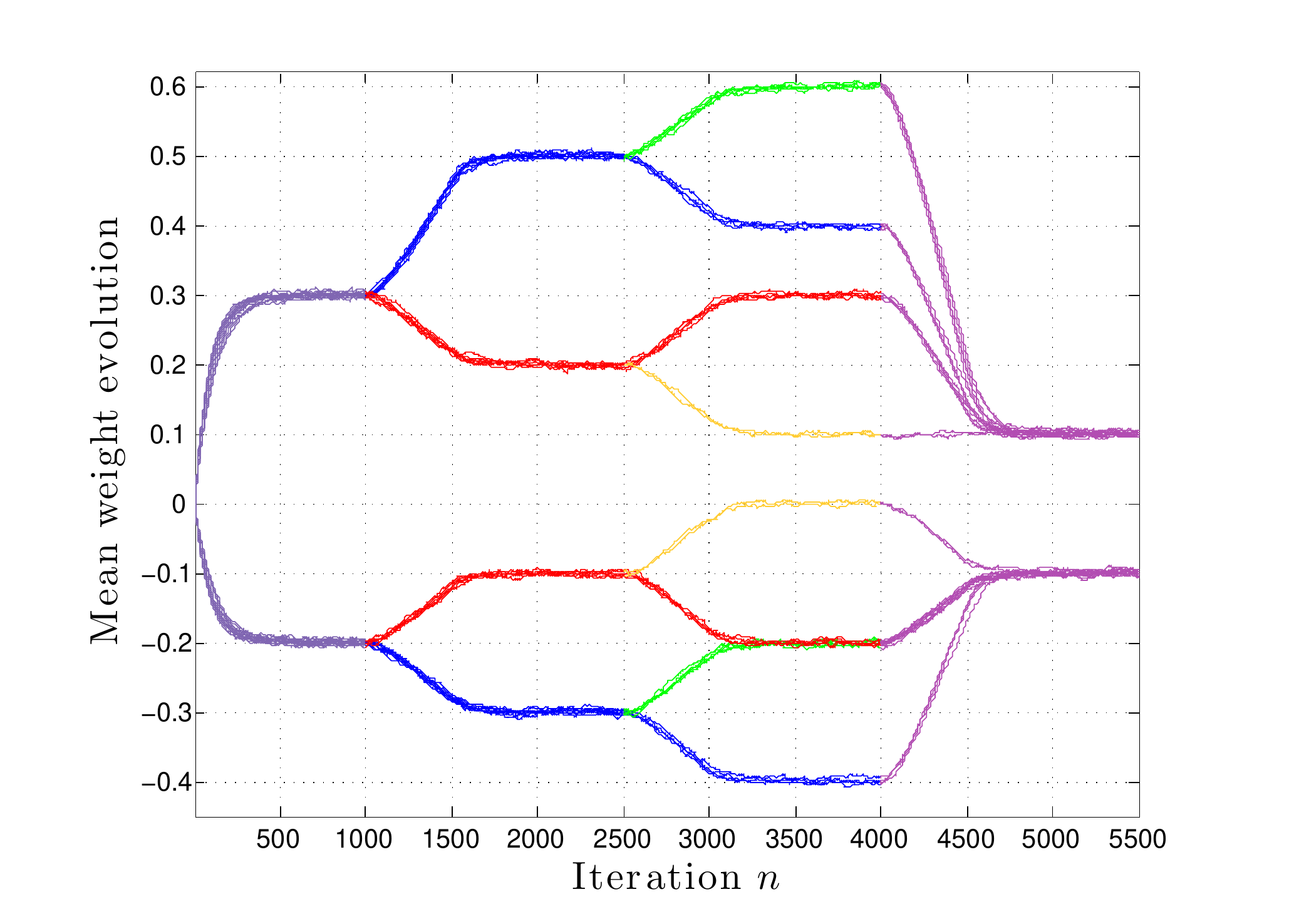} }
	\subfigure[Algorithm in~\cite{Zhao2012}. ]{\hspace{-1mm}
	\includegraphics[trim = 15mm -10mm 22mm 10mm, clip, scale=0.25]{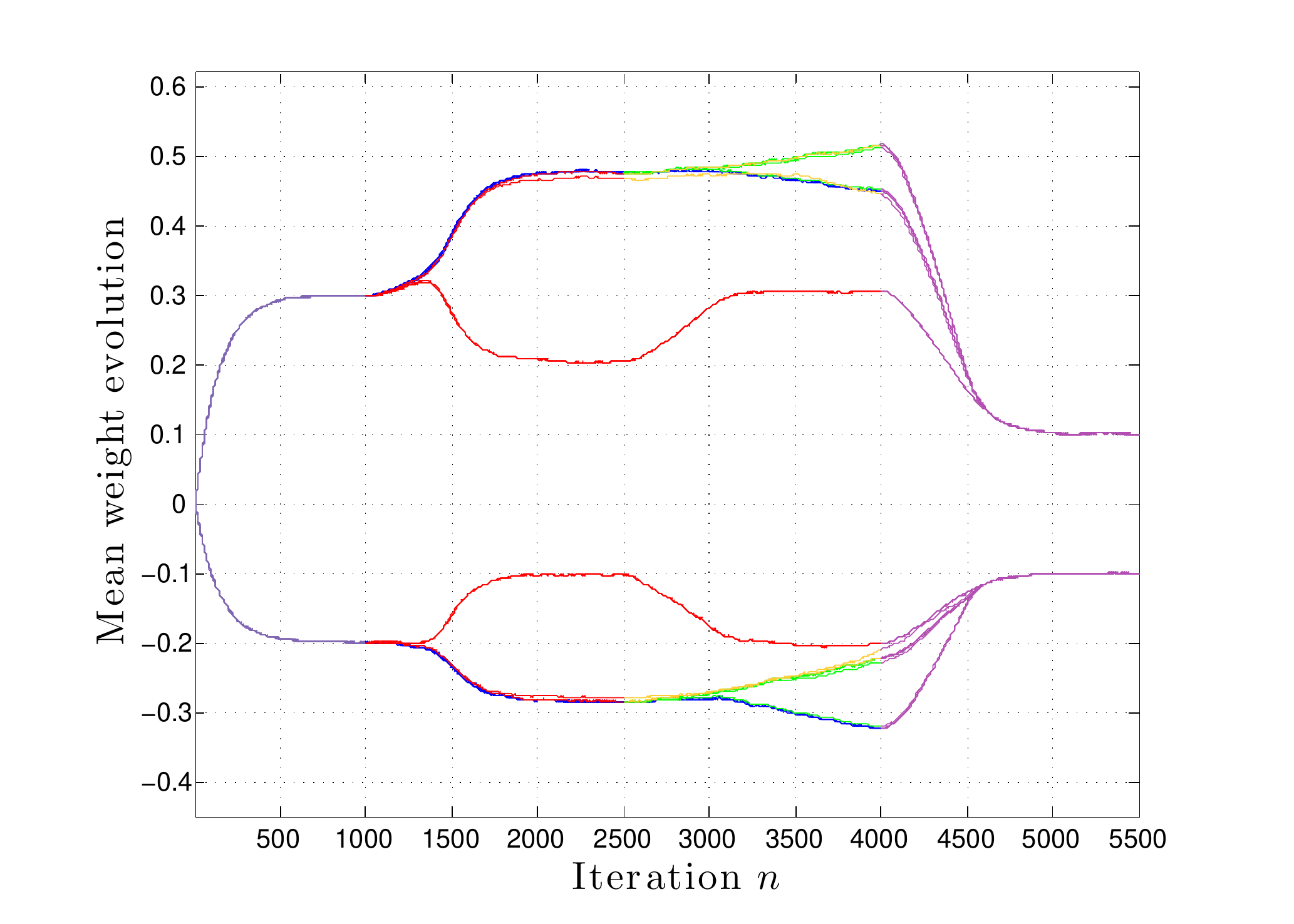} }
	\subfigure[Proposed with $\bC(n) = \bA^\top\!(n)$.]{\hspace{-1mm}
	\includegraphics[trim = 15mm -10mm 22mm 10mm, clip, scale=0.25]{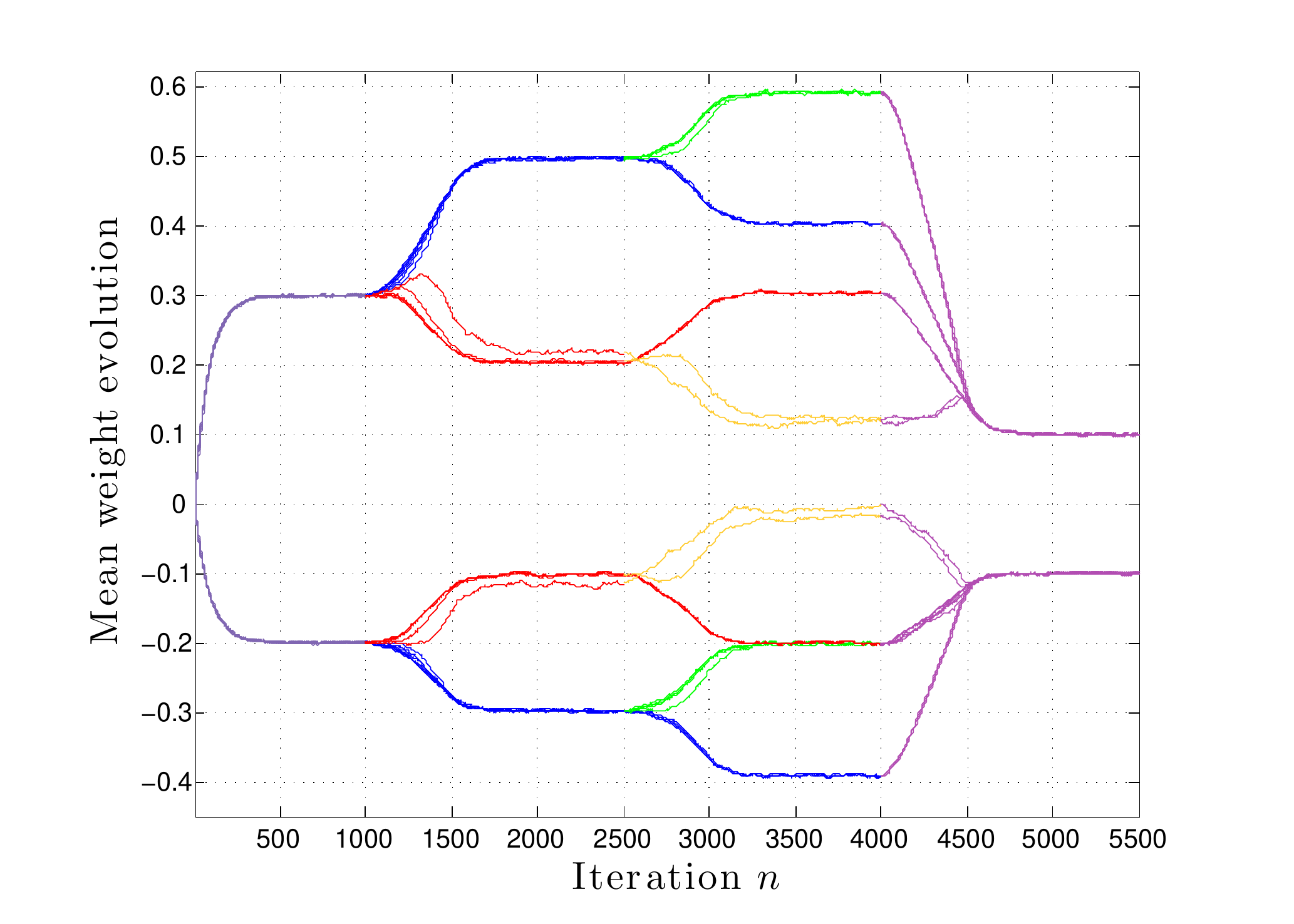}}
	\vspace{-3mm}
	\caption{Mean weight behavior of various algorithms in the non-stationary environment. Colors
	are consistent with cluster colors in Fig.~\ref{fig: multi_topo_real}.}
	\label{fig:Nst_weights_est}
	\vspace{-3mm}
\end{figure*}

\begin{figure}[!t]
\centering
	\! \includegraphics[trim = 5mm 0mm 0mm 0mm, clip, scale=0.5]{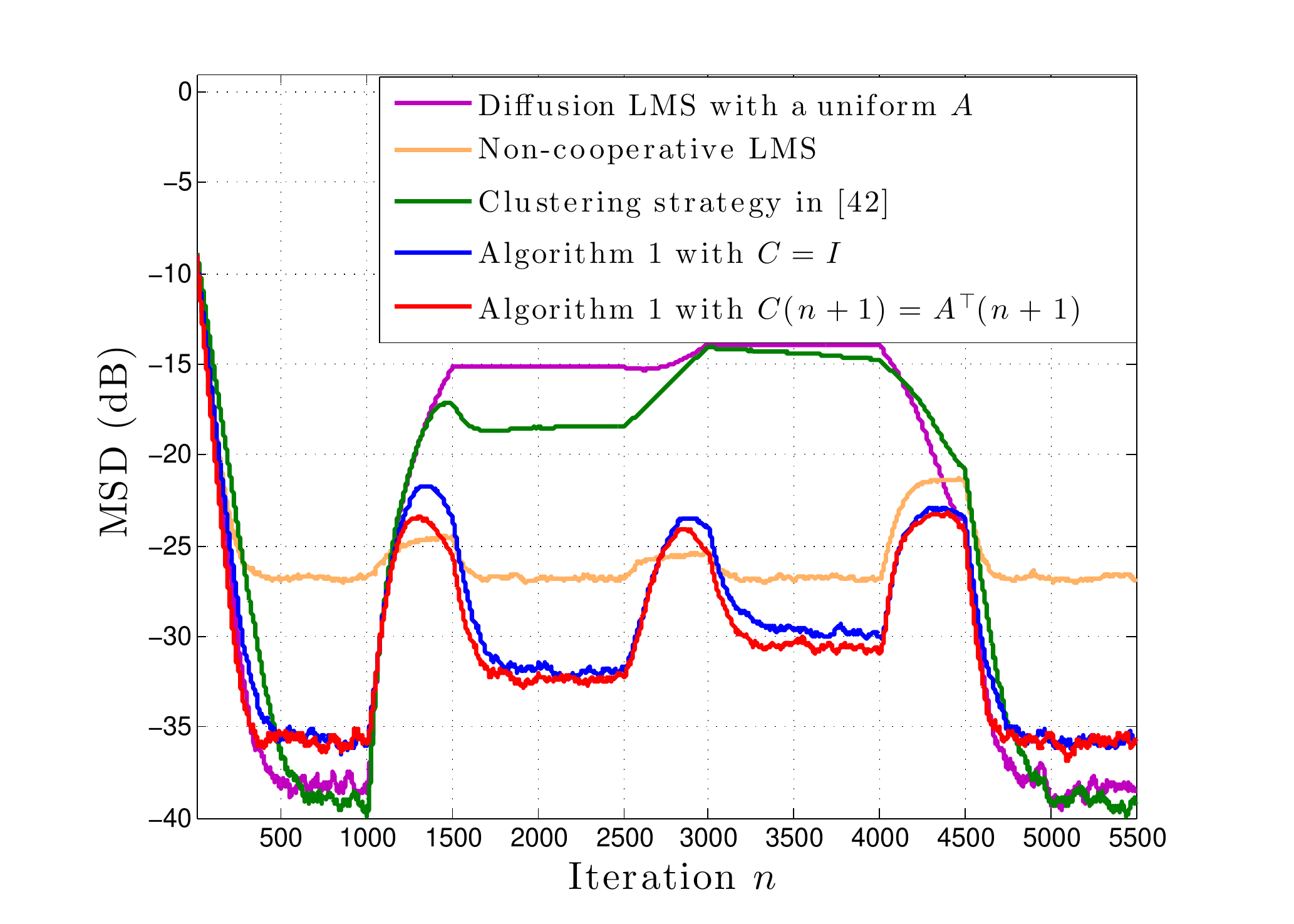}
	\caption{Network MSD behavior comparison in the time variant multitask environment.}
	\label{fig:Nst_MSD}
\end{figure}

\subsubsection{Large network and high-dimensional regressors}
{For the sake of simplicity, previous experiments were conducted with relatively small networks and low-dimensional optimum parameter vectors. A network consisting of two clusters with $50$ nodes in each cluster was randomly deployed in a given area, with physical connections defined by the connectivity matrix in Fig.~\ref{fig:clst100.cnt}. The optimum parameter vectors were set as follows: $\bw^\star_k = \cb{1}_{50}$ for $k = 1, \dots, 50$, and $\bw^\star_k = -\cb{1}_{50}$ for $k = 51, \dots, 100$. The regression inputs $\bx_{k}(n)$ were zero-mean $50\times 1$ random vectors governed by a Gaussian distribution with covariance matrices $\bR_{x,k} = \sigma_{x,k}^2\,\bI_L$. The background noises $z_k(n)$ were i.i.d. zero-mean Gaussian random variables, and independent of any other signal. The variances $\sigma_{x,k}^2$ and $\sigma_{z,k}^2$ were uniformly sampled in $[0.8,1.2]$ and $[0.018, 0.022]$, respectively. For all nodes, the step-sizes were set to $\mu_k = 0.01$. The same four algorithms as before were considered. Our algorithm was used with the normalized gradient $\cb{q}_k(n)/(\|\cb{q}_k(n)\|+\xi)$ and $\xi=0.01$. MSD learning curves are {shown} in Fig.~\ref{fig:clst100.MSD}, and the connectivity matrix determined by our algorithm is represented in Fig.~\ref{fig:clst100.w}. It can be observed that the performance of our algorithm is better than that of other methods.}
%

\begin{figure*}[!t]  \addtolength{\subfigcapskip}{-0.2in}
	\centering
          \subfigure[MSD learning curves.]{\hspace{-1mm}\label{fig:clst100.MSD}
	\includegraphics[trim = 170mm 100mm 180mm 110mm, clip, scale=0.33]{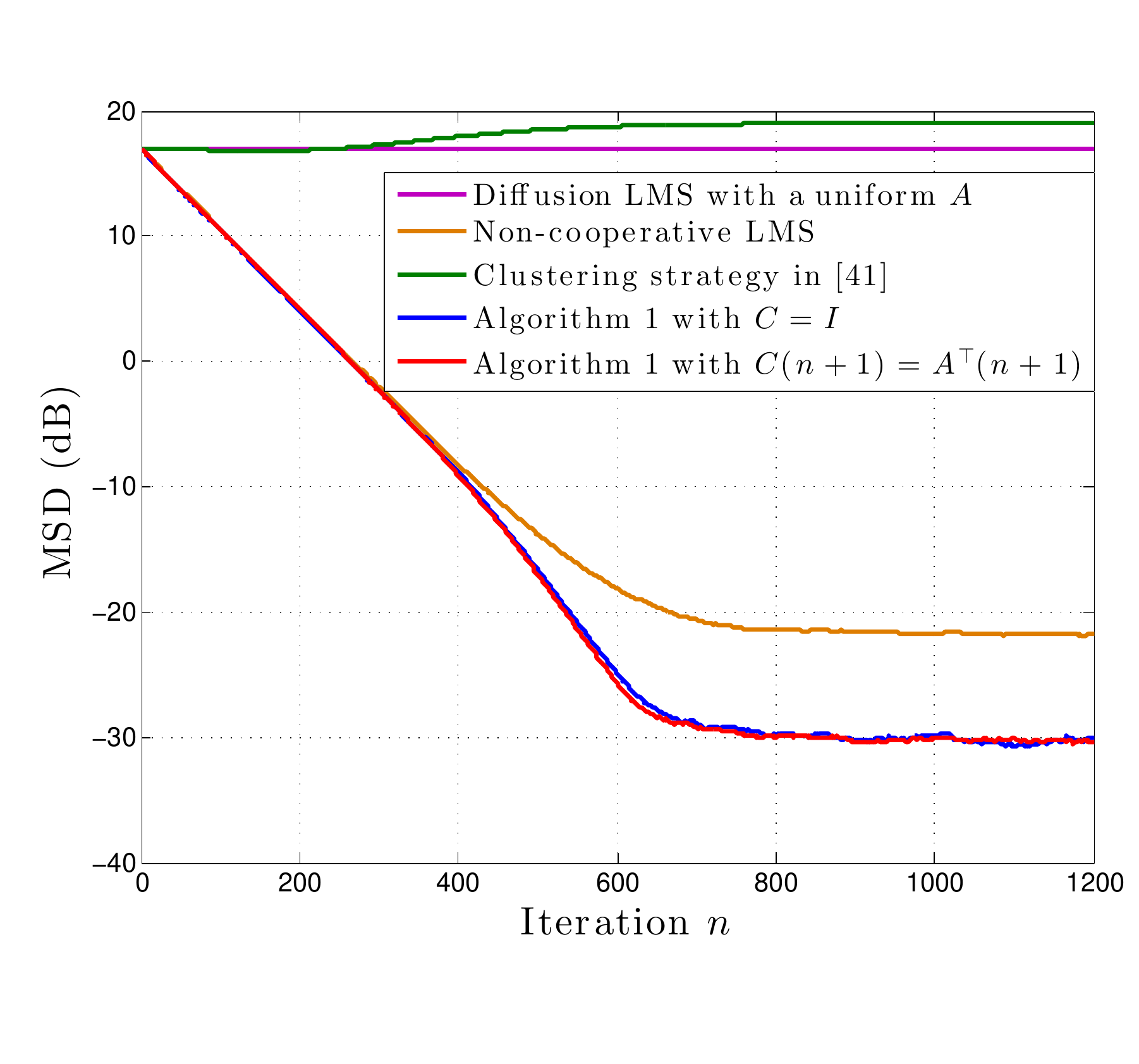}}
         \subfigure[Network physical connection matrix.]{\hspace{-1mm} \label{fig:clst100.cnt}
	\includegraphics[trim = 45mm 10mm 42mm 30mm, clip, scale=0.33]{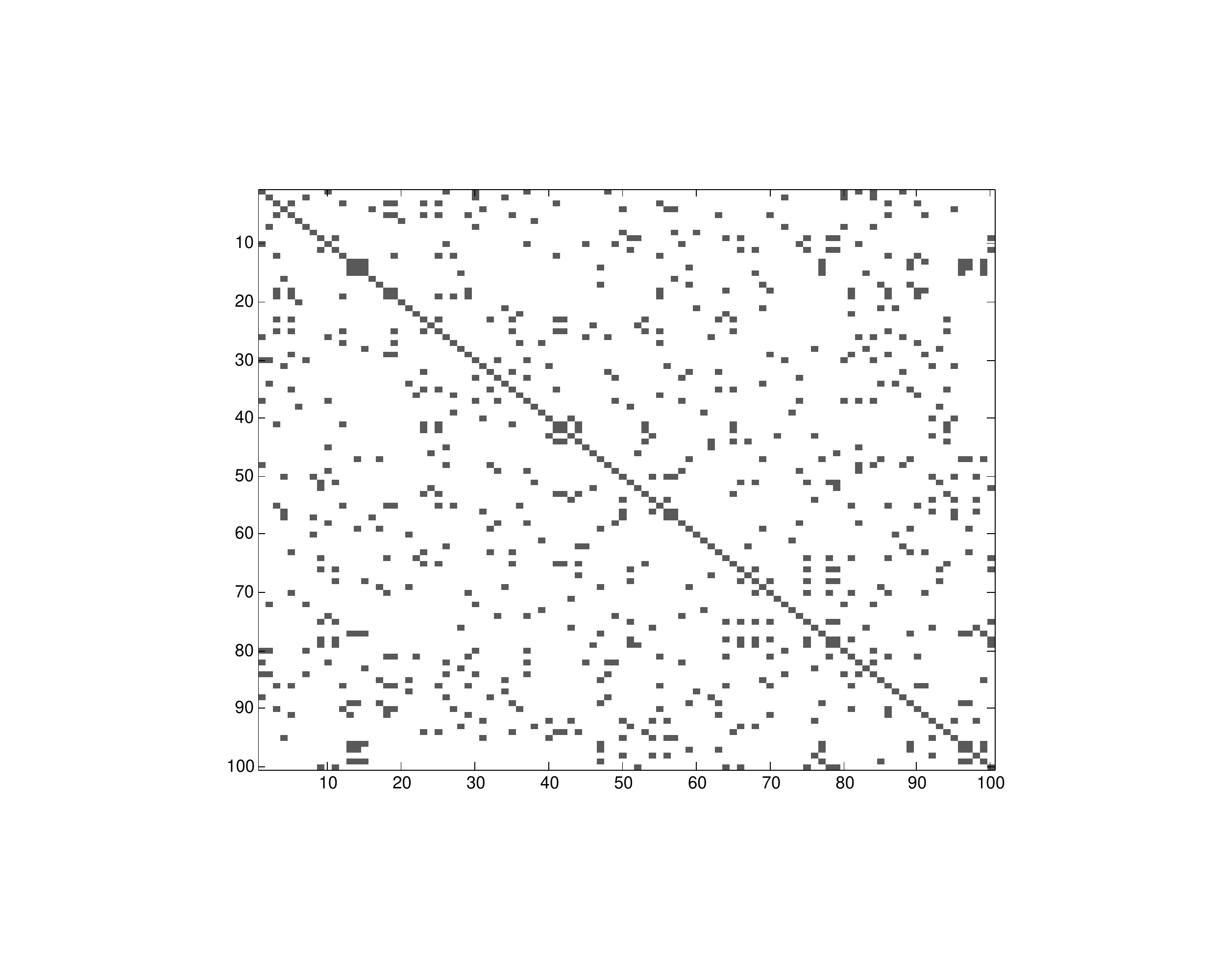} }
	\subfigure[Connection selected by the algorithm.]{\hspace{-1mm} \label{fig:clst100.w}
	\includegraphics[trim = 45mm 10mm 42mm 30mm, clip, scale=0.33]{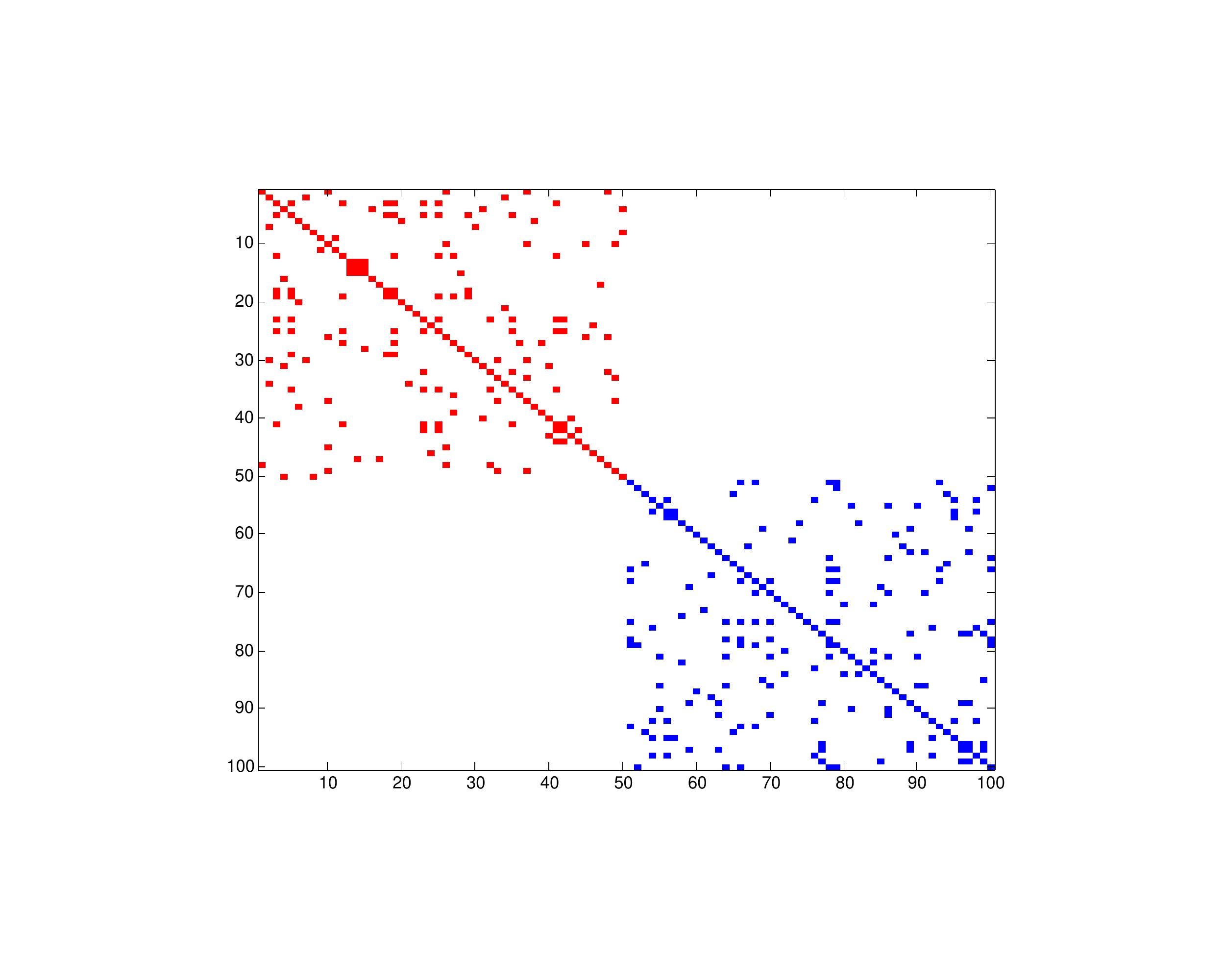} }
	\caption{Simulation results on a large network ($N=100$) with high-dimensional regressors $(L=50)$. 
	(a) Comparison of MSD learning curves.
	(b) Initial connectivity matrix, where gray elements represent physical connections. 
	(c) Connectivity matrix resulting from our clustering strategy.}
	\label{fig:clst100}
	\vspace{-3mm}
\end{figure*}

\subsection{{Collaborative target tracking over sensor networks}}

Consider now a target tracking problem to illustrate our adaptive clustering strategy with diffusion LMS. We focused on a scenario involving four targets, numbered from $i=1$ to~$4$, moving according to the state-transition equation
\begin{equation}
           \begin{split}
           &\bx_i(n+1) = \bT^\star_i\;\bx_i(n) + \cb{z}_i(n) \\ 
           &      \text{for }\; i = 1, \dots, 4,  \quad  n = 0,\dots, 100
           \end{split}
\end{equation}
where $\bx_i(n)$ is the 2-dimensional coordinates for target $i$ at instant $n$. Matrices $\bT_i$ are $2\times 2$  state-transition matrices that were set to
\begin{equation}
	\label{eq:transT}
	\begin{split}
	&\bT^\star_1 \!=\! \left(\begin{array}{cc} 1.0019 & -0.0129 \\ 0.0187 & \phantom{-}1.0034\end{array}\right), \,
	\bT^\star_2 \!=\! \left(\begin{array}{cc} 1.0149 & -0.0014 \\ 0.0033 & \phantom{-}1.0034\end{array}\right), \\
	&\bT^\star_3 \!=\! \left(\begin{array}{cc} 1.0128 & -0.0041 \\ 0.0156 & \phantom{-}1.0086\end{array}\right), \,
	\bT^\star_4 \!=\! \bT^\star_1
	\end{split}
\end{equation}
and $\cb{z}_i(n)$ is the modeling error with i.i.d. zero-mean Gaussian distribution with covariance matrix $\sigma_z^2\,\bI$. The standard deviation was set to $\sigma_z=0.01$.  The initial coordinates for the four targets were
\begin{equation}
       \label{eq:initx}
       \bx_1(0) = \bx_2(0)=\bx_3(0) = [1\,-1]^\top, \quad  \bx_4(0) = [1.2\,-1.2]^\top.
\end{equation}
{Figure~\ref{fig:trajectory} shows} the trajectories of the four targets from instant $n = 0$ to $100$. A network with $N=100$ nodes was randomly deployed in a given area, with physical connections defined by the connectivity matrix in Fig.~\ref{fig:CntMtx}.

We supposed that each node was able to track only one target during the experiment, with noisy observations $\tilde{\bx}_k(n)$:
\begin{equation}
      \label{eq:obs}
      \tilde{\bx}_k(n) = \bx_k(n) + \bu_k(n)  \qquad\text{for } \; k = 1, \dots, N
\end{equation}
with $\bu_k(n)$ an i.i.d. zero-mean Gaussian observation noise with covariance matrix $\sigma_u^2\bI$ and standard deviation $\sigma_u=0.01$. For ease of presentation,  we assumed that nodes $1$--$25$ tracked target $1$, nodes $26$--$50$ tracked target $2$, nodes $51$--$75$ tracked target $3$, and nodes $76$--$100$ tracked target $4$.

Considering $\tilde{\bx}(n)$ as input data and $\tilde{\bx}(n+1)$ as {the} desired output data for the learning algorithm, {each node was aimed} to track a target or, equivalently, to estimate its transition matrix given input-output noisy data. Without cooperation, this task can be performed by each node $k$ by minimizing the following cost function with respect to matrix $\bT_k$:
\begin{equation}
	J_k({\bT_k}) =   \mathbb{E} \| \tilde{\bx}_k(n+1) -{\bT_k}\,\tilde{\bx}_k(n)\|^2 \qquad \text{for }  k = 1,\dots, N.
\end{equation}
Collaboration among nodes may be beneficial as several nodes are conducting the same task, including nodes that track the same target and nodes that track distinct targets with the same state-transition matrix. Clearly, diffusion LMS with a uniform combination matrix is not suitable within this context since neighboring nodes may not have the same task to conduct. This problem requires adaptive clustering to automatically aggregate
nodes that perform a similar task.

{Algorithm~\ref{algo:diffLMS.mult} was run with $\bC = \bI_N$ and was initialized with $\bT_k(0) = \bI_2$.}
%
The step-size $\mu$ was set {equal} to $\mu=0.05$. Figure~\ref{fig:trackMSD} shows the MSD learning curves of transition matrices estimated by non-cooperative LMS, and diffusion LMS with adaptive clustering {strategy. 
 The} performance gain can be clearly seen from these figures. Figure~\ref{fig:EstMtx} shows the connectivity matrix determined by the clustering strategy at iteration $n=100$. {Gray elements in this figure represent the combination weights $a_{\ell k}$ that are larger than $0.05$.}  It can be seen that connections are distributed in $4$ blocks on the diagonal, each one corresponding to a target, and 2 other blocks (upper-right and lower-left ones) where nodes track two distinct targets with the same state-transition {matrix.} 

\begin{figure*}[!t]
	\centering
	\subfigure[Target trajectories.]{	\label{fig:trajectory}
		\includegraphics[trim = 0mm 78mm 0mm 70mm, clip, scale=0.38]{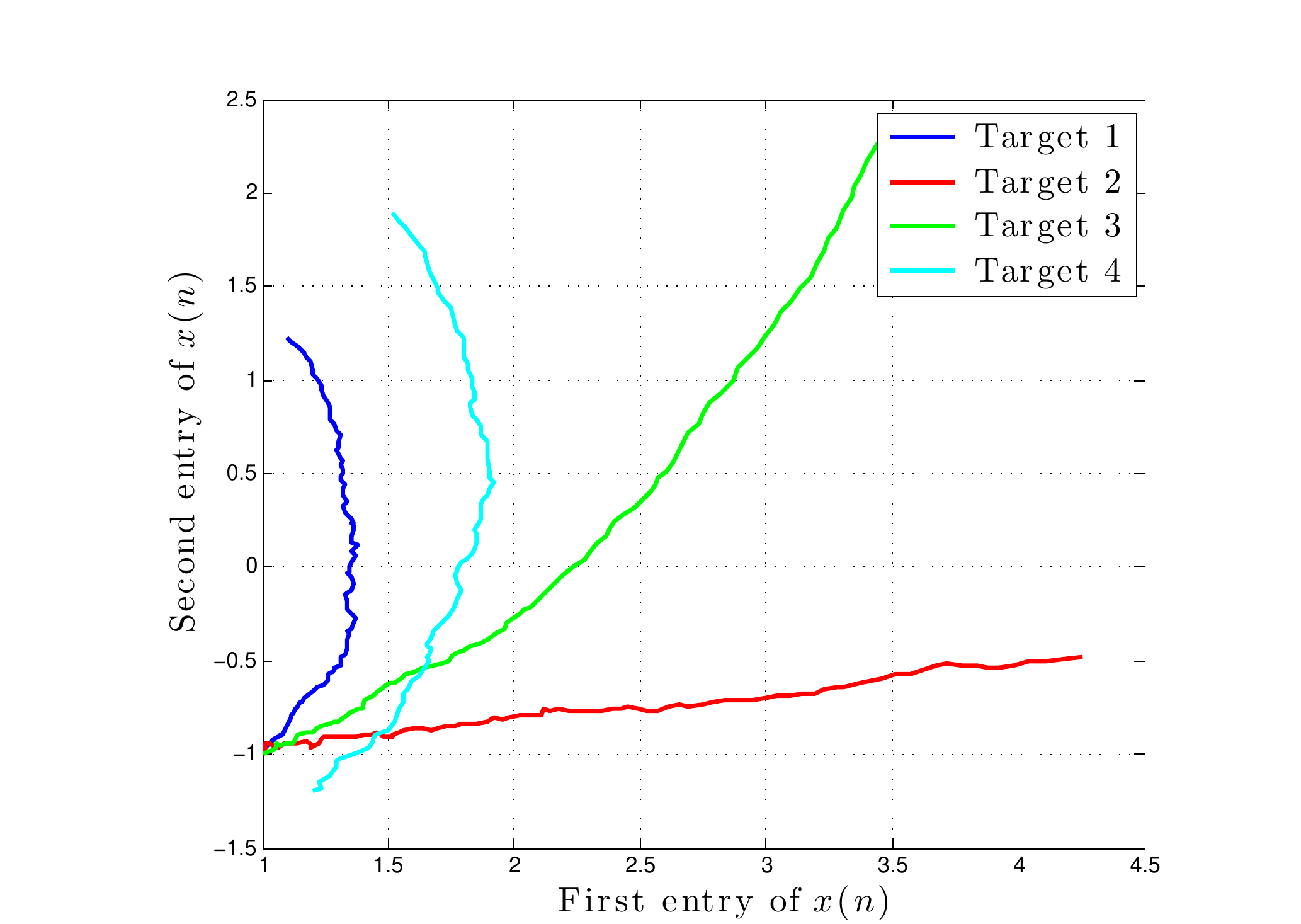}}
	\subfigure[MSD learning curves.]{	\label{fig:trackMSD}
		\includegraphics[trim = 0mm 78mm 0mm 70mm, clip, scale=0.38]{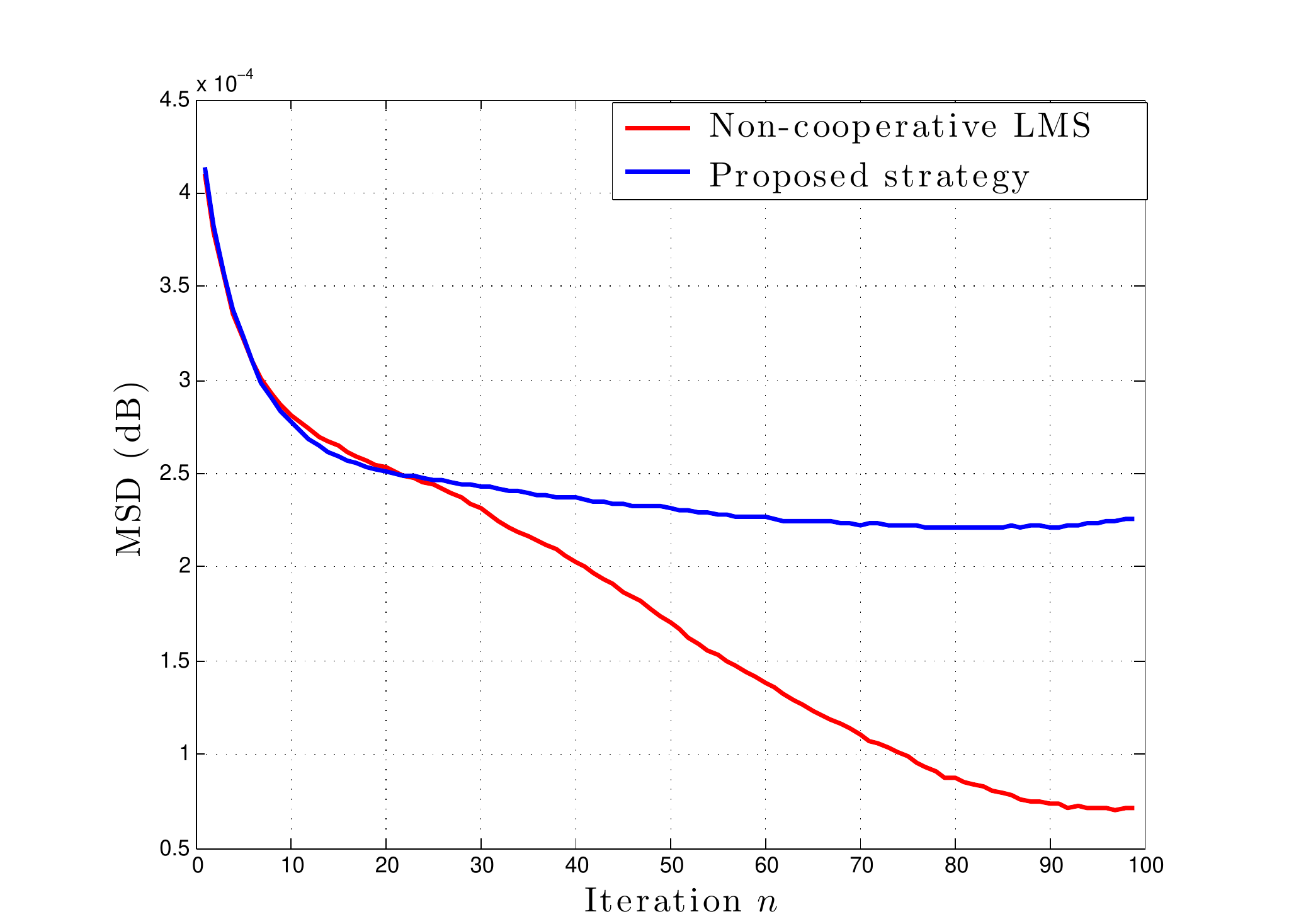}}
		\vspace{-3mm}
	\caption{(a) Trajectories of four targets with initial coordinates $\bx_1(0) = \bx_2(0)=\bx_3(0) = [1,  -1]^\top,  \bx_4(0) = [1.2,  -1.2]^\top$,
	from $n=0$ to $100$. (b) Comparison of MSD learning curves {for the estimate $\bT_k$}.}
\end{figure*}
\begin{figure*}[!t]
\centering
	\subfigure[Network physical connection matrix.]{	\label{fig:CntMtx}
		\includegraphics[trim = 0mm 10mm 10mm 10mm, clip, scale=0.38]{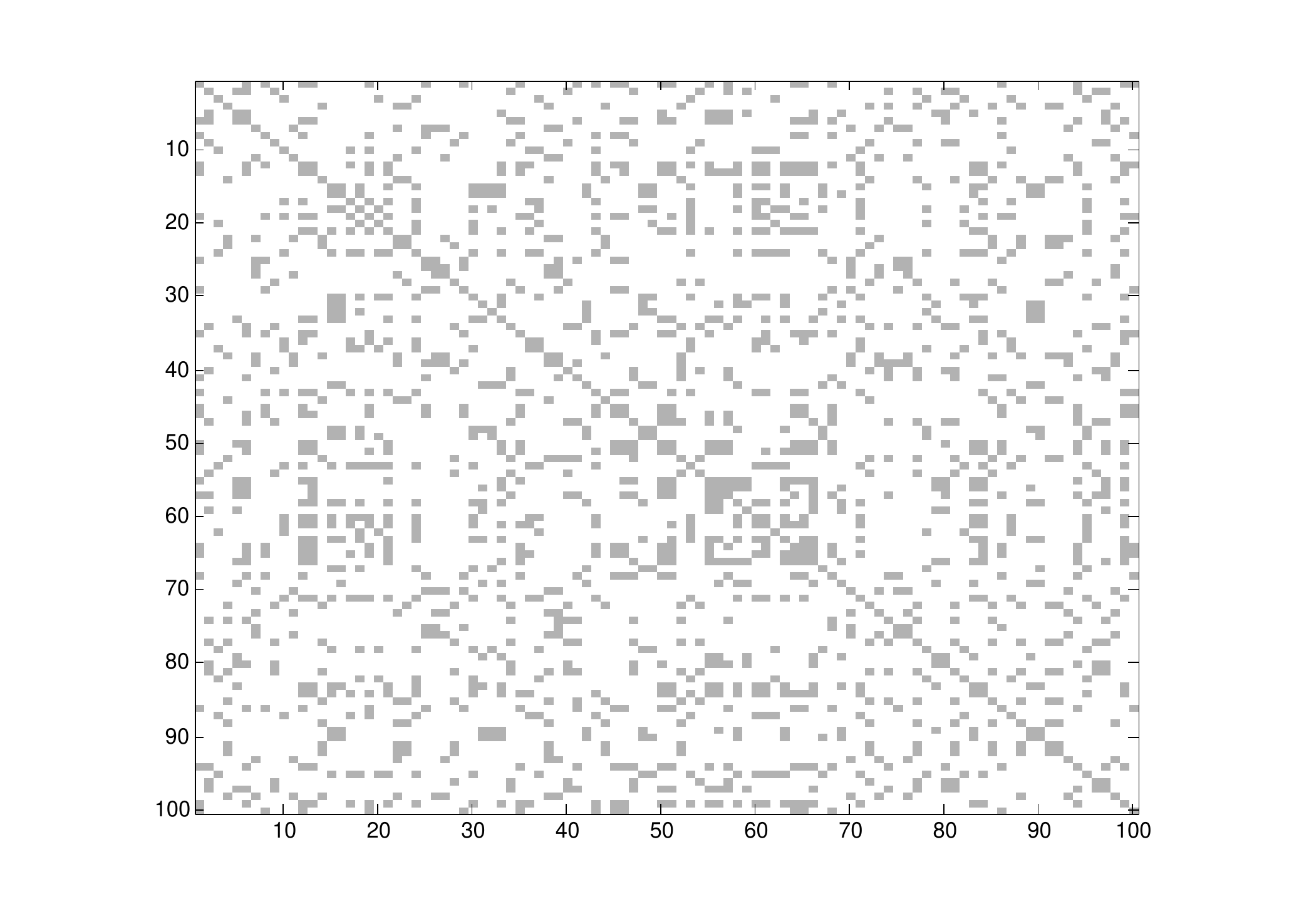}  }
	\subfigure[Connections selected collaboratively by the algorithm.]{\label{fig:EstMtx}
		\includegraphics[trim = 0mm 10mm 10mm 10mm, clip, scale=0.38]{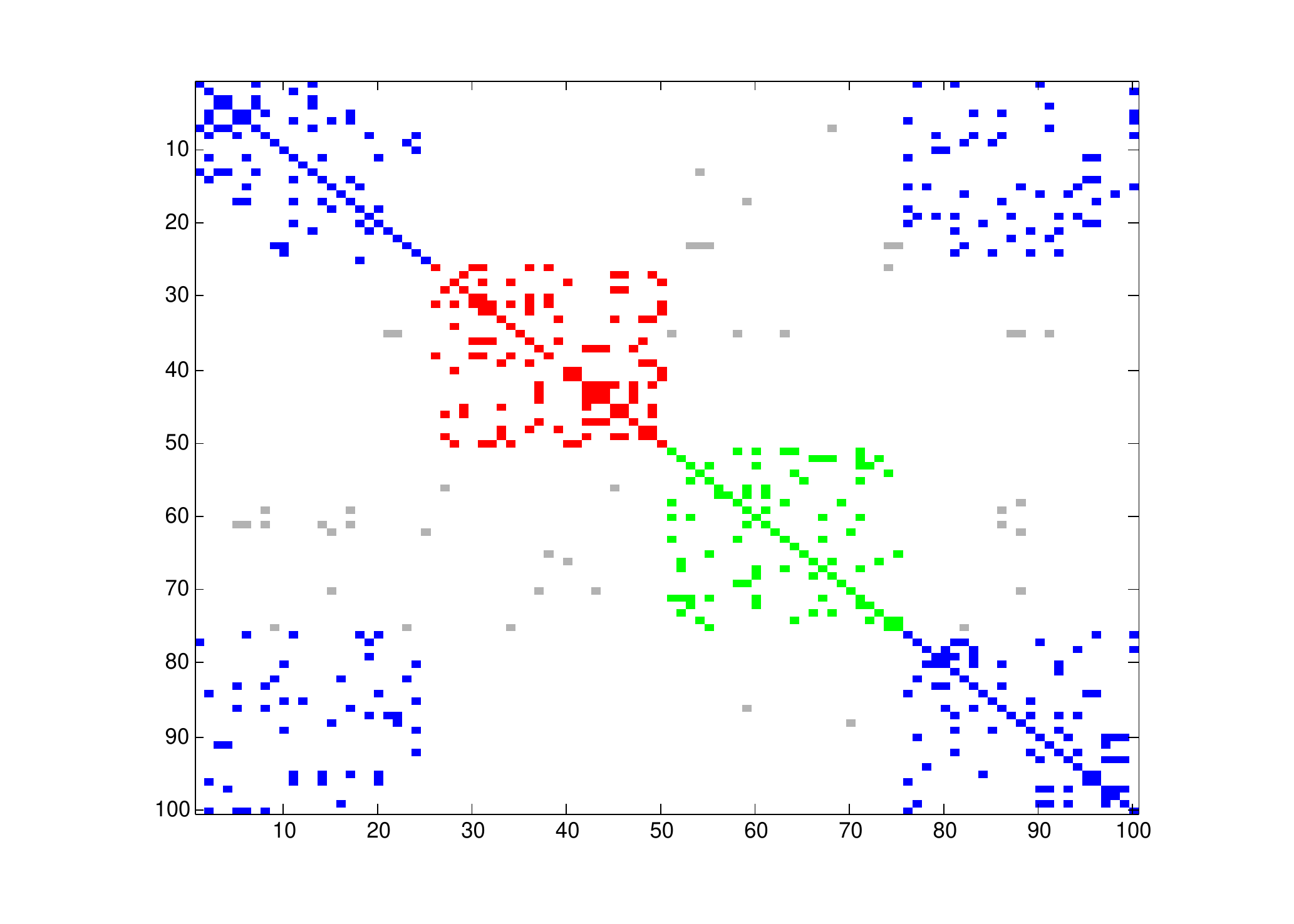} }
	\vspace{-2mm}	
	\caption{(a) Initial connectivity matrix, where gray elements represent physical connections. (b) Connectivity matrix resulting from our clustering
	strategy. Blue elements correspond to the connections used to estimate the transition matrices $\bT_1^\star=\bT_4^\star$. 
	Red and green elements correspond to the connections used to estimate the transition matrices $\bT_2^\star$ and $\bT_3^\star$, respectively.
	Gray elements can be considered as false connections because they involve nodes that do not estimate the same transition matrix.}
	 \vspace{-5mm}
\end{figure*}

\section{Conclusion and perspectives}

{Many practical} problems of interest happen to be multitask-oriented in the sense that there are multiple optimum parameter vectors to be inferred {simultaneously}. In this paper, we studied the performance of the {single-task} diffusion LMS algorithm when it is {run in} a multitask environment. Accurate mean weight behavior model and mean square deviation model were derived. Next, we proposed an unsupervised clustering strategy that allows each node to {select the} neighboring nodes with which it can collaborate to address a given task. Simulations were presented to  demonstrate the efficiency of the proposed clustering {strategy.}

\bibliographystyle{IEEEbib}
\bibliography{ref_abbr}

\end{document}